\def\av#1{\langle#1\rangle}          
\newcommand{\omitit}[1]{}

\documentclass[aps,pre,preprint, showpacs, showkeys,superscriptaddress]{revtex4}
\usepackage[dvips]{graphicx}
\usepackage{latexsym}
\usepackage{natbib}

\begin{document}
\begin{center}
{\bf \large Optimal Path and Minimal Spanning Trees in Random
Weighted Networks}
\end{center}
\tableofcontents
\newpage
\title{Optimal Path and Minimal Spanning Trees in Random
Weighted Networks}
\author{Lidia A. Braunstein}
\affiliation{Departamento de F\'{\i}sica, Facultad de Ciencias Exactas y
  Naturales, Universidad Nacional de Mar del Plata, Funes 3350, 7600 Mar del
  Plata, Argentina}
  \email{lbrauns@mdp.edu.ar}
\affiliation{Center for Polymer Studies, Boston University,
Boston,
  Massachusetts 02215, USA}
\author{Zhenhua Wu}
\affiliation{Center for Polymer Studies, Boston University, Boston,
  Massachusetts 02215, USA}
  \author{Yiping Chen}
\affiliation{Center for Polymer Studies, Boston University,
Boston,
  Massachusetts 02215, USA}
\author{Sergey V. Buldyrev}
\affiliation{Center for Polymer Studies, Boston University, Boston,
  Massachusetts 02215, USA}
\affiliation{Department of Physics Yeshiva University, 500 West
185th Street Room 1112,NY, 10033, USA}
\author{Tomer Kalisky}
\affiliation{Minerva Center and Department of Physics, Bar-Ilan
University, 52900 Ramat-Gan, Israel}
\author{Sameet Sreenivasan}
\affiliation{Center for Polymer Studies, Boston University, Boston,
  Massachusetts 02215, USA}
\author{Reuven Cohen}
\affiliation{ Dept. of Electrical and Computer Engineering, Boston
University, Boston,
  Massachusetts 02215, USA}
\affiliation{Minerva Center and Department of Physics, Bar-Ilan
University, 52900 Ramat-Gan, Israel}
\author{Eduardo L\'opez}
\affiliation{Center for Polymer Studies, Boston University, Boston,
  Massachusetts 02215, USA}
  \affiliation{Theoretical Division, Los Alamos National Laboratory,
Mail Stop B258, Los Alamos, NM 87545 USA}
\author{Shlomo Havlin}
\affiliation{Minerva Center and Department of Physics, Bar-Ilan
University, 52900 Ramat-Gan, Israel}
\affiliation{Center for
Polymer Studies, Boston University, Boston,
  Massachusetts 02215, USA}
\author{H. Eugene Stanley}
\affiliation{Center for Polymer Studies, Boston University, Boston,
  Massachusetts 02215, USA}

 \begin{abstract}

  We review results on the scaling of the optimal path length
  $\ell_{\mbox{\scriptsize opt}}$ in random networks with weighted links or
  nodes. We refer to such networks as ``weighted'' or ``disordered''
  networks.  The optimal path is the path with minimum sum of the weights.
  In strong disorder, where the maximal weight along the path dominates the
  sum, we find that $\ell_{\mbox{\scriptsize opt}}$ increases dramatically
  compared to the known small world result for the minimum distance
  $\ell_{\mbox{\scriptsize min}} \sim \log N$, where $N$ is the number of
  nodes.  For Erd\H{o}s-R\'enyi (ER) networks $\ell_{\mbox{\scriptsize
  opt}}\sim N^{1/3}$, while for scale free (SF) networks, with degree
  distribution $P(k) \sim k^{-\lambda}$, we find that
  $\ell_{\mbox{\scriptsize opt}}$ scales as $N^{(\lambda - 3)/(\lambda - 1)}$
  for $3<\lambda<4$ and as $N^{1/3}$ for $\lambda\geq 4$.  Thus, for these
  networks, the small-world nature is destroyed. For $2 < \lambda < 3$ in
  contrary, our numerical results suggest that $\ell_{\mbox{\scriptsize
  opt}}$ scales as $\ln^{\lambda-1}N$, representing still a small world.  We
  also find numerically that for weak disorder $\ell_{\mbox{\scriptsize
  opt}}\sim\ln N$ for ER models as well as for SF networks.  We also review
  the transition between the strong and weak disorder regimes in the scaling
  properties of $\ell_{\rm opt}$ for ER and SF networks and for a general
  distribution of weights $\tau$, $P(\tau)$. For a weight distribution of the
  form $P(\tau) = 1/(a\tau)$ with ($\tau_{\rm min}< \tau < \tau_{\rm max}$)
  and $a=\ln{\tau_{\rm max}}/{\tau_{\rm min}}$, we find that there is a
  crossover network size $N^* = N^*(a)$ at which the transition occurs. For
  $N \ll N^*$ the scaling behavior of $\ell_{\rm opt}$ is in the strong
  disorder regime, while for $N \gg N^*$ the scaling behavior is in the weak
  disorder regime.  The value of $N^*$ can be determined from the expression
  $\ell_\infty(N^*)=a p_c$, where $\ell_\infty$ is the optimal path length in the
  limit of strong disorder, $A\equiv a p_c\to \infty$ and $p_c$ is the
  percolation threshold of the network.  We suggest that for any $P(\tau)$
  the distribution of optimal path lengths has a universal form which is
  controlled by the scaling parameter $Z=\ell_{\infty}/A$ where $A \equiv
  p_c\tau_c/ \int_0^{\tau_c}\tau P(\tau) d \tau$ plays the role of the
  disorder strength and $\tau_c$ is defined by $\int_0^{\tau_c} P(\tau) d
  \tau =p_c$.  In case $P(\tau)\sim 1/( a\tau)$, the equation for $A$ is reduced
  to $A=a p_c$. The relation for $A$ is derived analytically and supported by
  numerical simulations for Erd\H{o}s-R\'enyi and scale-free graphs. We also
  determine which form of $P(\tau)$ can lead to strong disorder $A\to
  \infty$.  We then study the minimum spanning tree (MST), which is the
  subset of links of the network connecting all nodes of the network such
  that it minimizes the sum of their weights.  We show that the minimum
  spanning tree (MST) in the strong disorder limit is composed of percolation
  clusters, which we regard as "super-nodes", interconnected by a scale-free
  tree. The MST is also considered to be the skeleton of the network where
  the main transport occurs. We furthermore show that the MST can be
  partitioned into two distinct components, having significantly different
  transport properties, characterized by centrality --- number of times a
  node (or link) is used by transport paths. One component the {\it
  superhighways}, for which the nodes (or links) with high centrality
  dominate, corresponds to the largest cluster at the percolation threshold
  (incipient infinite percolation cluster) which is a subset of the MST.  The
  other component, {\it roads}, includes the remaining nodes, low centrality
  nodes dominate.  We find also that the distribution of the centrality for
  the incipient infinite percolation cluster satisfies a power law, with an
  exponent smaller than that for the entire MST. We demonstrate the
  significance identifying the superhighways by showing that one can improve
  significantly the global transport by improving a very small fraction of
  the network, the superhighways.

\end{abstract}
\maketitle
\section{Introduction}

Recently much attention has been focused on the topic of complex networks
which characterize many biological, social, and communication systems
\cite{Albert02,DM02,PastorXX}. The networks are represented by nodes
associated to individuals, organizations, or computers and by links
representing their interactions.  The classical model for random networks is
the Erd\H{o}s-R\'enyi (ER) model \cite{ER59,ER60,Bollobas}.  An important
quantity characterizing networks is the average distance (minimal hopping)
$\ell_{\mbox{\scriptsize min}}$ between two nodes in the network of total $N$
nodes. For the Erd\H{o}s-R\'enyi network $\ell_{\mbox{\scriptsize min}}$
scales as $\ln N$ \cite{Bollobas}, which leads to the concept of ``small
worlds'' or ``six degrees of separation''. For scale-free (SF)
\cite{Albert02} networks $\ell_{\mbox{\scriptsize min}}$ scales as $\ln \ln
N$, this leads to the concept of ultra small worlds \cite{Cohen,DM02}.

In most studies, all links in the network are regarded as
identical and thus a crucial parameter for information flow
including efficient routing, searching, and transport is
$\ell_{\mbox{\scriptsize min}}$. In practice, however, the weights
(e.g., the quality or cost) of links are usually not
equal~\cite{vespignani_pnas,report-latora}.

Thus the length of the optimal path $\ell_{\rm opt}$, minimizing
the sum of weights, is usually longer than
$\ell_{\mbox{\scriptsize min}}$. For example, the cost could be
the time required to transit the link.  There are often many
traffic routes from site A to site B with a set of transit time
$\tau_i$, associated with each link along the path. The fastest
(optimal) path is the one for which $\sum_i\tau_i$ is a minimum,
and often the optimal path has more links than the shortest path.
In many cases, the selection of the path is controlled by most of
the weights (e.g., total cost) contributing to the sum.  This case
corresponds to weak disorder (WD).  However, in other cases, for
example when the distribution of disorder is very broad a {\it
single} weight dominates the sum. This situation---in which one
link controls the selection of the path---is called the strong
disorder limit (SD).

For a recent
quantitative criterion for SD and WD, see Ref~\cite{yipingprl} and
Section \ref{sec.tws}(B) in this article.

The strong disorder is  relevant {\it e.g.} for computer and
traffic networks, since the slowest link in communication networks
determines the connection speed. An example for SD is when a
transmission at a constant high rate is needed (e.g., in
broadcasting video records over the Internet). In this case the
narrowest band link in the path between the transmitter and
receiver controls the rate of transmission. This limit is also
called the ``ultrametric'' limit and we refer to the optimal path
in this limit as the min-max path.

The SD limit is also related to the minimal spanning tree which
includes all optimal paths between all pairs of sites in the
network. The disorder on a network is usually implemented from a
distribution $P(\tau) \sim 1/(a \tau)$, where $1 < \tau <
e^a$~\cite{Porto99,Brauns01,Cieplak,Brauns03}. We assign to each
link of the network a random number $r$, uniformly distributed
between 0 and 1. The cost associated with link $i$ is then
$\tau_i\equiv\exp(ar_i)$ where $a$ is the parameter which controls
the broadness of the distribution of link costs. The parameter $a$
represents the strength of disorder. The limit
$a\rightarrow\infty$ is the strong disorder limit, since for this
case clearly only one link dominates the cost of the path. The
strong disorder limit (SD) can be implemented in a disordered
media by assigning to each link a potential barrier $\epsilon_i$
so that $\tau_i$ is the time to cross this barrier in a thermal
activation process. Thus $\tau_i=e^{\epsilon_i/K T}$, where $K$ is
the Boltzmann constant and $T$ is absolute temperature. The
optimal path corresponds to the minimum $(\sum_i\tau_i)$ over all
possible paths.  We can define disorder strength $a= 1/K T$.  When
$a \to\infty$, only the largest $\tau_i$ dominates the sum. Thus,
$T\to 0$ (very low temperature) corresponds to the strong disorder
limit.

There are distinct scaling relationships between the length of the
average optimal path $\ell_{\rm opt}$ and the network size (number
of nodes) $N$ depending on whether the network is strongly or
weakly disordered \cite{Porto99, Brauns03}. It was shown using
percolation arguments (See Section~\ref{sec.sd}) that for strong
disorder \cite{Brauns03}, $\ell_{\rm opt} \sim N^{\nu_{\rm opt}}$,
where $\nu_{\rm opt} = 1/3$ for Erd\H{o}s-R\'enyi (ER) random
networks \cite{ER59} and for scale-free (SF) \cite{Albert02}
networks with $\lambda > 4$, where $\lambda$ is the exponent
characterizing the power law decay of the degree distribution. For
SF networks with $3 < \lambda < 4$, $\nu_{\rm
opt}=(\lambda-3)/(\lambda-1)$. For $2<\lambda <3$, percolation
arguments do not work, but the numerical results suggest
$\ell_{\rm opt}\sim \ln^{\lambda-1} N$, which is again much larger
than the ultra small result for the shortest path $\ell_{\rm min}
\sim \ln \ln N$ found for $2<\lambda<3$ in Ref.\cite{Cohen3}. When
the weights are taken from a uniform distribution we are in the
weak disorder limit. In this case $\ell_{\rm opt} \sim \ln N$ for
both ER and SF for all the values of $\lambda$~\cite{Brauns03}.
For $2<\lambda<3$, this result is significantly different from the
ultra small world result found for unweighed networks.

Porto {\it et al} .~\cite{Porto99} considered the optimal path
transition from weak to strong disorder for 2-D and 3-D lattices,
and found a crossover in the scaling properties of the optimal
path that depends on the disorder strength $a$, as well as the
lattice size $L$ (see also~\cite{Buldyrev06}). Similar to regular
lattices, there exists for any finite $a$, a crossover network
size $N^*(a)$ such that for $N \ll N^*(a)$, the scaling properties
of the optimal path are in the strong disorder regime while for $N
\gg N^*(a)$, the network is in the weak disorder regime. The
function $N^*(a)$ was evaluated. Moreover, a general criterion to
determine which form of $P(\tau)$ can lead to strong disorder, and
a general condition when strong or weak disorder occurs was found
analytically~\cite{yipingprl}. The derivation was supported by
extensive simulations.

The study of the distribution of the length of the optimal paths in a network
was reported in Ref~\cite{tomerdistributions}. It was found that the
distribution has the scaling form $P(\ell_{\rm opt},N,a) \sim
\frac{1}{\ell_{\infty}} G \left( \frac{\ell_{\rm opt}}{\ell_{\infty}} ,
  \frac{1}{p_c}\frac{\ell_{\infty}}{a} \right)$, where $\ell_{\infty}$ is
$\ell_{\rm opt}$ for $a \to \infty$ and $p_c$ is the percolation
threshold. It was also shown that a single parameter $Z \equiv
\frac{1}{p_c}\frac{\ell_{\infty}}{a}$ determines the functional
form of the distribution. Importantly, it was found
\cite{yipingprl} that for all $P(\tau)$ that possess a strong-weak
disorder crossover, the distributions $P(\ell_{\rm opt} )$ of the
optimal path lengths display the same universal behavior.

Another interesting question is about a possible origin of scale-free degree
distribution with $\lambda=2.5$ in some real world networks.  Kalisky {\it et
  al.,}~\cite{Tomer_grey} introduced a simple process that generates random
scale-free networks with $\lambda=2.5$ from weighted Erd\"{o}s-R\'enyi
graphs~\cite{ER60}. They found that the minimum spanning tree (MST) on an
Erd\"{o}s-R\'enyi graph is composed of percolation clusters, which we regard
as ``super nodes'', interconnected by a scale-free tree with $\lambda = 2.5$.

Known as the tree with the minimum weight among all possible spanning tree,
the MST is also the union of all ``strong disorder'' optimal paths between
any two nodes~\cite{invpercbara, Dob, Cieplak, Porto99, Brauns03, zhenhua}.
As the global optimal tree, the MST plays a major role for transport process,
which is widely used in different fields, such as the design and operation of
communication networks, the travelling salesman problem, the protein
interaction problem, optimal traffic flow, and economic
networks~\cite{Khan_tech, Skiena_book_mst, Fredman_mst, Kruskal_mst,
macdonald-almaas-barabasi-2004:minimum-spanning-trees, mst_eco_1,
mst_eco_2}. One important question in network transport is how to identify
the nodes or links that are more important than others. A relevant quantity
that characterizes transport in networks is the betweenness centrality, $C$,
which is the number of times a node (or link) used by all optimal paths
between all pairs of nodes~\cite{Newman_centrality, Goh_load_prl,
Kim_bc}. For simplicity we call the ``betweenness centrality'' here
``centrality'' and we use the notation ``nodes'' but similar results have
been obtained for links. The centrality, $C$, quantifies the ``importance''
of a node for transport in the network.  Moreover, identifying the nodes with
high $C$ enables us to improve their transport capacity and thus improve the
global transport in the network. The probability density function (pdf) of
$C$ was studied on the MST for both SF~\cite{Barabasi_sf} and ER
~\cite{ER59,ER60} networks and found to satisfy a power law, ${\cal P}_{\rm
MST}(C) \sim C^{-\delta_{\rm MST}}$, with $\delta_{\rm MST}$ close to
$2$~\cite{Goh_centrality, Kim_bc}. However, Ref~\cite{superhighway_prl} found
that a sub-network of the MST~\cite{FN_IIC_isIn_MST}, the infinite incipient
percolation cluster (IIC) has a significantly higher average $C$ than the
entire MST --- i.e., the set of nodes inside the IIC are typically used by
transport paths more often than other nodes in the MST (See
Section~\ref{seq.shw}).  In this sense the IIC can be viewed as a set of {\it
superhighways} (SHW) in the MST.  The nodes on the MST which are not in the
IIC are called {\it roads}, due to their analogy with roads which are not
superhighways (usually used by local residents). Wu {\it et al.}
\cite{highway_prl} demonstrate the impact of this finding by showing that
improving the capacity of the superhighways (IIC) is significantly a better
strategy to enhance global transport compared to improving the same number of
links of the highest $C$ in the MST, although they have higher
$C$~\cite{FN_highest_bc_IIC}. This counterintuitive result shows the
advantage of identifying the IIC subsystem, which is very small and of oder
zero compared to the full network~\cite{FN_iic_mst_mass_ratio}.

\section{Algorithms}\label{sec.alg}
\subsection{Construction of the Networks}\label{sec.draw_net}
To construct an ER network of size $N$ with average node degree
$\langle k \rangle$, we start with $\langle k \rangle N/2$ edges and
randomly pick a pair of nodes from the total possible $N(N-1)/2$ pairs
to connect with an edge. The only condition we impose is that there
cannot be multiple edges between two nodes. When $\langle k \rangle
>1$ almost all nodes of the network will be connected with high
probability.


To generate scale-free (SF) graphs of size $N$, we employ the Molloy-Reed
algorithm \cite{Molloy} : initially the degree of each node is chosen
according to a scale-free distribution, where each node is given a number of
open links or "stubs" according to its degree. Then, stubs from all nodes of
the network are interconnected randomly to each other with two constraints
that there are no multiple edges between two nodes and that there are no
looped edges with identical ends. The exact form of the degree distribution
is usually taken to be
\begin{equation}\label{
Eq.sf} P(k) =c  k ^{- \lambda}   \hspace{1cm}\mbox{$k= m, \cdots
K$}
\end{equation}
where m and K are the minimal and maximal degrees, and $c \approx
(\lambda-1)m^{\lambda-1}$ is a normalization constant . For real
networks with finite size, the highest degree $K$ depends on
network size $N$: $K \approx m N^{1/(\lambda-1)}$, thus creating a
"natural" cutoff for the highest possible degree . When $m > 1$
there is a high probability that the network is fully connected.

\subsection{Dijkstra's algorithm}\label{sec.dij}
The Dijkstra's algorithm \cite{Cormen90} is used in general to find the optimal
path, when the weights are drawn from an arbitrary distribution.  The
search for the optimal path follows a procedure akin to ``burning''
where the ``fire'' starts from our chosen origin. At the beginning,
all nodes are given a distance $\infty$ except the origin which is
given a distance $0$.
At each step we choose the next unburned node which is nearest to the
origin, and ``burn'' it, while updating the optimal distance to all
its neighbors.  The optimal distance of a neighbor is updated only if
reaching it from the current burning node gives a total path length
that is shorter than its current distance.

\subsection{Ultrametric Optimization}\label{sec.um}
\begin{figure}[ht]
\begin{center}
\includegraphics[width=13cm,height=8cm]{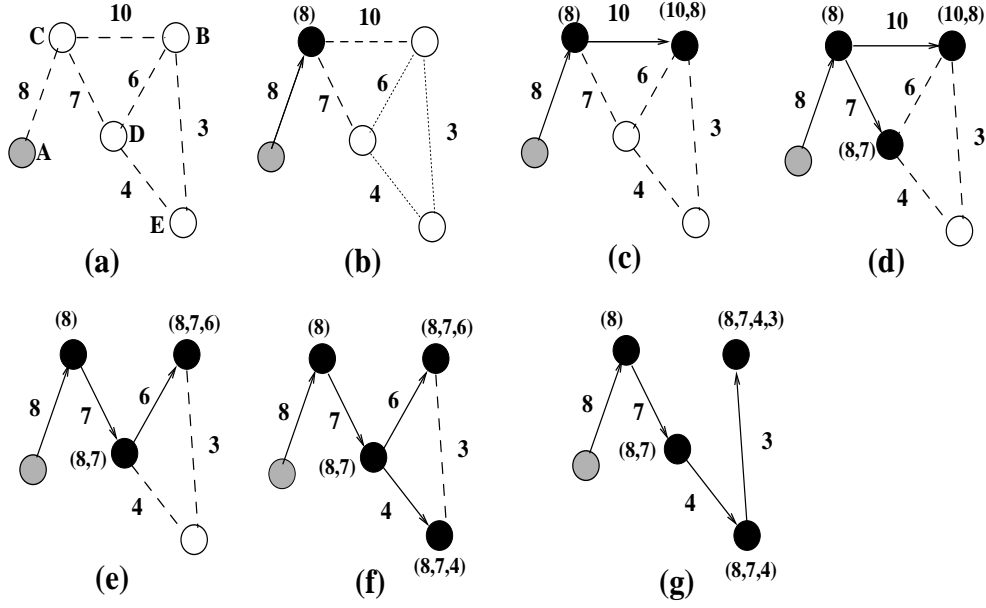}
\end{center}
\caption{In (a) we show schematically a network consisting of five nodes
  (A, B, C, D, and E). The links between them are shown in dashed
  lines. The origin (A) is marked in gray. All links were assigned
  random weights, shown beside the links. In (b) one node (C) has been
  visited for the first time (marked in black) and assigned the
  sequence (8) of length $\ell=1$. The path is marked by a solid
  arrow. Notice that there is no other path going from the origin (A)
  to this node (C) so $\ell_{\mbox{\scriptsize opt}}=1$ for that path.
  In (c) another node (B) is visited for the first time (marked in
  black) and assigned the sequence $(10,8)$ of length $2$. The
  sequence has the information of all the weights of that path
  arranged in decreasing order. In (d) another node (D) is visited for
  the first time and assigned the sequence $(8,7)$ of length $2$. In
  (e), node (B) visited in (c) with sequence $(10,8)$ is visited again
  with sequence $(8,7,6)$. The last sequence is smaller than the
  previous sequence $(10,8)$ so that node (B) is reassigned the
  sequence $(8,7,6)$ of length $3$ [See Eq.~(\ref{Eq.sent})]. The new path is shown as a solid
  line. In (f) a new node (E) is assigned with sequence $(8,7,4)$. In
  (g) node (B) is reached for the third time and reassigned the
  sequence $(8,7,4,3)$ of length $4$. The optimal path for this
  configuration from A to B is denoted by the solid arrows in
  (g) (After \cite{reviewstat}).\label{f.0}}
\end{figure}

Next we describe a numerical method for computing
$\ell_{\mbox{\scriptsize opt}}$ between any two nodes in strong
disorder \cite{Dob,Brauns01}. In this case the sum of the weights
must be completely dominated by the largest weight.  Sometimes
this condition is referred as ultrametric. We can satisfy this
condition assigning weights to all the links $\tau_i=\exp(a r_i)$
choosing $a$ to be so large, that any two links will have
different binary orders of magnitude. For example, if we can
select $0 \leq r_i < 1$ from a uniform distribution, using a
48-bit random number generator, there will be no two identical
values of $r_i$ in a system of any size that we study. In this
case $\Delta r_i\geq 2^{-48}$ and we can select $a\geq 2^{48}\ln
2$ to guarantee the strong disorder limit. To find the optimal
paths under the ultrametric condition, we start from one node (the
origin---see Fig.~\ref{f.0}) and visit all the other nodes
connected to the origin using a burning algorithm. If a node at
distance $\ell_0$ (from the origin) is being visited for the first
time, this node will be assigned a list $S_0$ of weights
$\tau_{0i}$, $i=1\cdots \ell_0$ of the links by which we reach
that node sorted in descending order. Since $\tau_{0i}=exp(a
r_{0i})$, we can use a list of random numbers $r_{0i}$ instead.
\begin{equation}
S_0 = \{r_{0 1}, r_{0 2} , r_{0 3} , ...r_{0 l_0}\},
\end{equation}
with $r_{0 j} > r_{0 j+1}$ for all $j$. If we reach a node for a
second time by another path of length $\ell_1$, we define for this path
a new list $S_1$,
\begin{equation}
S_1 = \{r_{1 1}, r_{1 2}, r_{1 3},  \dots r_{1 l_1}\},
\end{equation}
and compare it with $S_0$ previously defined for this node.

Different sequences can have weights in common because some paths have
links in common because of the loops, so it is not enough to identify
the sequence by its maximum weight; in this case it must also be
compared with the second maximum, the third maximum, etc. We define
$S_p<S_q$ if there exists a value $m$, $1\leq m\leq\min(\ell_p,\ell_q)$
such that
\begin{eqnarray}\label{Eq.sent}
\nonumber r_{pj} = r_{qj} \qquad &\mbox{for}& \qquad 1\leq j < m
     \qquad \mbox{and}\\ r_{pj} < r_{qj} \qquad &\mbox{for}&
     \qquad j = m,
\end{eqnarray}
or if $\ell_q>\ell_p$ and $r_{pj}=r_{qj}$ for all $j\leq \ell_p$.
If $S_1<S_0$, we replace $S_0$ by $S_1$. The procedure continues until
all paths have been explored and compared. At this point,
$S_0=S_{\mbox{\scriptsize opt}}$, where $S_{\mbox{\scriptsize opt}}$ is
the sequence of weights for the optimal path of length
$\ell_{\mbox{\scriptsize opt}}$.  A schematic representation of this
ultrametric algorithm is presented in Fig.~\ref{f.0}.  This algorithm is
slow and memory consuming since we have to keep track of  a sequence of
values and the rank. Using this method, we obtain systems of sizes up
to $2^{12}$ nodes, typically $10^5$ realizations of disorder.

\subsection{Bombing Optimization}\label{sec.bomb}
This algorithm allows to compute $\ell_{\mbox{\scriptsize opt}}$ (and other
relevant quantities) between any two nodes in strong disorder limit and
was introduced by Cieplak et. al. \cite{Cieplak}.  Basically the
algorithm does the following

\begin{enumerate}
\item
Sort the edges by descending weight.
\item
\label{loop}
If the removal of the highest weight edge will not disconnect $A$ from
$B$ --  remove it.
\item
Go back to step \ref{loop} until all edges have been processed.
\end{enumerate}
Since the edge weights are random, so is the ordering. Therefore, in
fact, one does not need even to select edge weights and ``bombing'' algorithm
can  be simplified by removing randomly chosen edges one at a time, provided
that their removal does not break the connectivity between the two nodes.
The bottleneck of this algorithm is checking the connectivity after each removal.
To speed it up, we first compute the minimal path between nodes A and B using
Dijkstra's algorithm. Then we must check the connectivity only if the removed bond
belongs to this path. In this case, we attempt to compute a new minimal path between A
and B on the subset of remaining bonds. If our attempt fails, it means that the removal of
this bond would destroy the connectivity between A and B. Therefore, we restore this bond and
exclude it from the list of bonds subject to random removal. With this improvement we could reach
systems of sizes up to $2^{16}$ nodes and $10^5$ realizations of weight disorder.

\subsection{The Minimum Spanning Tree (MST)}
\label{sec.mst}
The MST on a weighted graph is a tree that reaches all nodes of the graph and
for which the sum of the weights of all the links or nodes (total weight) is
minimal. Also, in the ``strong disorder'' limit, each path between two sites on the MST is the optimal path
\cite{Cieplak,Dob}, meaning that along this
path the maximum {\em barrier} (weight) is the smallest possible
\cite{Dob,Brauns03,Sameet04}.  Standard algorithms for finding the MST are
Prim's algorithm\cite{Cormen90} which resembles invasion
percolation~\cite{BH96} and Kruskal's algorithm \cite{Cormen90}. First we
explain the Prim's algorithm.
\begin{itemize}
\item[(a)] Create a tree containing a single vertex, chosen arbitrarily
from the graph.
\item[(b)] Create a set containing all the edges in the
graph.

\item[(c)] Remove from the set an edge with minimum weight that
connects a vertex in the tree with a vertex not in the tree.
\item[(d)]
Add that edge to the tree.
\item[(e)] Repeat steps (c-d) until every edge
in the set connects two vertices in the tree.
\end{itemize}
Note that two nodes in the tree cannot be connected again by a link, thus
forbidding loops to be formed.  Prim's algorithm essentially starts by
choosing a random node in the network, and then growing outward to the
"cheapest" link which is adjacent to the starting node. Each link which is
"invaded" is added to the growing cluster (tree), and the process is iterated
until every site has been reached. Bonds can only be invaded if they do not
produce a loop, so that the tree structure is maintained [20]. This process
resembles invasion percolation with trapping studied in the physics
literature \cite{invpercbara, Porto97}.  A direct consequence of the invasion
process is that a path between two sites A and B on the MST is the path whose
maximum weight is minimal, i.e., the minimal-barrier path. This is because if
there were another path with a smaller barrier (i.e. maximal weight link)
connecting A and B, the invasion process would have chosen that path to be on
the MST instead.  The minimal-barrier path is important in cases where the
"bottleneck" link is important. For example, in streaming video broadcast on
the Internet, it is important that each link along the path to the client
will have enough capacity to support the transmission rate, and even one link
with not enough bandwidth can become a bottleneck and block the transmission.
In this case we will choose the minimal-barrier path rather than the optimal
path.  An equivalent algorithm for generating the MST is the Kruskal's
algorithm:
\begin{itemize}
\item[(a)] Create a forest F (a set of trees), where each
vertex in the graph is a separate tree.
\item[(b)] Create a set S
containing all the edges in the graph.
\item[(c)] While S is nonempty: "
Remove an edge with minimum weight from S. "   If that edge
connects two different trees, then add it to the forest, combining
two trees into a single tree. "   Otherwise discard that edge.
Note that an edge cannot connect a tree to itself, thus forbidding
loops to be formed.
\end{itemize}
    Kruskal's algorithm resembles the percolation process because we add
    links to the forest according to increasing order of weights. The forest
    is actually the set of percolation clusters growing as the occupation
    probability is increasing.  It was noted by Dobrin {\it et al.} \cite{Dob} that the
    geometry of the MST depends only on the unique ordering of the links of
    the network according to their weights. It does not matter if the weights
    are nearly the same or wildly different, it is only their ordering that
    matters. Given a network with weights on the links, any transformation
    which preserves the ordering of the weights (e.g., the link which has the
    fiftieth largest energy is the same before and after the transformation)
    leaves the MST geometry unaltered. This property is termed "universality"
    of the MST. Thus, given a network with weights, represented by a random
    variable distributed uniformly, a monotonic transformation of the
    weights will leave the MST unchanged.

Another equivalent algorithm to find the MST
is the ``bombing optimization algorithm'' ~\cite{Brauns03}. Similar to the
one explained in Section~\ref{sec.bomb},  we start with the full network and
remove links in order of descending weights.  If the removal of a
link disconnects the graph, we restore the link
~\cite{Ios04}; otherwise the link  is removed.
The algorithm ends and the MST is obtained when no more links can be removed
without disconnecting the graph.

\subsection{The Incipient Infinite Cluster (IIC)}\label{sec.iic}
To find the IIC of ER and SF in uncorrelated weighted networks
\cite{footnote_uncorr}, we start with the fully connected network
and remove links in descending order of their weights. After each
removal of a link, we calculate the weighted average degree
$\kappa \equiv \langle k^2 \rangle
/\langle k \rangle$, which decreases with link removals. When
$\kappa < 2 $,  we stop
the process~\cite{CEBH00}. The meaning of this criterion is explained in the next section,
where its connection with the percolation threshold $p_c$ is established.
The largest remaining component is the IIC.
For the two dimensional (2D) square lattice we cut the links
(bonds) in descending order of their weights until we reach the
percolation threshold $p_c$ ($=0.5$). At that point the largest
remaining component is the IIC~\cite{BH96}.

\section{Optimal path in strong  disorder and percolation on the Cayley tree.}
\label{sec.Cayley}

In this section we review classical analytical methods for
exploring random networks based on percolation theory on a Cayley
tree \cite{stauffer,BH96}, or branching processes \cite{Harris}.
To obtain the optimal path in the strong disorder limit, we
present the following theoretical argument. It has been shown
\cite{Brauns01,Cieplak} that the optimal path for $a\to\infty$
between two nodes $A$ and $B$ on the network can be obtained by
the bombing algorithm described in Section~\ref{sec.bomb}.  This
algorithm is based on randomly removing links. Since randomly
removing links is a percolation process, the optimal path must be
on the percolation backbone connecting A and B. We can explore the
network starting with node A by Dijkstra's algorithm, sequentially
creating burning shells of chemical distance $n$ from the node A.
Alternatively we can think of the $n$-th shell as of $n$-th
generation of descendants of a parent A in a branching process.
The random network consisting of a large number of nodes $N\to
\infty$ and small average degree $\langle k\rangle \ll N$, has a
tree-like local structure with no loops, since the probability
that a node we randomly chose by an outgoing link has been
already visited is less than $\langle k\rangle^n/N$, which remains
negligible for $n<\ln N/\ln \langle k\rangle$.

As we remove links by the bombing
algorithm, the average degree of remaining nodes decreases, and
the role of loops decreases. Thus finite loops play no role in
determining the properties of the optimal path. In fact,
connecting the nodes A and B by an optimal path  is equivalent to
connecting each of them to a very distant shell on a corresponding
Cayley tree. As the fraction $q=1-p$ of remaining links decreases, we
reach the percolation threshold at which removal of a next link
destroys the connectivity with a very high probability. Note that
if we select weights of the links $\tau_i=\exp(a r_i)$, where
$r_i$ is uniformly distributed on $[0,1]$, the fraction of
remaining bonds, $p$, is equal to $r_i$ of the next link we will
remove.

\subsection{Distribution of the maximal weight on the optimal path}

In order to further develop this analogy, we will show that the
distribution of the maximal random number $r_{\mbox{\scriptsize
max}}$ along the optimal path \footnote{The maximal random number,
is the first random number in the bombing process that   we cannot
remove without breaking the connection between a pair of nodes. In
other words it is the value that dominate the sum of the costs in
the SD limit (See Ref \cite{Brauns03,santafe})} can be expressed
in terms of the order parameter $P_{\infty}(p)$ in the percolation
problem on the Cayley tree, where $P_{\infty}(p)$ is the
probability that a randomly chosen site on the Cayley tree has
infinite number of generations of descendants or, in other words,
belongs to the infinite cluster.

If the original graph
has a degree distribution $P(k)$, the probability that we reach a
node with a degree $k$ by following a randomly chosen link on the
graph, is equal to $k P(k) /\langle k \rangle$, where $\langle k
\rangle$ is the average degree. This is because the probability of
reaching a given node by following a randomly chosen link is
proportional to the number of links, $k$, of that node
and $\langle k \rangle$ comes from normalization. Also, if we
arrive at a node with degree $k$, the total number of outgoing
branches is $k - 1 $ .  Therefore, from the point of view of the
Cayley tree, the probability $p_{k-1}$ to arrive at a node with $k-1$
outgoing branches by following a randomly chosen link is
\begin{equation}
p_{k-1}=k P(k)/\langle k \rangle.
\label{eq:pk1}
\end{equation}
In the asymptotic limit, $N\to \infty$, when the optimal path between the two nodes is very
long, the probability distribution for the maximal weight link can be
obtained from the following analysis. Let us assume that the
probability of {\it not} reaching $n-th$ generation starting from a randomly chosen link
of the Cayley tree whose links exist with a probability $p$, is
$Q_n$. Suppose this link leads to a node whose outgoing degree is $2$. Then the
probability that starting from this link, we will not reach $n$ generations of
its descendants is the sum of three terms:
\begin{enumerate}
\item
  The probability that both outgoing links do not exist is equal to $(1-p)^2$
\item The probability that both outgoing links exist, but they do
not have $n - 1$ generations of descendants is equal to $p^2
Q_{n-1}^2$ \item The probability that only one of the two outgoing
links exist but it does not have $n-1$ generations of descendants
is equal to $2(1-p) p Q_{n-1}$
\end{enumerate}
Therefore, in this case 
\begin{equation}
Q_n = (1-p)^2 + p^2Q_{n-1}^2 + 2(1-p)p Q_{n-1},
\end{equation}
which on simplification becomes
\begin{equation}
Q_n = ((1-p) + pQ_{n-1})^2.
\end{equation}
Following this argument for the case when our link leads to a node with $m$ outgoing links,
the probability that starting from this node, we
can not reach $n$ generations, is
\begin{equation}\label{Eq.Qn}
Q_n = ((1-p) + pQ_{n-1})^{m}.
\end{equation}
In the case of a Cayley tree with a variable degree, we
must incorporate a factor $p_{k-1}$ given by Eq.(\ref{eq:pk1}) which accounts for the probability that the node
under consideration has $k-1$ outgoing edges and sum up over all possible values of $k$.
Thus for a conducting link on the Cayley tree, the probability that it does not have
descendants in generation $n$ can be obtained by applying a recursion
relation
\begin{equation}
Q_l = \sum_{k=1}^\infty P(k) k ((1-p)+p Q_{l-1})^{k-1}/\langle
k \rangle
\end{equation}
for $l=1,2,...,n$ with the initial condition $Q_0 = 0$, which
indicates that a given link is always present in generation zero
of its descendants.

For a random graph, a randomly chosen node has
$k$ outgoing edges with the original probability $P(k)$. Thus it
has a slightly different probability $Q_n(p)$ of not having
descendants in its $n$th generation:
\begin{equation}
 \tilde{Q}_n = \sum_{k=0}^\infty P(k) ((1-p)+p Q_{n-1})^{k}.
\end{equation}

It is convenient to introduce the generating function of the original degree distribution
\begin{equation}
\tilde{G}(x)\equiv  \sum_{k=1}^\infty P(k)x^k
\end{equation}
and the generating function of the degree distribution of the Cayley tree
\begin{equation}
G(x)\equiv \sum_{k=1}^\infty \frac{k P(k)}{\langle k \rangle} x^{k-1},
\end{equation}
where $x$ is an arbitrary complex variable. Using the normalization
conditions for the probabilities $\sum_{k=0}^\infty P(k)=1$, it easy to see
that $\tilde{G}(1)=1$.  Taking into account that $\langle k\rangle =
\sum_{k=0}^\infty k P(k)$ we have $\langle k
\rangle=d\tilde{G}/dx\vert_{x=1}=\tilde{G}'(1)$ and hence $\tilde {G}(x)$ and
$G(x)$ are connected by a relation
\begin{equation}
G(x) =\tilde{G}'(x)/\tilde{G}'(1).
\end{equation}
For any degree distribution $P(k)\to 0$, as $k\to\infty$ and thus
both functions are analytic functions of $x$ and have a
convergence radius $R \ge 1$. Since $P(k)>0$, these functions and
all their derivatives are monotonically increasing functions on an
interval $[0,1)$. For the ER networks, the degree distribution is
Poisson given by:  $P(k) = \langle k \rangle ^k \exp^{-
  \langle k \rangle}/k!$, hence $\tilde{G}(x)=G(x)=\exp[\langle k \rangle (x-1)]$.
For scale free distribution, $P(k)\sim k^{-\lambda}$, hence $\tilde{G}(x)$ is proportional to Riemann
$\zeta$-function, $\zeta_\lambda(x)$.

If we denote by $f_n(p)$, the probability that starting at a randomly chosen
conducting link we can reach, or survive up to, the $n$-th generation, then
\begin{equation}
f_n = 1 - Q_n(p)
\end{equation}
and by $\tilde{f}_n(p)$, the probability that a randomly chosen
node has at least  $n$ generation of descendants,
\begin{equation}
\tilde{f}_n = 1 - \tilde{Q}_n(p)
\end{equation}
then
\begin{equation}
f_n = 1 - G(1-p f_{n-1})
\label{eq:iterations}
\end{equation}
and
\begin{equation}
\tilde{f}_n = 1 - \tilde{G}(1-p f_{n-1}).
\end{equation}
\begin{figure}
\includegraphics[width=8cm,height=10cm,angle=270]{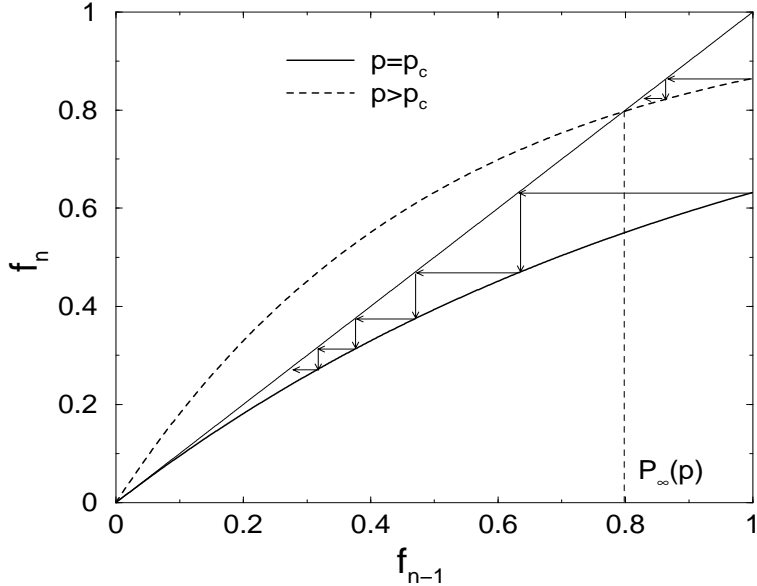}
\caption{ The iterative process of solving equation (\ref{eq:root}).
The thin straight line $y=x$ represents the left hand side. The bold curve represents
the right hand side (r.h.s) for $p=p_c$, at which the r.h.s. is tangential to $y=x$ at the origin.
The dashed curve represents r.h.s. for $p>p_c$. Both cases are computed for the Poisson degree distribution
with $\langle k \rangle=2$, so r.h.s of Eq. (\ref{eq:root}) is given by $1-\exp(-2px)$. The arrows represent iterations starting
from $f_0=1$ (the starting link belongs to generation 0). It is clear that the convergence
of the iterations is very fast (exponential) for $p\neq p_c$, while it is very slow
(power law) for $p=p_c$.}
\label{f:iterations}
\end{figure}
The sequence of iterations (\ref{eq:iterations}) is visualized (See Fig.~\ref{f:iterations})
as a process of solving the equation
\begin{equation}
x=1-G(1-px)
\label{eq:root}
\end{equation}
by an iteration method. Obviously, this equation has at least one
root $x_0=1$. But if the derivative of the right hand side,
$[1-G(1-px)]'\vert_{x=0}=pG'(1)>1$, we will have another root
$0<x_1\leq 1$. This root has a physical meaning of a probability
$P_\infty(p)$ that a randomly selected conducting link is
connected to infinity (See also \cite{CEBH00}). For $p>1/G'(1)$, the iterations will
converge to this root, while for $p\leq 1/G'(1)$, the iterations
will converge to $P_\infty(p)=0$. Thus
\begin{equation}
p_c\equiv {1\over G'(1)}={\langle k\rangle\over \langle k^2\rangle - \langle k \rangle}=\frac{1}{\kappa -1}
\label{eq:pc}
\end{equation}
has a meaning of the percolation threshold above which there is a
finite probability to reach the infinity. Using this equation we
can derive the condition $\kappa<2$ to stop bombing in the process
of obtaining IIC. Indeed $\kappa<2$ indicates that equation
(\ref{eq:root}) has only one trivial solution $x_0=0$ even for
$p=1$. This means that all the clusters in this network are
finite. If $\kappa>2$, $p_c<1$ accordingly $P_\infty(1)>0,$ i.e.
the infinite cluster does exist.  The condition $\kappa=2$
corresponds to $p_c=1$ which means that any further link removal
will produce a network in which $P_\infty(1)=0$ ,i.e. the network
with only finite clusters, while at $p=1$, the infinite cluster is
incipient.
\begin{figure}
\includegraphics[width=10cm,height=8cm,angle=0]{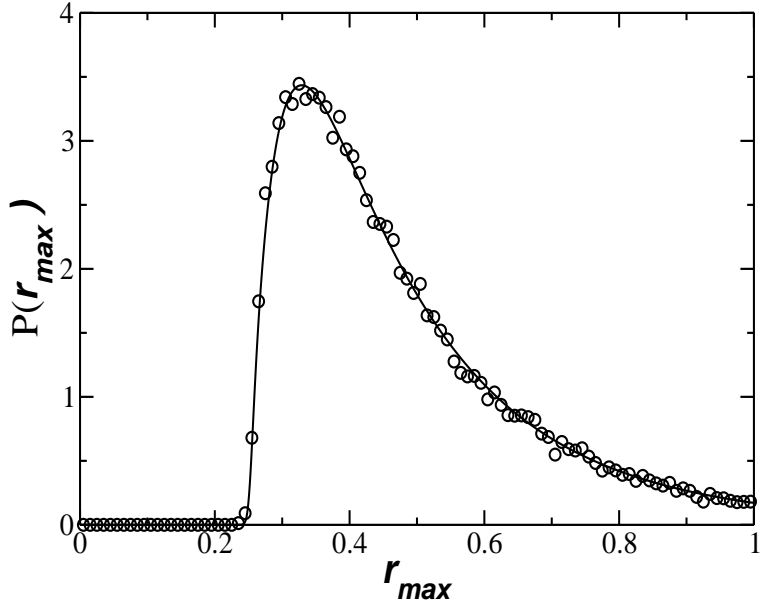}
\caption{ The probability distribution of the maximal random number
  $r_{\mbox{\scriptsize max}}$ along the optimal path obtained using
  simulations on a random graph with $\langle k \rangle = 4$ and using the
  analytical method on a Cayley tree with Poisson degree distribution and
  $\langle k \rangle = 4$. The simulations involve 100000 network
  realizations and are carried out on a network of 65536 nodes. The values of
  $l_{\mbox{\scriptsize opt}}$ for this network lie in the range $40 <
  l_{\mbox{\scriptsize opt}} < 120$ (After \cite{santafe}). \label{f.3}}
\end{figure}

The probability that a randomly chosen node is connected to infinity can be determined as
\begin{equation}
\tilde P_\infty(p)=1-\tilde{G}(1-p P_\infty(p)),
\end{equation}
where $ P_\infty(p)$ is a non-trivial solution of Eq. (\ref{eq:root}).
For some degree distributions including Poisson distribution,
$\tilde{P}_\infty(1)<1$. This indicates that a randomly chosen node on the original network
may not belong to the giant component of the network. In fact, the optimal path between
nodes A and B exists if both  belong to the giant component. Provided that A and B
both belong to the giant component, the probability that they are still connected when
the fraction $1-p$ of bonds is removed is
\begin{equation}
\Pi(p) = \left(\frac{\tilde{P}_\infty(p)}{\tilde{P}_\infty(1)}\right)^2.
\end{equation}
Translating this condition to the bombing algorithm of generating an optimal
path, $\Pi(p)$ is the probability that the maximum random number along the
optimal path $r_{\mbox{\scriptsize max}} \le p$. Indeed, $\Pi(p)$ is the
probability that when only a fraction $p$ of links remains the connectivity
between A and B still exists. Hence $r_{\mbox{\scriptsize max}}\leq p$. Thus,
$\Pi(r_{\mbox{\scriptsize max}})$ is the cumulative distribution of
$r_{\mbox{\scriptsize max}}$. The probability density of
$r_{\mbox{\scriptsize max}}$ is thus equal to the derivative of $\Pi(p)$ with
respect to $p$:
\begin{equation}
  P(r_{\mbox{\scriptsize max}})={d\over dp}\Pi(p)\vert_{p= r_{\mbox{\scriptsize max}}}.
\label{prob_vi}
\end{equation}

In Fig.~\ref{f.3} we plot two curves. The curve with symbols is the
probability distribution of $r_{\mbox{\scriptsize max}}$ in a strongly
disordered ER graph with $\langle k \rangle = 4$ obtained by simulations.
The line shows the same probability distribution obtained using (Eq.~(\ref{prob_vi})) for a Poisson degree
distribution with $\langle k\rangle = 4$. The curves coincide very well, indicating the excellent
agreement between the theoretical analysis and simulations.

\subsection{Distribution of the cluster chemical length at percolation threshold}

Figure~\ref{f:iterations} illustrates the convergence of the
probability of the random link to have descendants in the $n$-th
generations. The difference $P(n)=f_n -f_{n+1}$ is the probability
that the last generation of the descendants of this link is $n$.
In percolation language, it is the probability distribution of the
cluster chemical length $\ell=n$. In order to find, how $f_n\to 0$
when $p=p_c$, we can expand equation (\ref{eq:iterations}) in Taylor series at
$f_n=0$. For ER networks, $G(x)$ has all the derivatives at $x=1$,
thus (\ref{eq:iterations}) can be presented as
\begin{equation}
f_n=p G'(1)f_{n-1}-{1\over 2}p^2G''(1)f^2_{n-1} +O(f^2_{n-1}). \label{eq:T1}
\end{equation}
For SF graphs with $\lambda >4$, $G''(1)$ also exists, thus the
above equation holds. For $3<\lambda<4$ the second derivative does
not exist, however using the Tauberian theorem which relates the speed
of the decay of the coefficients $P(k)\sim k^{-\mu}$ of the power
series and the behavior of its singularity at the convergence
radius: $G_s(x)\sim (1-x)^{\mu-1}$ we can write:
\begin{equation}
f_n=p G'(1)f_{n-1}-c f^{\lambda-2}_{n-1} +O(f^{\lambda-2}_{n-1}),
\label{eq:T2}
\end{equation}
where $c$ is some positive coefficient.

As $\lambda\to 3$, $G'(1)\to \infty$ and hence, according to
Eq.(\ref{eq:pc}) $p_c\to 0$. This means that for SF networks with
$\lambda \le 3$, percolation approach breaks down. However, for
finite networks, it is unlikely to have a degree larger than
$N^{1/(\lambda -1)}$. This fact is obvious since when one
generates random degrees with probability distribution
$P(k)$, one produces random numbers $x$ uniformly distributed on a
interval between 0 and 1, and compute $k=f(x)$, where $f(x)$
satisfies the equation $x=\sum_{f(x)}^\infty P(k) \sim
f(x)^{-\lambda+1}$. Thus the largest $k$ corresponds to the
smallest $x$. Generating $N$ random numbers is equivalent to
throwing $N$ points on an interval $[0,1]$ which divide this
interval into $N+1$ segments whose lengths are identically
distributed with an exponential distribution. Thus the average
value of the smallest $x$ is equal to $1/(N+1)$. Accordingly the
average value of the largest $k$ can be approximated as $k_{\rm
max}=f(1/(N+1))\sim N^{1/(\lambda-1)}$ \cite{CEBH00}. Thus replacing summation
by integration up to $k_{\rm max}$ in the expression for
$G'(1)\approx \int^{k_{\rm max}}k^{-\lambda+2} dk \sim
k_{max}^{3-\lambda}= N^{(3-\lambda)/(\lambda-1)}$. Hence for
$2<\lambda<3$ \cite{CEBH00}
\begin{equation}
p_c \sim N^{(\lambda-3)/(\lambda-1)}.
\label{eq:cohen}
\end{equation}

When $p<p_c$, $f_n \sim (p/p_c)^n$, i.e. the convergence is
exponential. When $p=p_c$, we will seek the solution of the above
recursion relations in a power law form: $f_n \sim n^{-\theta}$.
Expanding them in powers of $n^{-1}$, and equating the leading
powers, we have $\theta+1=\theta(\lambda-2)$, from which we obtain
 \begin{equation}
f_n \sim n^{-1/(\lambda-3)},
\end{equation}
or
\begin{equation}
P(\ell)=f_{\ell}-f_{\ell+1} \sim \ell^{-\tau_\ell},
\end{equation}
where \cite{Cohen,Cohen02}
\begin{eqnarray}
\tau_\ell =
\cases {
    \begin{array}{rll}
        2,  & \lambda>4 & {\rm ER} \\
        \frac{1}{(\lambda-3)}+1, &  3<\lambda \leq 4 &
    \end{array}
    }\,.
\label{eq:tau_ell}
\end{eqnarray}
The probability that a randomly selected {\it node} has exactly $\ell$ generations of descendants is
equal to
\begin{equation}
\tilde{P}(\ell)=\tilde{f}_\ell -\tilde{f}_{\ell+1} =\tilde{G}(1-pf_\ell)-\tilde{G}(1-pf_{\ell-1})\sim
\langle k\rangle p(f_\ell -f_{\ell-1}).
\end{equation}
Thus it is characterized by the same $\tau_\ell$ as $P(\ell)$.

Taylor expansions (\ref{eq:T1}) and (\ref{eq:T2}) can be used to derive the behavior of $P_\infty(p)$ as $p\to p_c$
by letting $f_n=f_{n-1}=P_\infty(p)$ and solving the resulting equations with a leading term accuracy:
\begin{equation}
P_\infty(p)=(p-p_c)^\beta,
\end{equation}
where \cite{CEBH00}
\begin{equation}
\beta =
\cases {
    \begin{array}{rll}
        1,  & \lambda>4 & {\rm ER} \\
        \lambda-3 ,  &  3<\lambda \leq 4 &
    \end{array}
    }\,.
\label{eq:beat}
\end{equation}

\subsection{Distribution of the cluster sizes at percolation threshold}

Using the generating functions
\cite{Cohen,Cohen02,newman-callaway-2000:networks_robustness}, one can also
find the distribution of the clusters sizes, $P(s)$, connected to a randomly
selected {\it link}.  For simplicity, let us again consider a link
(conducting with probability p) leading to a node of a degree $k=3$, so it
has only two outgoing links. The probability that this link is connected to a
cluster consisting of $s$ nodes obeys the following relations
\begin{equation}
P(s)=p\sum_{k+l=s-1}P(k)P(l)
\end{equation}
for $s>0$ and $P(0)=1-p$.
Introducing the generating function of the cluster size distribution $H(x)=\sum_0^\infty P(s)x^s$,
we have: $H(x)=1-p +xpH^2(x)$. In a general Cayley tree with an arbitrary degree distribution
we have:
\begin{equation}
H(x)=1-p +xpG(H(x)).
\label{eq:H}
\end{equation}
This equation defines the behavior of $H(x)$ for $x\to 1$, and thus via the Tauberian theorem defines
the asymptotic behavior of its coefficients $P(s)$.
Note that $H(1)$ is the cumulative probability of all finite clusters. Thus $(1-H(1))=p P_\infty(p)$
is the probability that a randomly selected link conducting with probability $p$ is connected to the infinity
and Eq.(\ref{eq:H}) becomes equivalent to Eq. (\ref{eq:root}) for  $P_\infty(p)$.

Introducing $\delta_x=1-x$ and $\delta_H=1-H(x)$ and expanding
$G(x)$ around $x=1$ at percolation threshold $p=1 /G'(1)$, we have
$\delta_H\delta_x +p \delta_x = c x \delta_H^{\lambda-2}
+O(\delta_H^{\lambda-2})$ which yields $\delta_H\sim
\delta_x^{1/(\lambda-2)}$. Using the Tauberian theorem we conclude
\cite{Cohen,Cohen02}:
\begin{equation}
P(s) \sim s^{-\tau_s},
\end{equation}
where
\begin{equation}
\tau_s =
\cases {
    \begin{array}{rll}
        3/2,  & \lambda>4 &{\rm  ER} \\
        \frac{1}{\lambda-2}+1,  &  3<\lambda \leq 4 &
    \end{array}
    }\,.
\label{eq:tau_s}
\end{equation}
Analogous considerations  suggest that the probabilities $\tilde{P}(s)$ that a
randomly selected {\it node} belongs to the cluster of size $s$
produce the generating function $\tilde{H}(x)=\tilde{G}(H(x))$.  Since
for $\lambda>3$, $\tilde{G}''(1)<\infty$, the singularity of
$\tilde{H}(x)$ for $x \to 1$ is of the same order as the
singularity of $H(x)$ and thus its coefficients, $\tilde P(s)$,
also decay as $s^{-\tau_s}$.

Following \cite{stauffer}, we will show that the
distribution of all the disconnected clusters
in a network scales as $P_{all}(s)=\tilde P(s)/s\sim s^{-\tau_s+1}$. Indeed, let us select a random node in this network.
The number of nodes belonging to the clusters of size $s$ is $N s P_{all}(s)/\sum_1^\infty s P_{all}(s)=NsP_{all}(s)/\langle s\rangle$.
Thus, $\tilde{P}(s)=sP_{all}(s)/\langle s\rangle$.

If we have a network of $N$ nodes, the size of the largest cluster $S$ is determined by the
relation $\sum_{s=S}^{\infty} P_{all}(s) \sim 1/N$, which becomes clear if we describe a concrete realization of
the cluster sizes by throwing $N/\langle s \rangle $ random points representing clusters under
the curve $P_{all}(s)$.  The average area corresponding to each of these points is $1/N$ and the area corresponding
to the rightmost point representing the largest cluster is $\sum_{s=S}^{\infty}P_{all}(s) \sim S^{-\tau_s}$.
Thus the largest cluster (which coincides with IIC) in the network of $N$ nodes scales as
\begin{equation}
S\sim N^{1/\tau_s}.
\label{eq:ER59}
\end{equation}
For ER graphs, the relation $S\sim N^{2/3}$ has been derived in a classical work \cite{ER59}.

\section{Scaling of the length of the optimal path in Strong  Disorder}
\label{sec.sd}

The relations obtained in the previous subsections allow us to
determine the scaling of the average optimal path length in a
network of $N$ nodes. When during bombing, we reach percolation
threshold, we have targeted only a tiny fraction of links (or
nodes) on the optimal path, with $r_{max}>p_c$ which we have to
restore, because their removal would destroy the connectivity. The
majority of the links on the optimal path remains intact. All of
them belong to the remaining percolation clusters which at
percolation threshold has a tree like structure with no loops. At
this point, the optimal path coincides with the shortest path,
which is uniquely determined. We will describe this situation in
detail in Section\ref{sec.tws}. With high probability, the optimal
path between any two nodes A and B goes through the largest
cluster at the percolation threshold. Thus its length must scale
as the chemical length of the largest percolation cluster
\cite{Brauns03}.  Assuming a power law relation between the
cluster size $s$ and its chemical dimension $\ell$,
$s=\ell^{d_\ell}$, and using the fact that both of the quantities
have power law distributions $P(\ell)d\ell=\tilde P(s)ds$, we have
$\ell^{-\tau_\ell}=\ell^{-d_\ell\tau_s+ d_\ell -1}$. Thus
\cite{vespignani_pnas}
\begin{equation}\label{eq:d_ell}
d_\ell=(\tau_l-1)/(\tau_s-1).
\end{equation}
Therefore, $S \sim \ell_{opt}^{d_{\ell}}$ and using
(\ref{eq:ER59}) we have $\ell_{opt}\sim S^{1/d_\ell}\sim
N^{\nu_{opt}}$, where
\begin{equation}\label{eq:nu_opt}
\nu_{opt}=1/(d_\ell\tau_s).
\end{equation}
Using Eqs. (\ref{eq:tau_s}) and (\ref{eq:tau_ell}) for $\tau_s$
and $\tau_\ell$ respectively, we have
\begin{equation}
\nu_{opt} = \cases {
    \begin{array}{rll}
        1/3,  & \lambda>4, & {\rm ER} \\
        (\lambda-3)/(\lambda-1),  &  3<\lambda \leq 4 &
    \end{array}
    }\,.
\end{equation}
Note that $\lambda=4$ corresponds to the special case when
$G''(1)$ diverges, in this case the Tauberian theorem predicts
logarithmic corrections, and hence we expect $\ell_{opt} \sim
N^{1/3} / \ln N$ for $\lambda =4$.

\begin{figure}[bht]
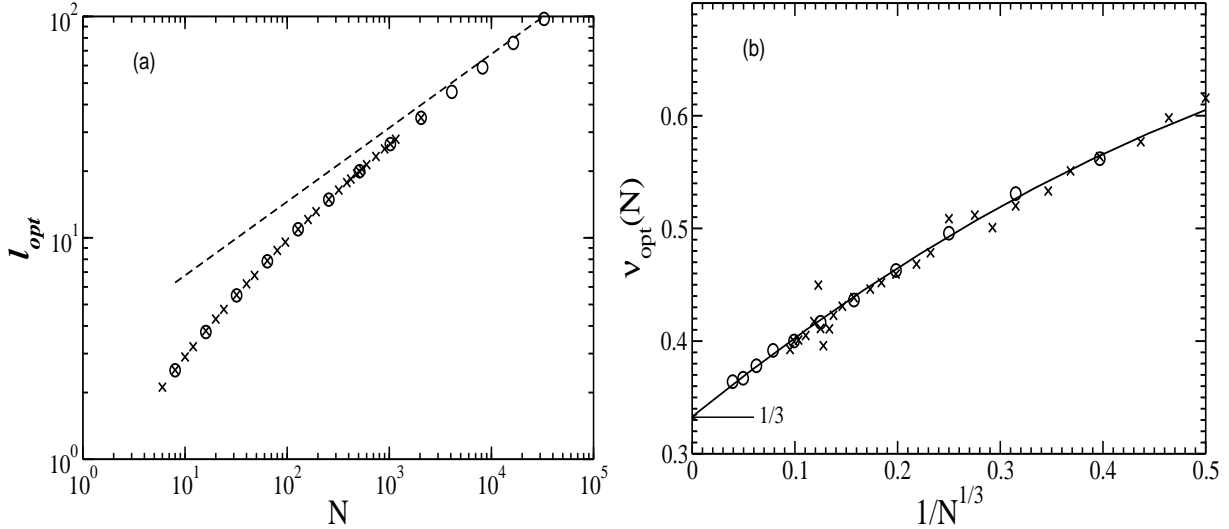
\label{fl.1}
\begin{center}
\includegraphics[width=8cm,height=7cm]{bbchs-fig1.eps}
\includegraphics[width=8cm,height=7cm]{bbchs-fig2.eps}
\end{center}
\caption{(a) Plot of $\ell_{\mbox{\scriptsize opt}}$ as a function of $N$ in
  double logarithmic scale for the optimal path length in strong
  disorder using the two numerical methods discussed in the text: (i)
  results obtained using the ``bombing'' approach ($\circ$) and (ii)
  results obtained using the ultrametric approach ($\times$). The
  dashed line shows the slope $1/3$.
  (b) Successive slopes $\nu_{\mbox{\scriptsize opt}} (N)$
  as a function of $1/N^{1/3}$ for the optimal path length in strong
  disorder using the two methods described in the text.
  The symbols denote the same as in (a). 
  The dashed line is the
  quadratic fitting of the results showing that the extrapolated value
  of the effective exponent in the limit $N\to\infty$ approaches
  $1/3$. This result coincides with our theoretical value
  $\nu_{opt}=1/3$ asymptotically (After \cite{Brauns03,reviewstat}).
\label{f.1}}
\end{figure}

We review above the exact results for the Cayley tree, from which
using heuristic arguments we have derived the scaling relation
between the average length of the optimal path and the number of
nodes in the network. Now we will show how the same predictions
can be obtained using general percolation theory. We will also
present numerical data supporting our heuristic arguments. We
begin by considering the ER graph. At criticality, it is
equivalent to percolation on the Cayley tree or percolation at the
upper critical dimension $d_c=6$. For the ER graph, we derived
above that the mass of the IIC, $S$, scales as $N^{2/3}$
\cite{ER59}. This result can also be obtained in the framework of
percolation theory for $d_c=6$. Since $S\sim R^{d_f}$ and $N\sim
R^d$ (where $d_f$ is the fractal dimension and $R$ the spatial
diameter of the cluster), it follows that $S\sim N^{d_f/d}$ and
for $d_c=6$, $d_f=4$ \cite{BH96} we obtain $S\sim N^{2/3}$
\cite{Watts03}.

It is also known \cite{BH96} that, at criticality, at the upper critical
dimension, the average shortest path length $\ell_{\mbox{\scriptsize min}} \sim R^2$, like a random walk and therefore
$S\sim\ell_{\mbox{\scriptsize min}}^{d_\ell}$ with
$d_\ell=2$. Thus
\begin{equation}
\ell_{\mbox{\scriptsize min}} \sim \ell_{\mbox{\scriptsize opt}}\sim
S^{1/d_\ell}\sim N^{2/3d_\ell} \sim N^{\nu_{\mbox{\scriptsize opt}}},
\label{E.1}
\end{equation}
where $\nu_{\mbox{\scriptsize opt}}=2/3d_\ell=1/3$.

For SF networks, we can also use the percolation results at
criticality. It was found \cite{Cohen,Cohen02} (see Sec. \ref{sec.Cayley}) that $d_\ell=2$ for $\lambda>4$,
$d_\ell=(\lambda-2)/(\lambda-3)$ for $3<\lambda<4$, $S\sim N^{2/3}$ for
$\lambda>4$, and $S\sim N^{(\lambda-2)/(\lambda-1)}$ for $3<\lambda\leq
4$. Hence, we conclude that
\begin{equation}
\ell_{\mbox{\scriptsize min}}\sim \ell_{\mbox{\scriptsize
opt}}\sim\cases{ N^{1/3} & $\lambda>4$ \cr N^{(\lambda-3)/(\lambda-1)} &
$3<\lambda\leq 4$}.
\label{E.2x}
\end{equation}
Thus $\nu_{opt} =1/3$ for ER and SF with $\lambda >4$, and $\nu_{opt}
=(\lambda-3)/(\lambda-1)$ for SF with $3 < \lambda < 4 $.  Since for
SF networks with $\lambda >4$ the scaling behavior of
$\ell_{\mbox{\scriptsize opt}}$ is the same as for ER graphs and for
$\lambda < 4$ the scaling is different, we can regard SF networks as a
generalization of ER graphs.

Next we describe the details of the numerical simulations and show that
the results agree with the above theoretical predictions.  We perform
numerical simulations in the strong disorder limit by the  method
described in Section~\ref{sec.bomb} for ER and SF networks.  We also
perform additional simulations for the case of strong disorder on ER
networks using the ultrametric optimization algorithm (see
Section~\ref{sec.um}) and find results identical to the results obtained
by randomly removing links.
In Fig.~\ref{f.1}(a) we show a double logarithmic plot of
$\ell_{\mbox{\scriptsize opt}}$ as a function of $N$ for ER graphs. To
evaluate the asymptotic value for $\nu_{\mbox{\scriptsize opt}}$ we use for
both approaches successive slopes, defined as the successive slopes
\cite{Brauns01} of the values on Fig.~\ref{f.1}. One can see from
Fig.~\ref{f.1}(b) that their value approaches $1/3$ when $N \gg 1$,
supporting Eq.~(\ref{E.1}).

\begin{figure}[ht]
\begin{center}
\includegraphics[width=6cm,height=8cm,angle=270]{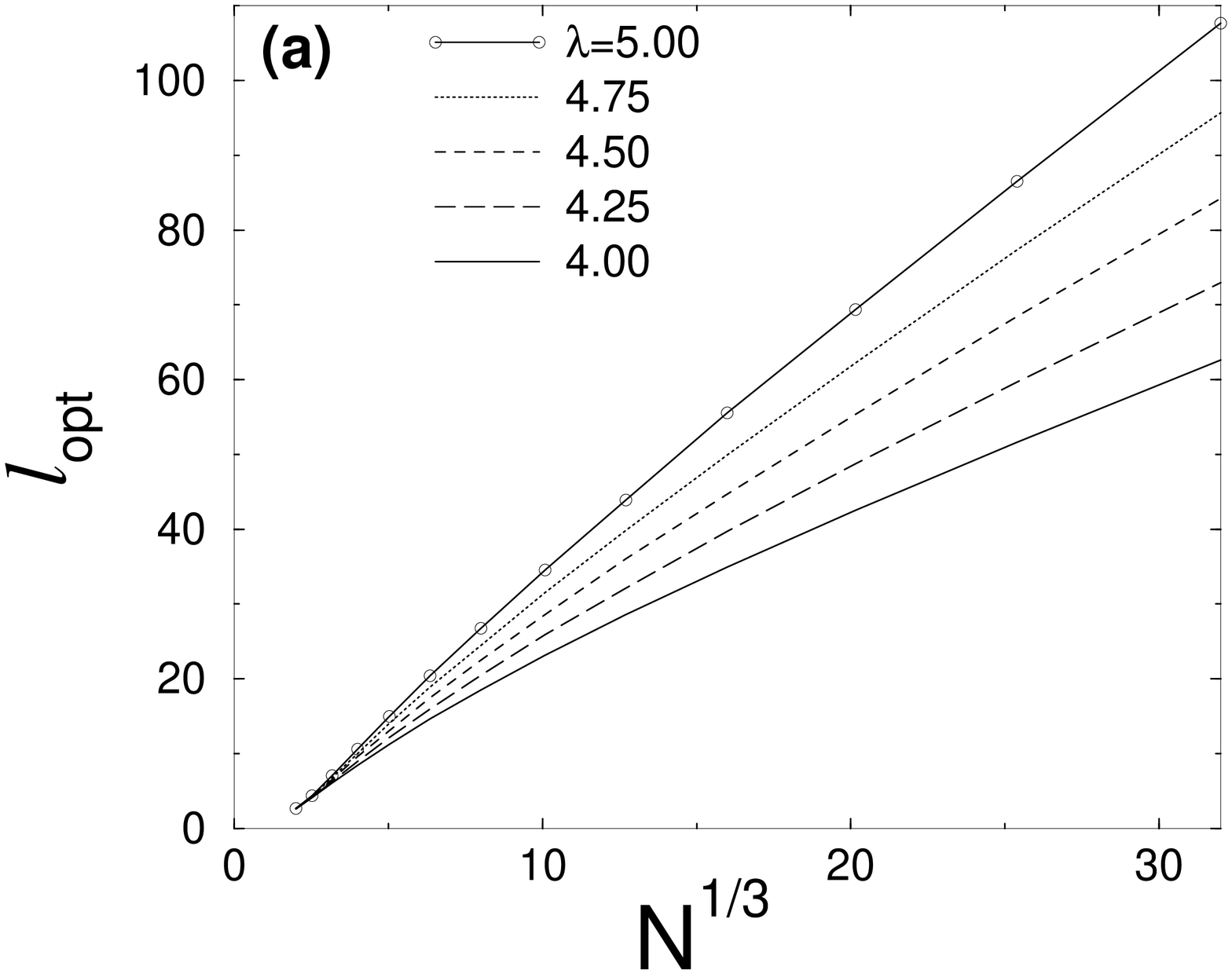}
\includegraphics[width=6cm,height=8cm,angle=270]{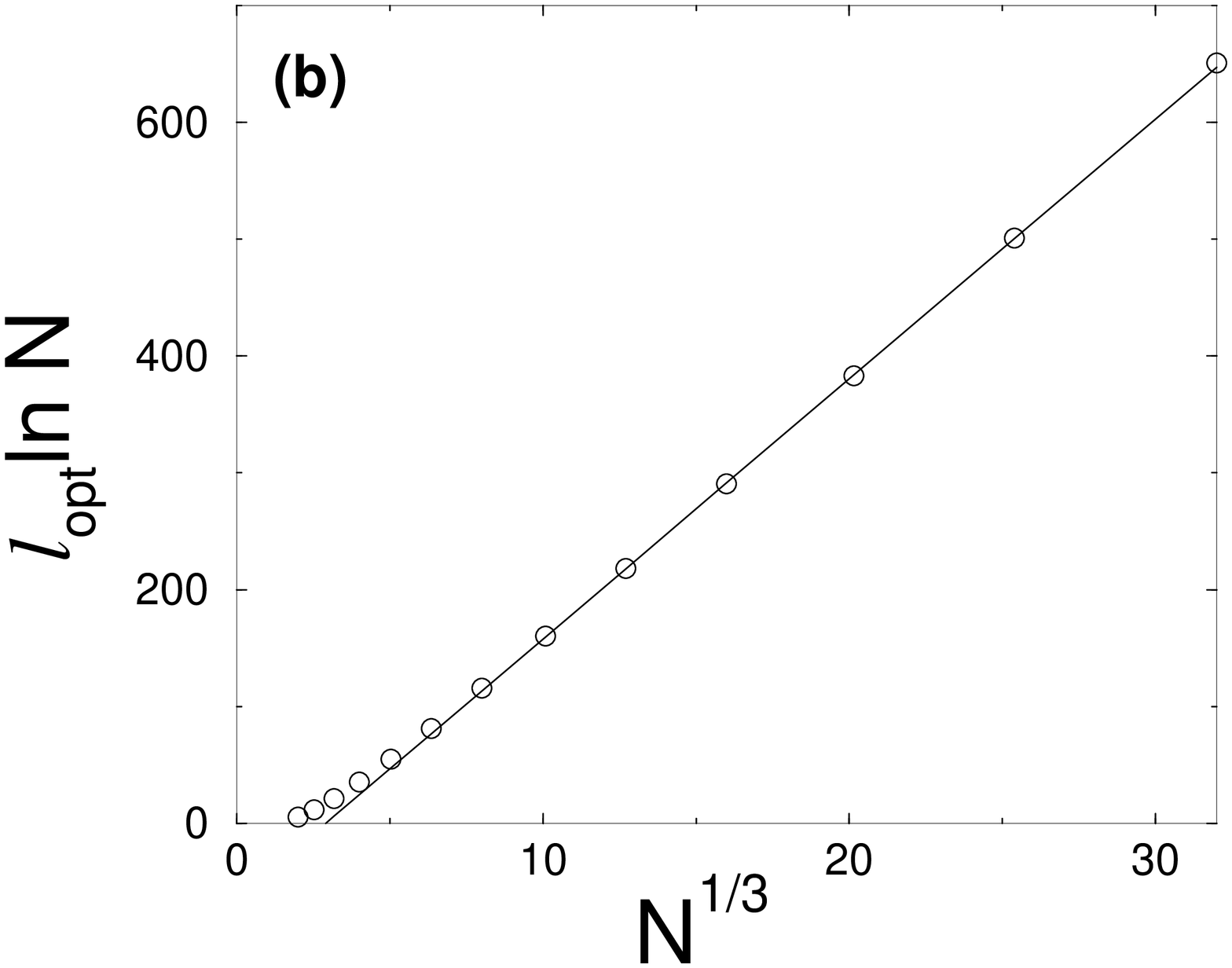}
\end{center}
\begin{center}
\includegraphics[width=7cm,height=6cm]{SF-fig4c.eps}
\includegraphics[width=7cm,height=6cm]{SF-fig4d.eps}
\end{center}
\caption{Results of numerical simulations.  (a) The dependence of
  $\ell_{\mbox{\scriptsize opt}}$ on $N^{1/3}$ for $\lambda\geq 4$.
  (b) The dependence of $\ell_{\mbox{\scriptsize opt}}/\;\ln N$ on
  $N^{1/3}$ for $\lambda =4$.  (c) The dependence of
  $\ell_{\mbox{\scriptsize opt}}$ on $N^{(\lambda-3)/(\lambda - 1)}$
  for $3< \lambda <4$.  (d) The dependence of $\ell_{\mbox{\scriptsize
      opt}}$ on $\ln N$ for $\lambda\leq 3$ (After \cite{Brauns03,reviewstat}).
\label{f.SF}}
\end{figure}

The theoretical considerations [Eqs.~(\ref{E.1}) and (\ref{E.2x})] predict
that SF graphs with $\lambda>4$, are similar to ER with
$\ell_{\mbox{\scriptsize opt}} \sim N^{1/3}$, while for SF graphs with
$3<\lambda<4$, $\ell_{\mbox{\scriptsize opt}}\sim N^{(\lambda - 3)/(\lambda -
  1)}$.  Figure~\ref{f.SF}a shows data from numerical simulations supporting
the linear behavior of $\ell_{\mbox{\scriptsize opt}}$ versus $N^{1/3}$ for
$\lambda\geq 4$.  The quality of the linear fit becomes poor for $\lambda
\rightarrow 4$. At this value, there are corrections probably due to
logarithmic divergence of the second moment of the degree distribution, i.e.,
$\ell_{\mbox{\scriptsize opt}} \sim N^{1/3}/\ln N$ (see Fig.~\ref{f.SF}b).
Figure~\ref{f.SF}c shows results of simulations supporting the asymptotic
linear behavior of $\ell_{\mbox{\scriptsize opt}}$ versus $N^{(\lambda -
  3)/(\lambda - 1)}$ for $3<\lambda \leq 4$.  Theoretically, as $\lambda
\rightarrow 3$, $\nu_{\mbox{\scriptsize opt}} =(\lambda - 3)/(\lambda -1)
\rightarrow 0$, and thus one can expect for $\lambda=3$ a logarithmic $N$
dependence of $\ell_{\mbox{\scriptsize opt}}$. Indeed, for $2<\lambda<3$ our
numerical results for the strong disorder limit suggest that
$\ell_{\mbox{\scriptsize opt}}$ scales slower than a power law with $N$ but
slightly faster than $\ln N$. The numerical results can be fit to
$\ell_{\mbox{\scriptsize opt}}\sim(\ln N)^{\lambda-1}$ (see
Fig.~\ref{f.SF}d).  Note that the correct asymptotic behavior may be
different and this result may represents only a crossover regime. The exact
nature of the percolation cluster at $\lambda<3$ is not clear yet, since in
this regime the transition does not occur at a finite (non zero) critical
threshold~\cite{CEBH00}. We obtain similar results for SF networks where the
weights are associated with nodes instead of links.

\section{Scaling of the length of the optimal path in Weak Disorder}\label{wd}
\begin{figure}[h]
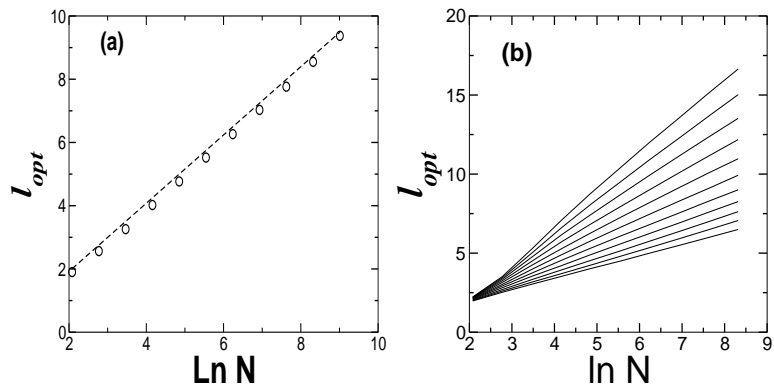

\begin{center}
\includegraphics[width=5cm,height=5cm,angle=0]{weak-ER.eps}
\includegraphics[width=5cm,height=5cm,angle=0]{fig3a.eps}
\caption{ Results of numerical simulations. (a) The linear dependence of
  $\ell_{\mbox{\scriptsize opt}}$ on $\ln N$ for ER graphs in the weak
  disorder case for $<k>=4$. The dashed line is used as a guide to show the
  linear dependence. (b) The dependence of $\ell_{\mbox{\scriptsize opt}}$ on
  $\ln N$ for SF graphs in the weak disorder case for various values of
  $\lambda$. The different curves represent different values of $\lambda$
  from $2.5$ (bottom) to $5$ (top) (After \cite{Sameet04,kalkotasameet}).
\label{f.wd-3}}
\end{center}
\end{figure}

When $a=1/kT\to 0$, all the $\tau_i$ essentially contribute to the total
cost. Thus $T\to \infty$ (very high temperatures) corresponds to weak
disorder limit. We expect that the optimal path length in the weak
disorder case will not be considerably different from the shortest
path, as found also for regular lattices \cite{RednerXX} and random
graphs \cite{verderHofstad01}. Thus we expect that the scaling for the
shortest path will also be valid for the optimal path in weak
disorder, but with a different prefactor depending on the details of
the graph and on the type of disorder. We simulate weak disorder by
selecting $0\leq\tau_i<1$ from a uniform distribution. To compute
$\ell_{\mbox{\scriptsize opt}}$ we use the Dijkstra algorithm (See
Section~\ref{sec.dij})\cite{Cormen90}.  The scaling of the length of
the optimal path in WD for ER, is shown in Fig.~\ref{f.wd-3}(a).  Here
we plot $\ell_{\mbox{\scriptsize opt}}$ as a function of $\ln N$ for
$\langle k \rangle =4$. The weak disorder does not change the scaling
behavior of $\ell_{\mbox{\scriptsize opt}}$ on ER compared to
$\ell_{min}$, only the prefactor.

\bigskip

For SF networks, the behavior of the optimal path in the weak disorder
limit is shown in Fig.~\ref{f.wd-3}(b) for different degree distribution
exponents $\lambda$. Here we plot $\ell_{\mbox{\scriptsize opt}}$ as a
function of $\ln N$. All the curves seem to have linear asymptotes.
This result is analogous to the behavior of the shortest path
$\ell_{\mbox{\scriptsize min}}\sim\ln N$ for $3<\lambda <4$ and
ER. Note, however, that for $2<\lambda<3$,
$\ell_{\mbox{\scriptsize min}}$ scale as $\ln\ln N$~\cite{Cohen3}.
Thus, $\ell_{\mbox{\scriptsize opt}}$ is significantly larger and
scales as $\ln N$ (Fig.~3b). Thus, weak disorder does not change
the universality class of the length of the optimal path except in
the case of ``ultra-small'' worlds $2<\lambda<3$, where
$\ell_{\mbox{\scriptsize opt}} \sim \exp(\ell_{min})$, and the
networks become small worlds.

\section{Crossover from Weak to Strong Disorder}
\label{sec.tws}
\subsection{Exponential Disorder}
Consider the case of finite $a$ ($T >0$). In this case we expect a crossover
in the length of the optimal path (or the system size $N$) from strong
disorder behavior to weak disorder depending on the value of $a$. In order to
study this crossover we have to use an implementation of disorder that can be
tuned to realize narrow distributions of link weights (WD) as well as broad
distributions of link weights (SD). The procedure that we adopt to implement
the disorder is as follows \cite{Cieplak,Porto99,Brauns01,Brauns03} (See
Sec.~\ref{sec.sd} A). Assign to each link $i$ of the network a random number
$r_{i}$, uniformly distributed between 0 and 1. For the analogy with the
thermally activated process described in Sec.~\ref{sec.sd} the $r_i$ play the
role of the energy barriers. The transit time or cost associated
with link $i$ is then $\tau_i\equiv\exp(a r_i)$, where $a$ controls the
strength of disorder i.e., the broadness of the distribution of link weights.
The limit $a\rightarrow\infty$ is the strong disorder limit, where a single
link dominates the cost of the path.  For $d$-dimensional lattices of size
$L$, the crossover is found \cite{Cieplak,Porto99} to behave as

\begin{eqnarray}
\ell_{opt} \sim \left\{%
\begin{array}{ll}
    L^{d_{\rm opt}}, & \hbox{$L \ll a^\nu$;} \\
    L, & \hbox{$L \gg a^\nu$.} \\
\end{array}%
\right.
\end{eqnarray}
where $\nu$ is the percolation correlation exponent~\cite{strelniker,
  zhenhua}. For $d=2$, $d_{\rm opt} \approx 1.22$ and for $d=3$, $d_{\rm opt}
\approx 1.44$ \cite{Cieplak,Porto99}. Here we show \cite{Sameet04}
that for any network of size $N$ and any finite $a$, there exists a crossover
network size $N^*(a)$ such that for $N \ll N^*(a)$ the scaling properties of
the optimal path are in the strong disorder regime, while for $N \gg N^*(a)$
the typical optimal paths are in the weak disorder regime. We evaluate below
the function $N^*(a)$.

\begin{figure}[h]
\includegraphics[width=11.0cm,height=5.0cm,angle=0]{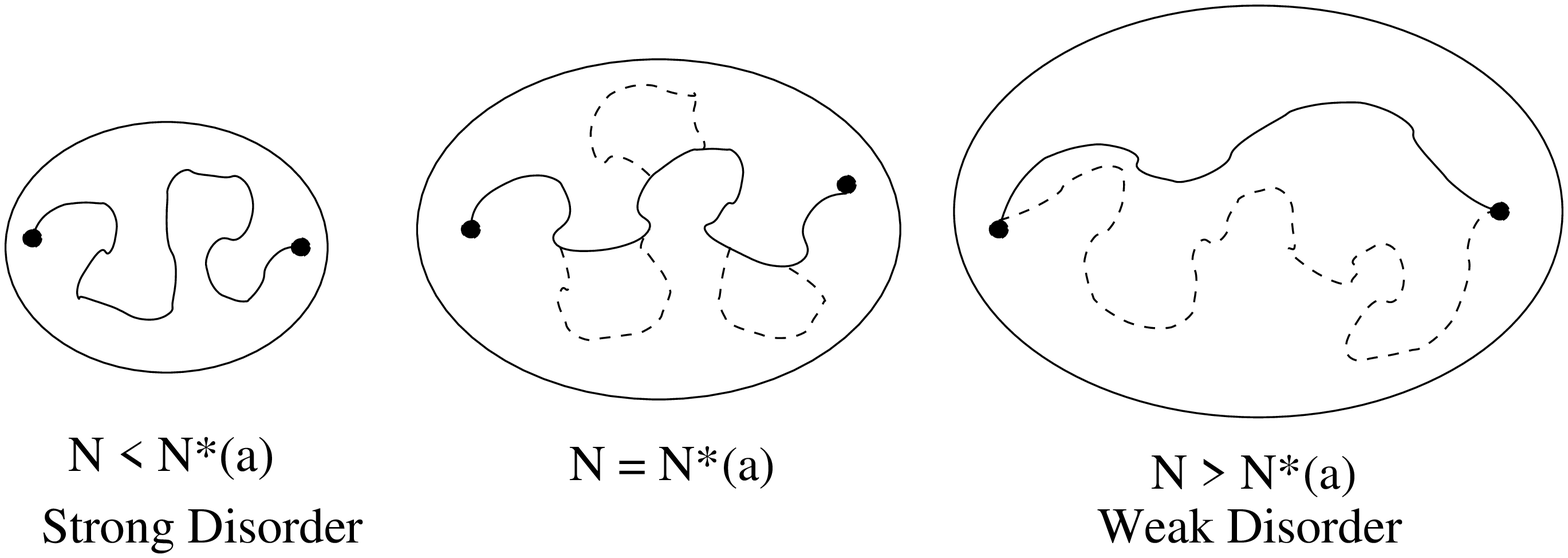}
\caption{Schematic representation of the transition in the topology of the
  optimal path with system size $N$ for a given disorder strength $a$. The
  solid line shows the optimal path at a finite value of $a$ connecting two
  nodes indicated by the filled circles. The portion of the min-max path that
  is distinct from the optimal path is indicated by the dashed line. (a) For
  $N \ll N^*(a)$ (i.e. $\ell_{\infty} \ll \ell^*(a)$), the optimal path
  coincides with the min-max path, and we expect the statistics of the SD
  limit. (b) For $N = N^*(a)$ (i.e. $\ell_{\infty} = \ell^*(a)$), the optimal
  path starts deviating from the min-max path. (c) For $N \gg N^*(a)$ (i.e.
  $\ell_{\infty} \gg \ell^*(a)$), the optimal path has almost no links in
  common with the min-max path, and we expect the statistics of the WD
  limit (After \cite{Sameet04,kalkotasameet}).}
\label{fs.1}
\end{figure}

In general, the average optimal path length $\ell_{opt}(a)$ in a
weighted network depends on $a$ as well as on $N$. In the
following we use instead of $N$ the min-max path length
$\ell_{\infty}$ which is related to $N$ as
$\ell_\infty\equiv\ell_{\rm opt}(\infty)\sim N^{\nu_{\rm opt}}$
[Eqs.(\ref{E.1}) and (\ref{E.2x})] and hence $N$ can be expressed
in terms of $\ell_{\infty}$,
\begin{equation}
N\sim\ell_\infty^{1/\nu_{\rm opt}}. \label{e1}
\end{equation}
Thus, for finite $a$, $\ell_{\rm opt}(a)$ depends on both $a$ and
$\ell_{\infty}$. We expect a crossover length $\ell^*(a)$, which
corresponds to the crossover network size $N^*(a)$, such that (i)
for $\ell_\infty\ll\ell^*(a)$, the scaling properties of
$\ell_{\rm opt}(a)$ are of the strong disorder regime, and (ii)
for $\ell_\infty\gg\ell^*(a)$, the scaling properties of
$\ell_{\rm opt}(a)$ are of the weak disorder regime. In
Fig.~\ref{fs.1}, we show a schematic representation of the changes
of the optimal path as the network size increases.

In order to study the transition from strong to weak disorder, we
introduce a measure which indicates how close or far the
disordered network is from the limit of strong disorder. A natural
measure is the ratio
\begin{equation}
W(a)\equiv\frac{\ell_{\rm opt}(a)}{\ell_\infty}. \label{equation1}
\end{equation}
Using the scaling relationships between $\ell_{\rm opt}(a)$ and
$N$ in both regimes, and $\ell_{\infty} \sim N^{\nu_{\rm opt}}$, we get
\begin{equation}
\ell_{\rm opt}(a)\sim\cases{
 \ell_\infty\sim N^{\nu_{\rm opt}} & [SD]\cr
\ln\ell_\infty\sim\ln N            & [WD].}
\label{equation2}
\end{equation}
From Eq.~(\ref{equation1}) and  Eq.~(\ref{equation2}) it follows,
\begin{equation}
W(a)\sim\cases{
\mbox{const.}                    & [SD]\cr
\ln\ell_\infty/\ell_\infty & [WD].}
\label{equation3}
\end{equation}

We propose the following scaling Ansatz for $W(a)$,
\begin{equation}
W(a)=F\left({\ell_\infty\over\ell^*(a)}\right), \label{equation4}
\end{equation}
where
\begin{equation}
F(u)\sim\cases{
\mbox{const.}  & $u\ll 1$\cr
      \ln(u)/u & $u\gg 1$},
\label{equation5}
\end{equation}

with
\begin{equation}
\label{e2} u\equiv{\ell_\infty\over\ell^{*}(a)}.
\end{equation}

We now develop analytic arguments~\cite{Sameet04} to obtain the
dependence of the crossover length $\ell^*$ on the disorder strength
$a$.
These arguments will also give a clearer picture about the nature
of the transition of the optimal path with disorder strength.

\begin{figure}[h]
\includegraphics[width=5.0cm,height=6.0cm,angle=270]{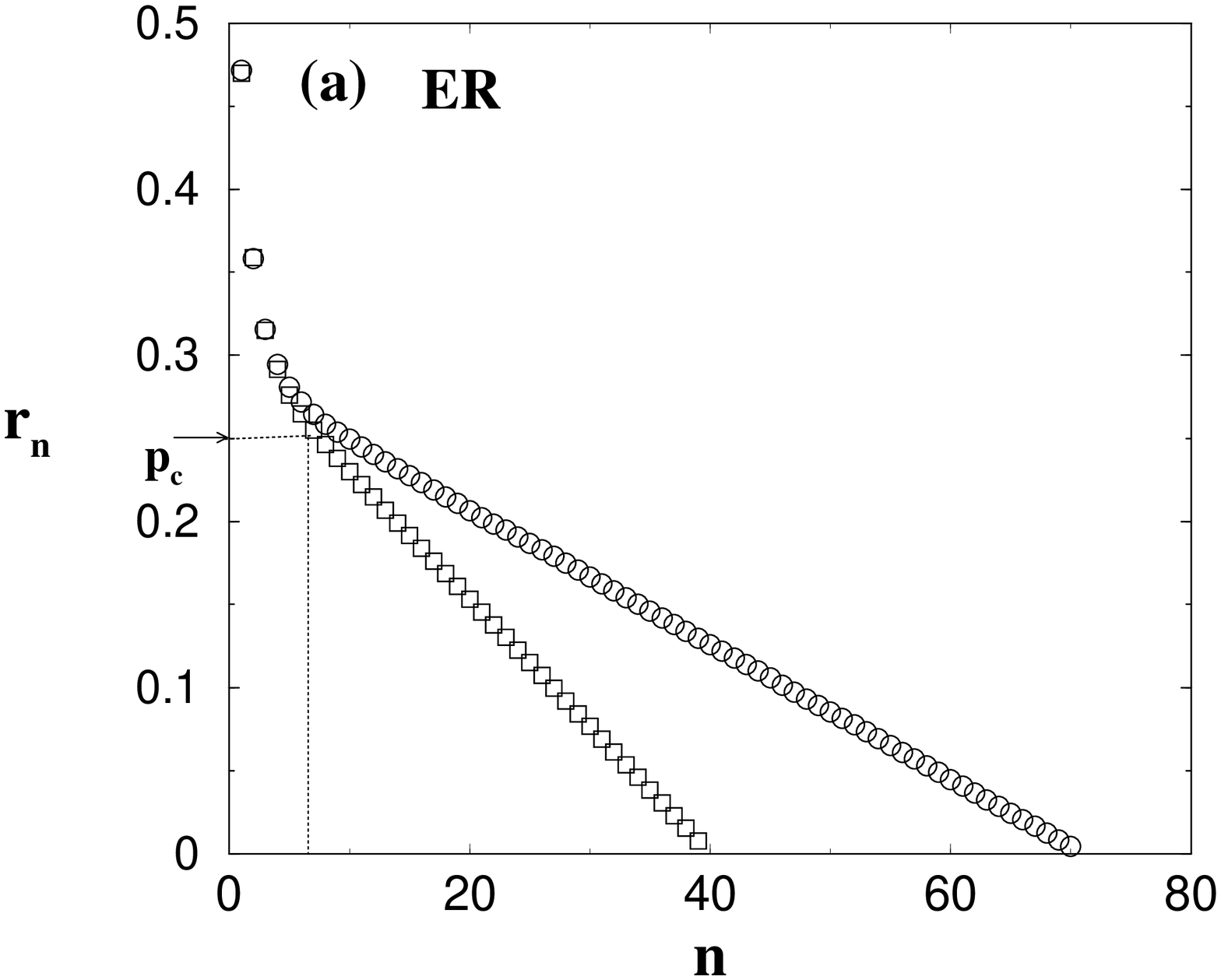}
\includegraphics[width=5.0cm,height=6.0cm,angle=270]{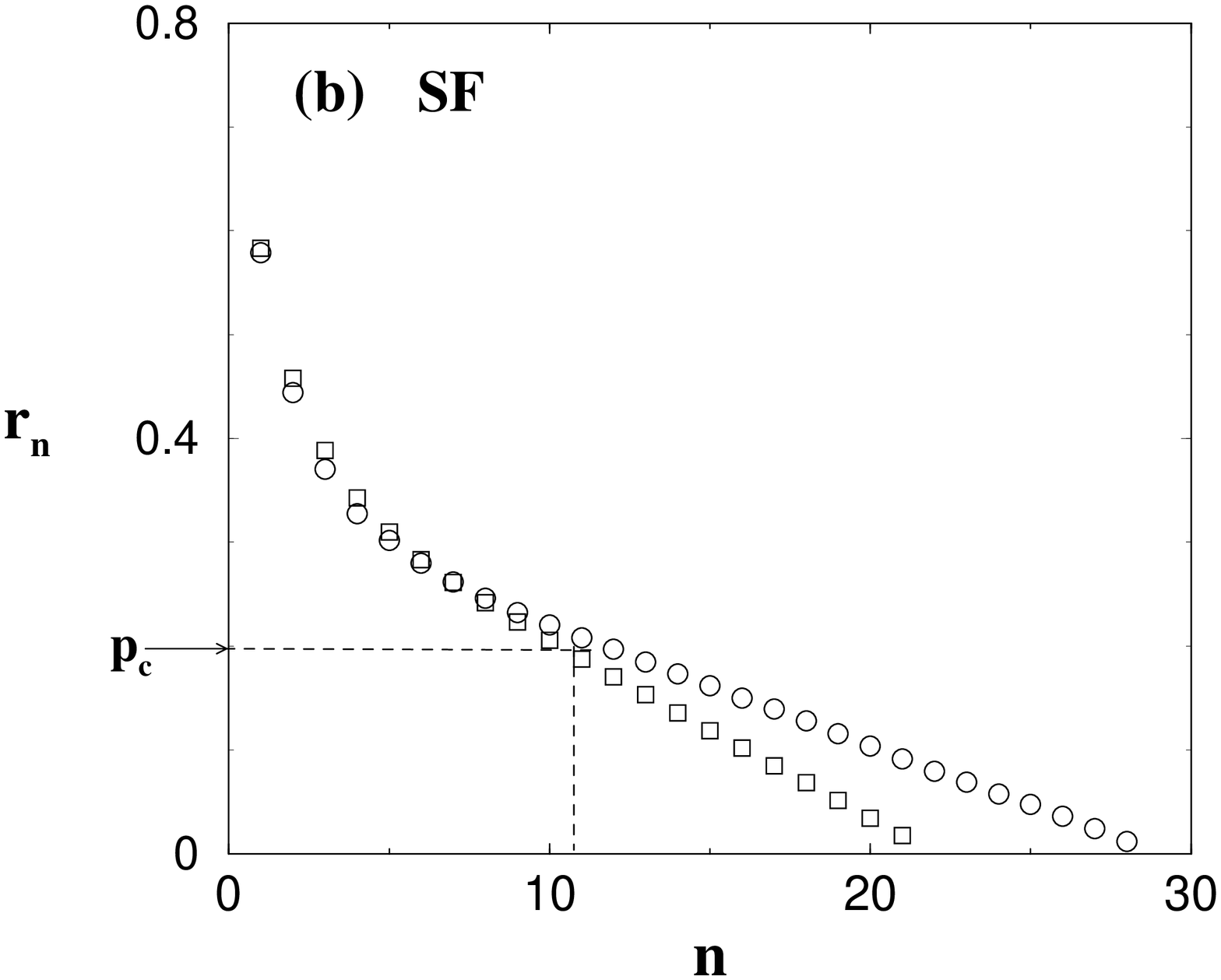}
\caption{Dependence on rank $n$ of the average values of the
random
  numbers $r_n$ along the most probable optimal path for (a) ER random
  networks of two different sizes $N = 4096$ ($\Box$) and $N = 16384$
  ($\circ$) and, (b) SF random networks (After \cite{Sameet04,kalkotasameet}).}
\label{fs.2}
\end{figure}

We begin by making few observations about the min-max path. In
Fig.~\ref{fs.2} we plot the average value of the random numbers
$r_n$ on the min-max path as a function of their rank $n$ ($1 \le
n \le\ell_{\infty}$) for ER networks with $\langle k \rangle = 4$
and for SF networks with $\lambda = 3.5$. This can be done for a
min-max path of any length but in order to get good statistics we
use the most probable min-max path length. We call links with $r
\le p_c$ ``black'' links, and links with $r>p_c$ ``gray'' links,
following the terminology of Ioselevich and Lyubshin \cite{Ios04}
where $p_c$ is the percolation threshold of the network
\cite{CEBH00}.

\begin{figure}[h]
\includegraphics[width=5.0cm,height=7.0cm,angle=270]{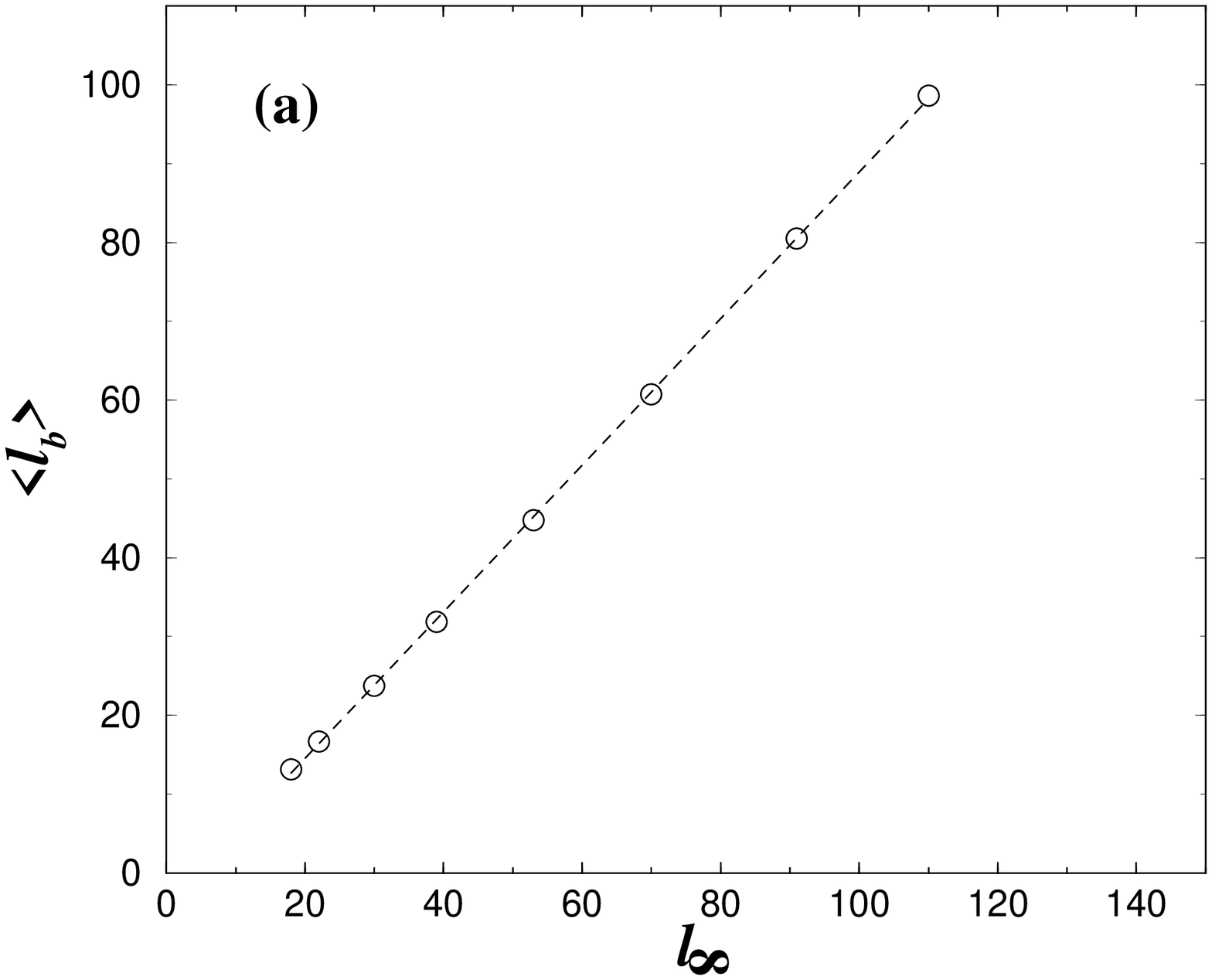}
\includegraphics[width=5.0cm,height=7.0cm,angle=270]{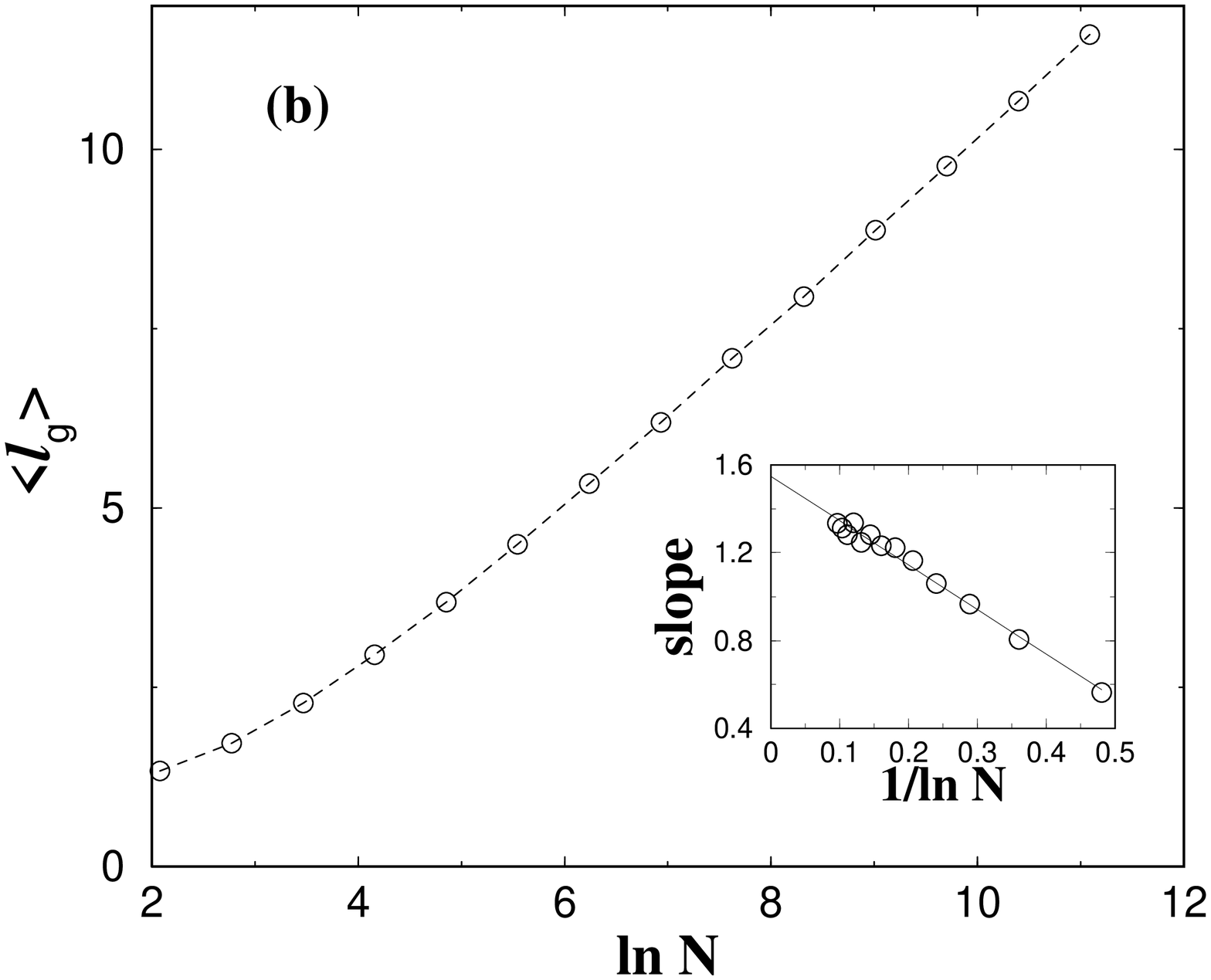}
{\caption{(a) The average number of links $\langle\ell_b \rangle$
with random number values $r \le p_c$ on the min-max path plotted as a
  function of its length $\ell_\infty$ for an ER network, showing that
  $\langle\ell_b \rangle$ grows linearly with $\ell_\infty$. (b) The
  average number of links $\langle\ell_g\rangle$ with random number
  values $r>p_c$ on the min-max path versus $\ln N$ for an ER
  network, showing that $\langle\ell_g\rangle\sim\ln N$. The
  inset shows the successive slopes, indicating
  that in the asymptotic limit $\langle\ell_g\rangle \approx 1.55\ln N$ (After \cite{Sameet04,kalkotasameet}).
\label{fs.3}}}
\end{figure}

We make the following observations regarding the min-max path:

\begin{itemize}

\item[{(i)}] For $r_n<p_c$, the values of $r_n$ decrease linearly
with rank $n$, implying that the values of $r$ for black links are
     uniformly distributed between $0$ and $p_c$, consistent with the
     results of Ref.~\cite{Szabo03}. This is shown in Fig.~\ref{fs.2}.

\item[{(ii)}] The average number of black links, $\langle\ell_b
     \rangle$, along the min-max path increases linearly with the
     average path length $\ell_\infty$. This is shown in Fig.~\ref{fs.3}a.

\item[{(iii)}] The average number of gray links $\langle\ell_g
\rangle$
     along the min-max path increases logarithmically with the average
     path length $\ell_\infty$ or, equivalently, with the network size
     $N$. This is shown in Fig.~\ref{fs.3}b.

\end{itemize}
\noindent The simulation results presented in Fig.~\ref{fs.3} are
for ER networks; however, we have confirmed that the observations
(ii) and (iii) are also valid for SF networks with $\lambda >3$ \cite{Sameet04,Tomer_grey}.

Next we discuss our observations using the concept of the
MST. The path on the MST between any two nodes $A$ and $B$, is the
optimal path between the nodes in the strong disorder limit---i.e,
the min-max path.

In order to construct the MST we use the bombing algorithm (See
Section~\ref{sec.alg} D). At the point that one cannot remove more
links without disconnecting the graph, the number of remaining
black links is
\begin{equation}
N_b={N\langle k\rangle p_c\over 2}, \label{equation6}
\end{equation}
where $\langle k \rangle$ is the average degree of the original
graph and $p_c$ is given by \cite{CEBH00}
\begin{equation}
p_c={\langle k \rangle\over\langle k^2 -k\rangle}.
\label{equation7}
\end{equation}

\begin{figure}[h]
\includegraphics[width=10.0cm,height=8.0cm,angle=0]{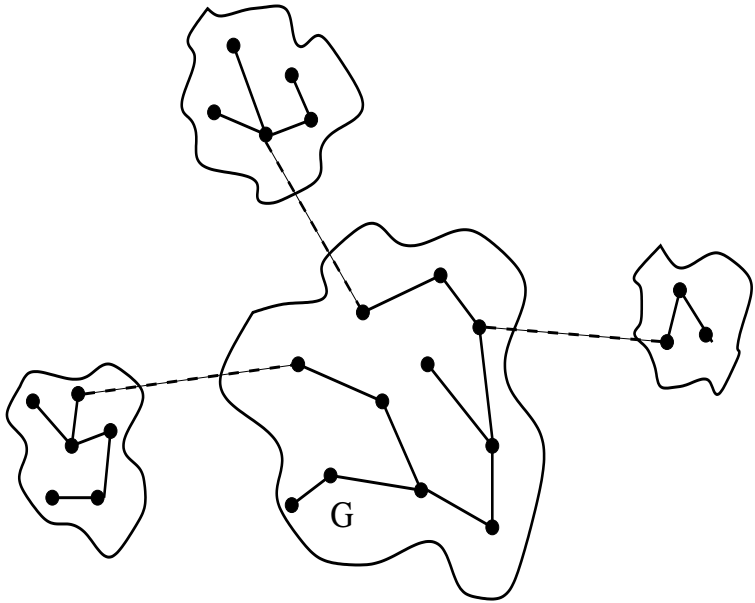}
\caption{Schematic representation of the structure of the minimal
  spanning tree, at the percolation threshold, with G being the giant
  component. Inside each cluster, the nodes are connected by black links
  to form a tree. The dotted lines represent the gray links which
  connect the finite clusters to form the gray tree. In this example
  $N_c = 4$ and the number of gray links equals $N_c - 1 = 3$ (After \cite{Sameet04,kalkotasameet}).}
\label{fs.4}
\end{figure}

The black links give rise to $N_c$ disconnected clusters. One of
these is a spanning cluster, called the {\it giant component} or IIC (see Section \ref{sec.alg} F).
 The
$N_c$ clusters are linked together into a connected tree by
exactly $N_c-1$ gray links (see Fig.~\ref{fs.4}). Each of the
$N_c$ clusters is itself a tree, since a random graph can be
regarded as an infinite dimensional system, and at the percolation
threshold in an infinite dimensional system the clusters can be
regarded as trees. Thus the $N_c$ clusters containing $N_b$ black
links, together with $N_c-1$ gray links form a spanning tree
consisting of $N_b + N_c -1$ links.

Thus the MST provides {\it all} min-max path between any two sites
on the graph. Since the MST connects all $N$ nodes, the number of
links on this tree must be $N-1$, so
\begin{equation}
N_b + N_c = N. \label{equation8}
\end{equation}
From Eq.~(\ref{equation6}) and Eq.~(\ref{equation8}) it follows
that
\begin{equation}
N_c=N\left(1-{\langle k\rangle p_c\over 2}\right).
\label{equation9}
\end{equation}
Therefore $N_c$ is proportional to $N$.

A path between two nodes on the MST consists of $\ell_b$ black
links. Since the black links are the links that remain after
removing all links with $r>p_c$, the random number values $r$ on
the black links are uniformly distributed between $0$ and $p_c$ in
agreement with observation (i) and Ref.~\cite{Szabo03}.

Since there are $N_c$ clusters which include clusters of nodes
connected by black links as well as isolated nodes, the MST can be
described as an effective tree of $N_c$ ``super'' nodes, each representing a
cluster, and $N_c - 1$ gray links. We call this tree the ``gray
tree'' (see Fig.~\ref{fs.4}). This tree is in fact a scale free
tree \cite{Tomer_grey,exp_1} with degree exponent $\lambda_g =
2.5$ for ER networks and for scale for networks with $\lambda \ge
4$, and  $\lambda_g = (2\lambda-3)/(\lambda-2)$ for SF networks
with $3 < \lambda < 4$. If we take two nodes $A$ and $B$ on the
original network, they will most likely lie on two distinct
effective nodes of the gray tree. The number of gray links
encountered on the min-max path connecting these two nodes will
therefore equal the number of links separating the effective nodes
on the gray tree. Hence the average number of gray links
$\langle\ell_g\rangle$ encountered on the min-max path between an
arbitrary pair of nodes on the network is simply the average
diameter of the gray tree. Our simulation results (see
Fig.~\ref{fs.3}b) indicate that
\begin{equation}
\label{equation9x} \langle\ell_g \rangle \sim \ln N.
\end{equation}

Since $\langle\ell_g\rangle\sim\ln\ell_\infty\ll\ell_\infty$, the
average number of black links $\langle\ell_b\rangle$ on the
min-max path scales as $\ell_\infty$ in the limit of large
$\ell_\infty$ in agreement
with observation (2) as shown in Fig.~\ref{fs.3}a.

Next we discuss the implications of our findings for the crossover from
strong to weak disorder. From observations (i) and (ii), it follows that for
the portion of the path belonging to the giant component, the distribution of
random values $r$ is uniform. Hence we can approximate the sum of weights
by~\cite{tomerdistributions},
\begin{equation}
\sum^{\ell_b}_{k=1}\exp(a r_k) \approx \frac{\ell_b}{p_c}
\int_0^{p_c} \exp{ a r} \;dr = \frac{\ell_b}{a p_c}  \left(\exp(a
p_c)-1\right)\equiv\exp(ar^\ast) , \label{equation10}
\end{equation}
where $r^{\ast} \approx p_c+ (1/a) \ln (\langle \ell_b \rangle / a
p_c)$. Since $\langle\ell_b\rangle\approx\ell_\infty$,
\begin{equation}
r^{\ast}\approx p_c+ \frac{1}{a} \ln \left(\frac{\ell_{\infty}}{a
p_c} \right). \label{equation11}
\end{equation}
Thus restoring a short-cut link between two nodes on the optimal
path with $p_c < r < r^{\ast}$ may drastically reduce the length
of the optimal path. When $ap_c \gg \ell_{\infty}$,  $r^{\ast} <
p_c$ and such a link does not exists, if $\ell_{\infty} >
ap_c$, the probability that such a link exist becomes positive.
Hence when the min-max path is of length $\ell_{\infty}
\approx ap_c$, the optimal path starts deviating from the min-max
path. The length of the min-max path at which the deviation first
occurs is precisely the crossover length $\ell^*(a)$, and
therefore $\ell^{\ast}(a) \sim ap_c$. In the case of a network
with an arbitrary degree distribution we can write using
Eq.~(\ref{equation7}), $\ell^{\ast}(a) \sim a {\langle k
\rangle\over\langle k^2 -k\rangle}$.

Note that in the case of SF networks, as $\lambda \to 3^+$, $p_c$ approaches
zero and consequently $\ell^*(a)\to 0$. This suggests that for any finite
value of disorder strength $a$, a SF network with $\lambda \le 3$ is in the
weak disorder regime. We perform numerical simulations and show that the
results agree with our theoretical predictions. For the details of our
simulation methods see Section~\ref{sec.alg}.

\begin{figure}[h]
\includegraphics[width=6.5cm,height=7.5cm,angle=270]{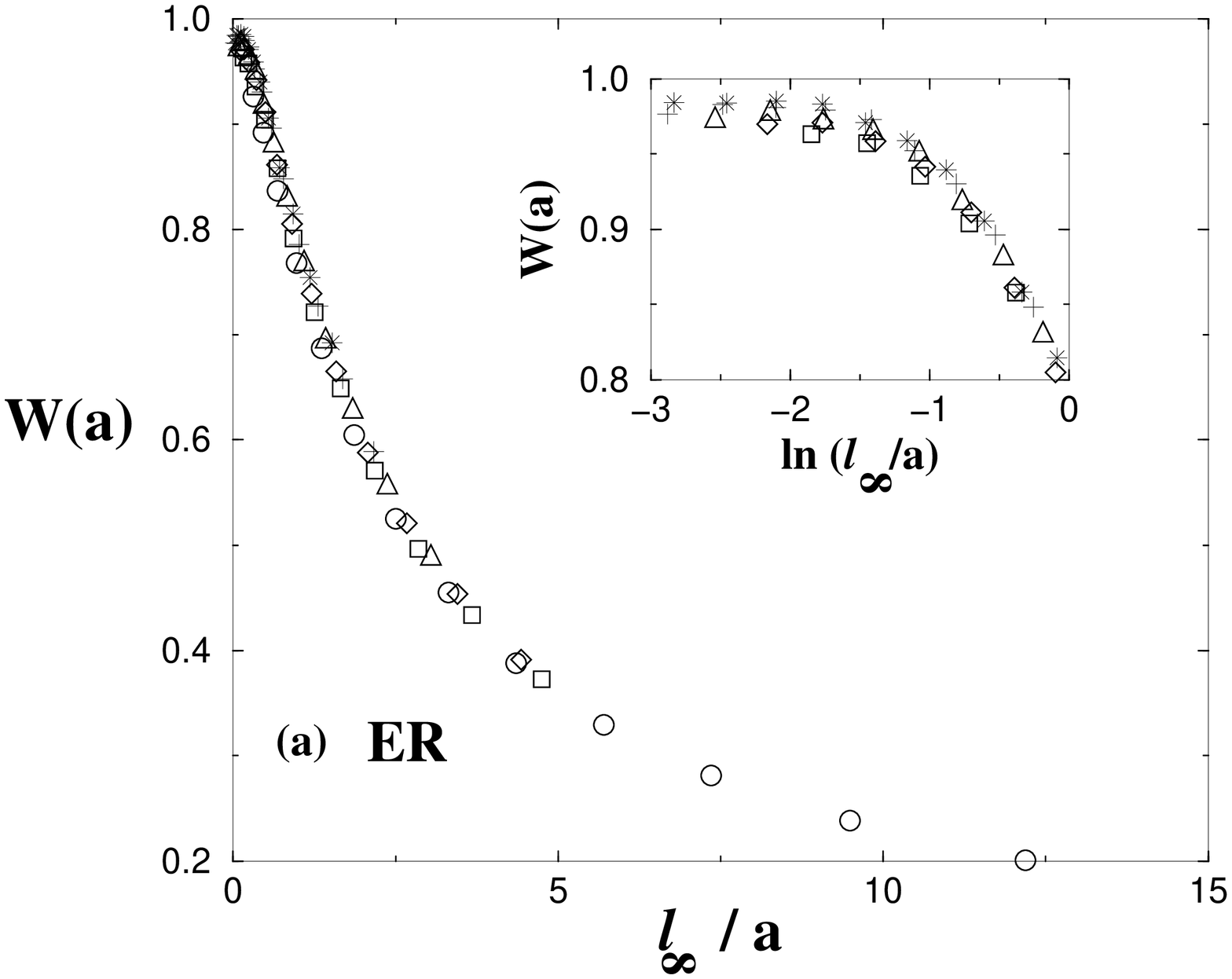}
\includegraphics[width=6.5cm,height=7.5cm,angle=270]{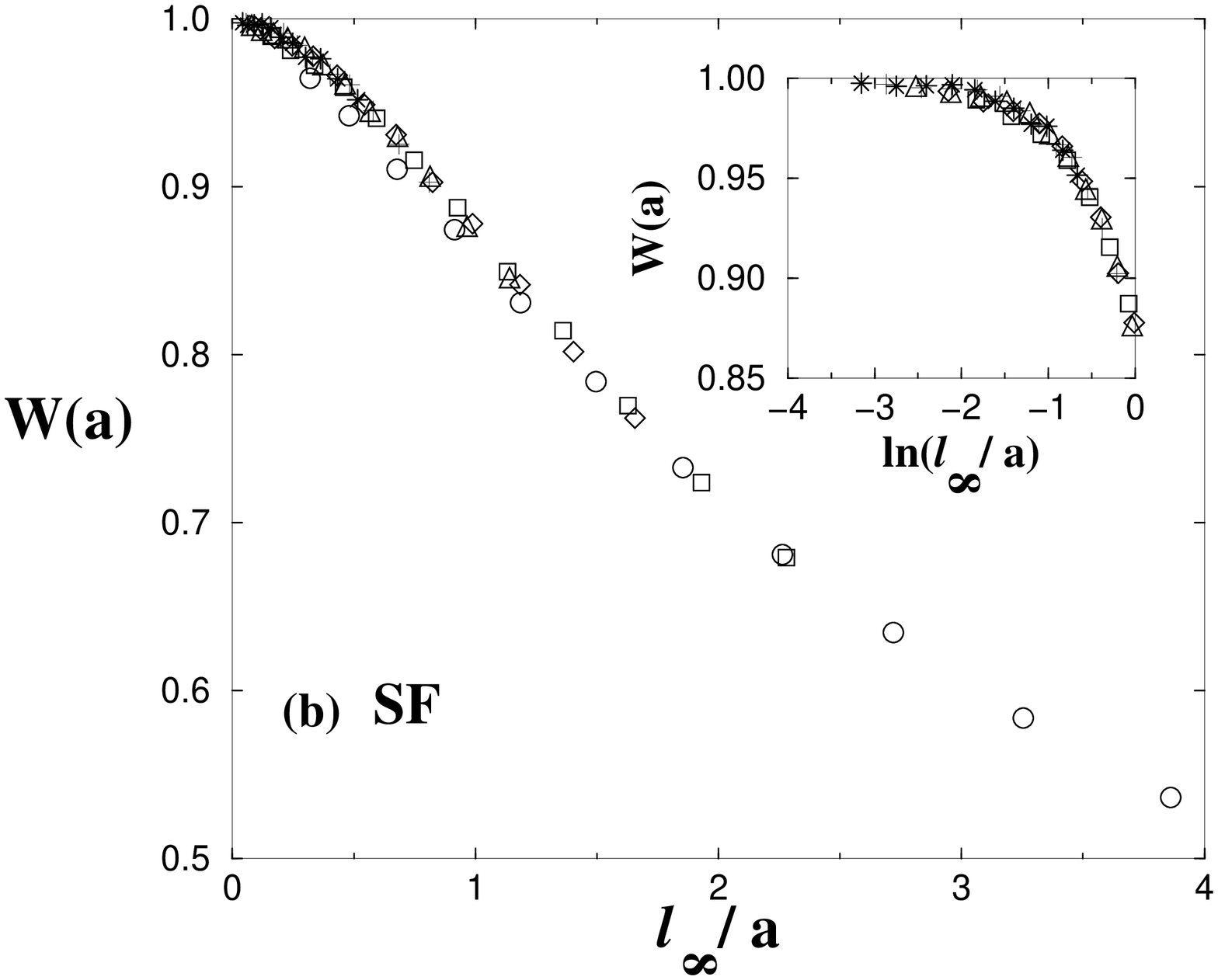}
\caption{Test of Eqs.~(\ref{equation4}) and (\ref{equation5}). (a)
   $W(a)$ plotted as a function of $\ell_{\infty}/a$ for different
   values of $a$ for ER networks with $\langle k \rangle = 4$. The
   different symbols represent different $a$ values: $a = 8$($\circ$),
   $a = 16$($\Box$), $a = 22$($\diamond$), $a = 32$($\bigtriangleup$),
   $a = 45$($+$), and $a = 64$($\ast$). (b) Same for SF networks with
   $\lambda = 3.5$. The symbols correspond to the same values of
   disorder as in (a). The insets show $W(a)$ plotted against
   $\log(l_{\infty}/a)$, and indicate for $\ell_\infty\ll a$, $W(a)$
   approaches a constant in agreement with Eq.~(\ref{equation5}) (After \cite{Sameet04,kalkotasameet}).}
\label{fs.5}
\end{figure}

\begin{figure}[h]
\includegraphics[width=6.5cm,height=7.0cm,angle=270]{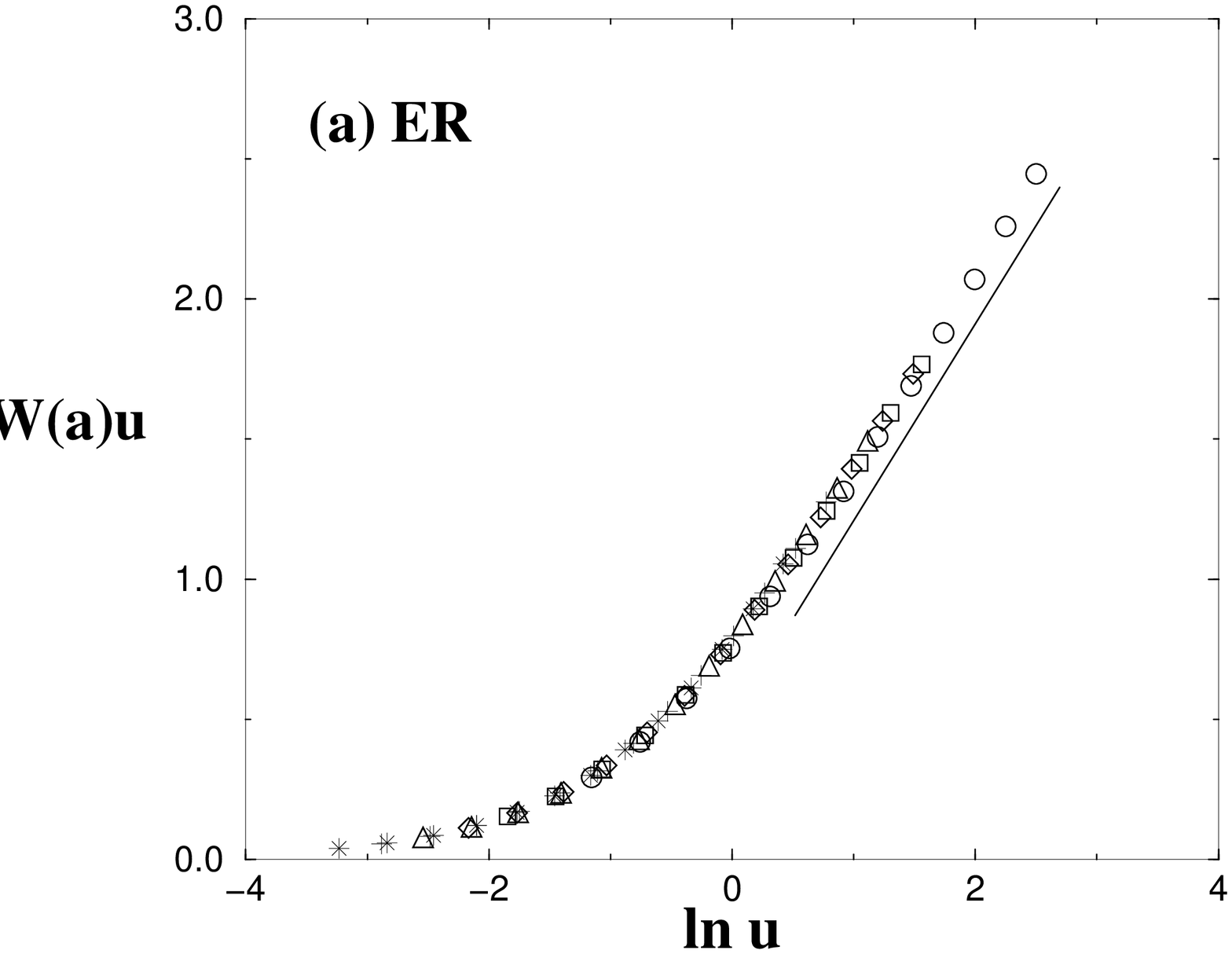}
\hspace{1cm}\includegraphics[width=6.5cm,height=7.0cm,angle=270]{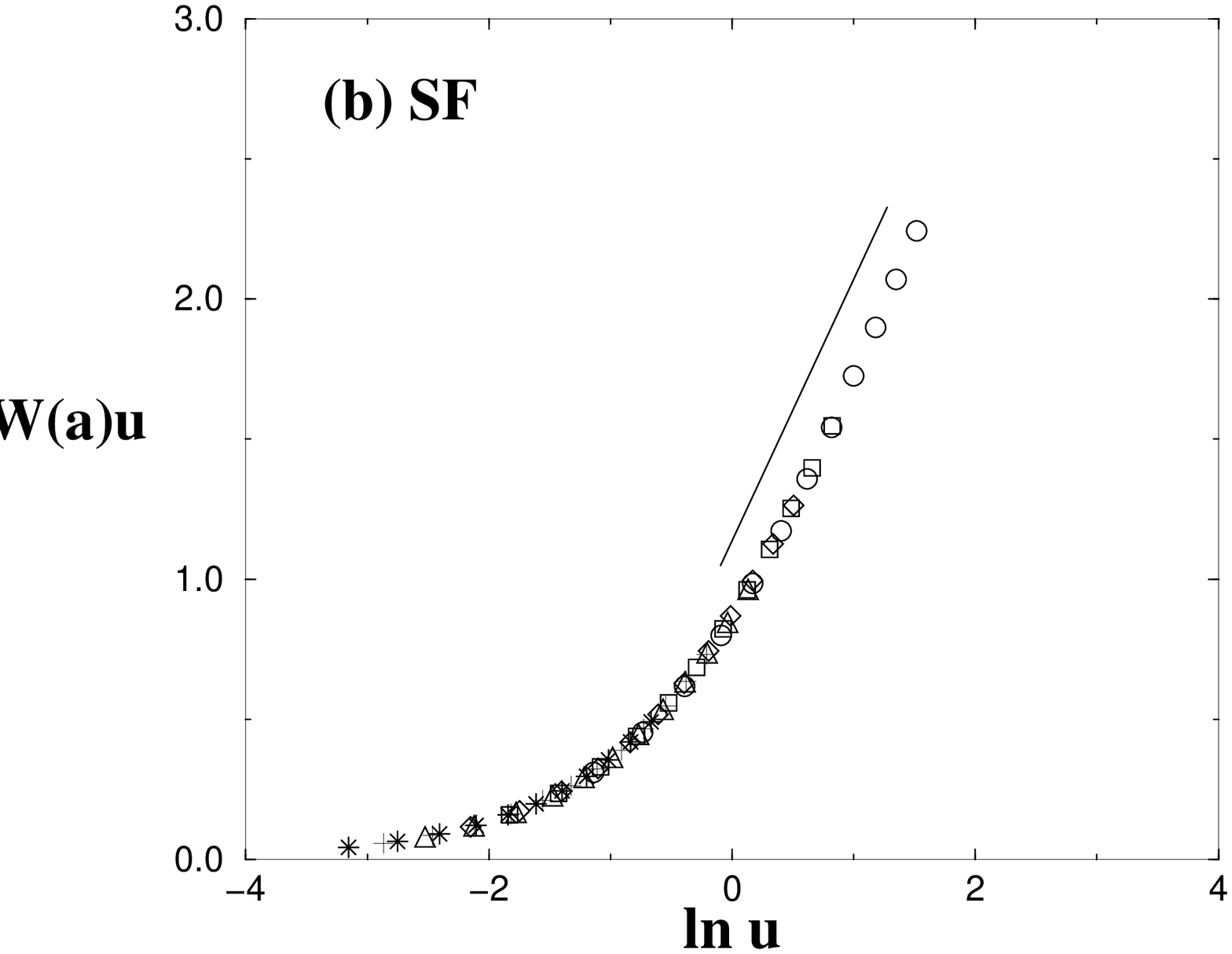}
\caption{(a) Plot of $W(a)u = \ell_{\rm opt}(a)/\ell^*(a) =
\ell_{\rm
  opt}(a)/a $ vs $\ln u \equiv \ln (\ell_{\infty}/\ell^*(a))=
  \ln(\ell_{\infty}/a) $ for ER networks with $\langle k \rangle = 4$
  for different values of $a$. (b) Plot of $W(a)u = \ell_{\rm
  opt}(a)/\ell^*(a) = \ell_{\rm opt}(a)/a $ vs $\ln u= \ln
  (\ell_{\infty}/\ell^*(a))= \ln(\ell_{\infty}/a)$ for SF networks with
  $\lambda = 3.5$. The values of $a$ represented by the symbols in (a)
  and (b) are the same as in Fig.~(\ref{fs.2}) (After \cite{Sameet04,kalkotasameet}).}
\label{fs.6}
\end{figure}

From our theoretical
arguments, $\ell^{\ast}(a) \sim a$ and therefore, from
Eq.~(\ref{equation4}), $W(a)$ must be a function of
$\ell_\infty/a$. In Fig.~\ref{fs.5} we show the ratio $W(a)$ for
different values of $a$ plotted against
$\ell_\infty/\ell^*(a)\equiv\ell_\infty/a$ for ER networks with
$\langle k \rangle = 4$ and for SF networks with $\lambda=3.5$.
The excellent data collapse is consistent with the scaling
relations Eq.~(\ref{equation4}). Fig.~\ref{fs.6} shows the scaled
quantities $W(a)u=\ell_{\rm opt}(a)/\ell^*(a)$ vs. $\ln
u\equiv\ln(\ell_\infty/\ell^*(a))\equiv\ln(\ell_\infty/a)$, for
both ER networks with $\langle k \rangle = 4$ and for SF networks
with $\lambda=3.5$. The curves are linear at large
$u\equiv\ell_\infty/\ell^*(a)$, supporting the validity of the
logarithmic term in Eq.~(\ref{equation5}) for large $u$.

To summarize, for both ER random
networks and SF networks we obtain a scaling function for the
crossover from weak disorder characteristics to strong disorder
characteristics. We show that the crossover occurs when the
min-max path reaches a crossover length $\ell^*(a)$ and
$\ell^*(a)\sim a$. Equivalently, the crossover occurs when the
network size $N$ reaches a crossover size $N^*(a)$, where
$N^*(a)\sim a^3$ for ER networks and for SF networks with $\lambda
\ge 4$ and $N^*(a)\sim a^{\frac{\lambda - 1}{\lambda -3}}$ for SF
networks with $3 < \lambda < 4$.

\subsection{General Disorder: Criterion for SD, WD crossovers}

Until now we considered a specific form of $P(\tau) \equiv P(\tau,a)= 1/ (a \tau)$ with $1 <
\tau < e^a$. The question is what happens for other distributions of weights and what is
the general criterion to determine which form of $P(\tau)$ can lead to strong
disorder, and what is the general condition for strong or weak disorder
crossover. We present analytical results~\cite{yipingprl} for such a
criterion which are supported by extensive simulations.  Using this criterion
we show that certain power law distributions and lognormal distributions,
$P(\tau,a)$, where $a$ is a parameter determining the broadness of the
distribution, can lead to strong disorder and to a weak-strong disorder
crossover~\cite{Porto99, Brauns03, Sameet04}. We also show that for
$P(\tau,a)$ uniform, Poisson or Gaussian, only weak disorder occurs
regardless of the broadness of $P(\tau,a)$.  Importantly, we find that for
all $P(\tau,a)$ that possess a strong-weak disorder crossover, the
distributions of the optimal path lengths display the same universal
behavior.

If we express $\tau$ in terms of a random variable $r$ uniformly distributed
 in $[0,1)$, we can use the same gray and black link formalism as in the
 previous section. This can be achieved by defining $r(\tau)$ by a relation:
\begin{equation}
  r(\tau) = \int_0^\tau P(\tau',a)d\tau'.
  \label{fx1}
\end{equation}
Solving this equation with respect to $\tau$ gives us $\tau(r,a)=f(r,a)$, where $f(r,a)$ satisfies the relation
\begin{equation}
  r =\int_0^{f(r,a)} P(\tau',a)d\tau'.
  \label{fx}
\end{equation}
For a strong disorder regime, the sum of the weights of the black links on the IIC
must be smaller than the smallest weight of the removed link
$\tau_c=f(p_c,a)$:
\begin{equation}
 \sum_{i=1}^{\ell_b}\tau_i= \sum_{i=1}^{\ell_b}f(r_i,a) <\tau_c,
\end{equation}
where $r_i$ are independent random variables uniformly distributed on
$[0,p_c]$. As we shown above, $\ell_b \approx \ell_\infty$ so in the
following we will replace $\ell_b$ by the average path length in the strong
disorder limit, $\ell_\infty$.  The transition to weak disorder begins when
the probability that this sum is greater than $\tau_c$ becomes substantial.
The investigation of this condition belongs to the realm of pure mathematics
and can be answered explicitly for any functional form $f(r,a)$.  This
condition is satisfied when the mathematical expectation of the sum is
greater than $\tau_c$.
\begin{equation}
\frac{\ell_\infty}{p_c} \int_0^{p_c}f(r,a)dr >\tau_c.
\end{equation}
Thus the crossover to weak disorder happens if
\begin{equation}
\ell_\infty >A \equiv \frac{f(p_c,a) p_c}{\int_0^{p_c}f(r,a)dr}=\frac{\tau_c p_c}{\int_0^{\tau_c}\tau P(\tau,a)d\tau},
\label{eq:A}
\end{equation}
where $A$ plays the role of the disorder strength and $\tau_c$ satisfies the
equation $p_c=\int_0^{\tau_c}P(\tau,a)d\tau$.  In order for the strong
disorder to exist for any network size $N$, the disorder strength must
diverge together with the parameter $a$ of the weight distribution $a\to\infty$. In
order to determine if a network exhibits a strong disorder behavior it is
useful to introduce a scaling variable
\begin{equation}
Z \equiv \ell_\infty/A,
\label{eq:Z}
\end{equation}
so that if $Z\gg 1$ the network is in the weak disorder regime and if $Z\ll 1$, the network is in the strong disorder regime.

Note that if $f'/f > A_0$ on the entire interval $[0,1]$, then
$A>A_0 p_c$. Thus another sufficient condition for a strong
disorder to exist is
\begin{equation}
f'/f>\ell_\infty/p_c.
\label{eq:fpc}
\end{equation}
For the exponential disorder function $\tau=\exp(ar)$, we have
$f'/f=a$ and thus Eq.(\ref{eq:fpc}) coincides with the condition
of strong disorder $a p_c >\ell_\infty$ derived in the previous
Section.

In the following, we will show how the above condition is related
to the strong to weak crossover condition for the optimal path on
lattices  \cite{Cieplak, Porto99,Buldyrev06}. For the optimal path in the strong
disorder limit connecting the opposite sides of the lattice of
linear size $L$, the largest random number $r_1$ follows a
distribution characterized by a width which scales as
$L^{-1/\nu}$, where $\nu$ is the percolation connectivity length
exponent~\cite{stauffer,BH96,coniglio-1982:cluster_structure,TomerandReuven}.
The transition to  weak disorder starts when the optimal path may
prefer to go through a slightly larger value $r_2$, taken from the
same distribution and thus $r_2-r_1 \sim p_c L^{-1/\nu}$. The
condition for this to happen is $[f(r_2)-f(r_1)]/f(r_1) <1$, which
is equivalent to
\begin{equation}
f'/f <L^{1/\nu}/p_c.
\label{eq:fpc2}
\end{equation}
Now we will show that this condition is equivalent to (\ref{eq:fpc}).
 Percolation on Erd\H{o}s-R\'{e}nyi (ER)
networks  is equivalent to percolation on a lattice at the upper critical dimension
$d_c=6$~\cite{BH96,Cohen02}. For $d=6$, $L \sim N^{1/6}$, and $\nu=1/2$.
Thus indeed $L^{1/\nu}\sim N^{1/(d_c\nu)} \sim \ell_\infty$ \cite{Brauns03}.

Following similar arguments for a scale-free network with degree distribution
$P(k)\sim k^{-\lambda}$ and $3<\lambda<4$, we can replace $L^{-1/\nu}$ by
$N^{-(\lambda-3)/(\lambda-1)}$ since
$d_c=2(\lambda-1)/(\lambda-3)$~\cite{Cohen02}. Thus, due to Eq. (\ref{eq:nu_opt}) $L^{1/\nu} \sim \ell_\infty$
and we can introduce the analogous scaling parameter $Z$ for lattices:
\begin{equation}
Z={L^{1/\nu} \over p_c f'/f}.
\label{eq:Zlat}
\end{equation}

Next we calculate $A$ for several specific weight distributions $P(\tau)$
\cite{yipingprl}. We begin with the well-studied exponential disorder
function $f(x)=e^{a r}$, where $r$ is a random number between $0$ and
$1$~\cite{strelniker,Cieplak}.  From Eq.~(\ref{fx}) follows that
$P(\tau,a)=1/(a\tau)$, where $\tau \in [1,e^a]$. Using Eq.~(\ref{eq:A}) we
have
\begin{equation}
  A=ap_c\tau_c/(\tau_c-1)\sim ap_c;
  \label{exp}
\end{equation}
For fixed $A$, but different $a$ and $p_c$, we expect to obtain the same optimal path behavior. Indeed,
this has been shown to be valid~\cite{zhenhua, strelniker,tomerdistributions,yipingprl}.

Next we study $A$ for the disorder function $f(r,a)=r^{a}$, with $r$ between
$0$ and $1$ where $a>0$ ~\cite{alex}. For this case the disorder distribution
is a power law $P(\tau,a)=a^{-1}\tau^{1/a-1}$. Following Eq.~(\ref{eq:A}) we
obtain
\begin{equation}
  A = a+1 \sim a.
\end{equation}
Note that here $a$ plays a similar role as $a$ in Eq.~(\ref{exp}), but now
$A$ is independent of $p_c$, which means that networks with different $p_c$,
such as ER networks with different average degree $<k>=1/p_c$, yield the
same optimal path behavior.

For the power law distribution with negative exponent
$f(r)=(1-r)^{-a}$ ($a >0$), we have $P(\tau,a)=a^{-1}\tau^{-1-1/a}$
and
\begin{equation}
  A = \frac{(a-1)p_c (1-p_c)^{-a}}{(1-p_c)^{1-a}-1}\sim \frac {ap_c}{1-p_c}.
\end{equation}

We further generalize the power law distribution with the disorder function
$f(r,a)=r^a$ by introducing the parameter $0\le\Delta\le 1$ which is defined as
the lower bound of the uniformly distributed random number $r$, i.e.,
$1-\Delta\le r\le 1$ \cite{alex}.  Under this condition, the distribution becomes
\begin{equation}
  P(\tau,a)=\frac{\tau^{1/a-1}}{\vert a\vert \Delta}  \qquad   \tau\in [(1-\Delta)^a,1].
\end{equation}
Again using Eq. (\ref{eq:A}), we obtain
\begin{equation}
  A= \frac{ a p_c \Delta}{p_c\Delta+1-\Delta}.
\end{equation}

Table~\ref{table1} shows the results of similar analysis for the lognormal,
Gaussian, uniform and exponential distributions $P(\tau,a)$. From
Table~\ref{table1}, we see that for exponential function, power law and
lognormal distributions, $A$ is proportional to $a$ and thus can become large. However for
uniform, Gaussian and exponential distributions, $A$ is limited to a value of
order $1$, so $Z \gg 1$ for large $N$ and the
optimal path is always in the weak disorder regime.
Note that for these distributions, $A$ is independent of $a$.

In general, one can prove that $A\to \infty$ for a given
distribution $P(\tau,a)$ as $a\to \infty$ if there exist a
normalization function $c(a)$ and a cutoff function $\tau(a)$ such
that for $\forall \epsilon >0,\forall {\rm E}>0$, $\exists M>0$ such that for $a>M$ and $\tau \in
[\tau(a),\tau(a){\rm E}]$, $\vert \tau P(\tau,a)c(a)-1\vert
<\epsilon$. We will call such functions $P(\tau,a)$ ``quasi-$1/\tau$''
distributions because they behave as $1/\tau$ in a wide range of
$\tau$. Obviously that exponential, power-law and lognormal
distributions are quasi-$1/\tau$ functions, so for them, for large
enough $a$ we can observe a strong disorder.

\begin{table}[h!]
\caption{Parameters controlling the optimal paths on networks for
  various distributions of disorder (After \cite{yipingprl}),}
\begin{center}
\begin{tabular}{|c|c|c|c|c|}\hline
  Name & Function & Distribution & Domain &$A$ \\
  \hline
  Inverse & $e^{ax}$ & $\frac{1}{a \tau}$ & $\tau \in [1,e^a]$ & $ap_c$\\
 \hline
  Power Law & $x^a$ & $\frac{\tau^{1/a-1}}{a}$ & $\tau \in (0,1]$ & $a$\\
 \hline
  Power Law & $(1-x)^a$ & $\frac{1}{a}\tau^{1+1/a}$ & $\tau \in [1,\infty]$ & $a\frac{p_c}{1-p_c}$\\
 \hline
  Power Law & $x^a$ & $\frac{\tau^{1/a-1}}{\vert a\vert\Delta}$  & $\tau \in [(1-\Delta)^a,1]$ & $a\frac{ p_c\Delta }{1-(1-p_c)\Delta}$\\
 \hline
  Lognormal & $e^{\sqrt{2}a \mathrm{erf^{-1}}(2x-1)}$  & $\frac{e^{-(\mathrm{ln}\tau)^2/2 a^2}}{\tau a \sqrt{2\pi}}$
 & $\tau \in (0,\infty)$  & $a\frac{\sqrt{2\pi} p_c}{e^{-[\mathrm{erf^{-1}}(2p_c-1)]^2}}$\\
  \hline
  Uniform & $ax$ & $1/a$  & $\tau \in [0,a]$& $2$\\
  \hline
  Gaussian & $\sqrt{2}a\mathrm{erf^{-1}}(x)$ &
  $\frac{2e^{-\tau^2/(2 a^2)}}{a\sqrt{2\pi}}$
  &  $\tau \in [0,\infty)$ & $\frac{\sqrt{2\pi}p_c\mathrm{erf^{-1}}
  }{1-e^{-[\mathrm{erf^{-1}}(p_c)]^2}}$\\
  \hline
  Exponential & $-a\mathrm{ln}(1-x)$  & $ \frac{e^{-\tau/a}}{a}$ & $\tau \in [0,\infty)$ &
  $\frac{-p_c\mathrm{ln}(1-p_c)}{p_c+(1-p_c)\mathrm{ln}(1-p_c)}$\\
  \hline
\end{tabular}
\end{center}
\label{table1}
\end{table}
To test the validity of our theory, we perform simulations of optimal paths
in $2d$ square lattices and ER networks. Random weights from different
disorder functions were assigned to the bonds. For an $L\times L$ square
lattice, we calculate the average length $\ell_{opt}$ of the optimal path from one
lattice edge to the opposite. For an ER network of $N$ nodes, we calculate
$\ell_{opt}$ between two randomly selected nodes.

\begin{figure}[h]
\begin{center}
\includegraphics[width=5.0cm,height=5.0cm,angle=0]{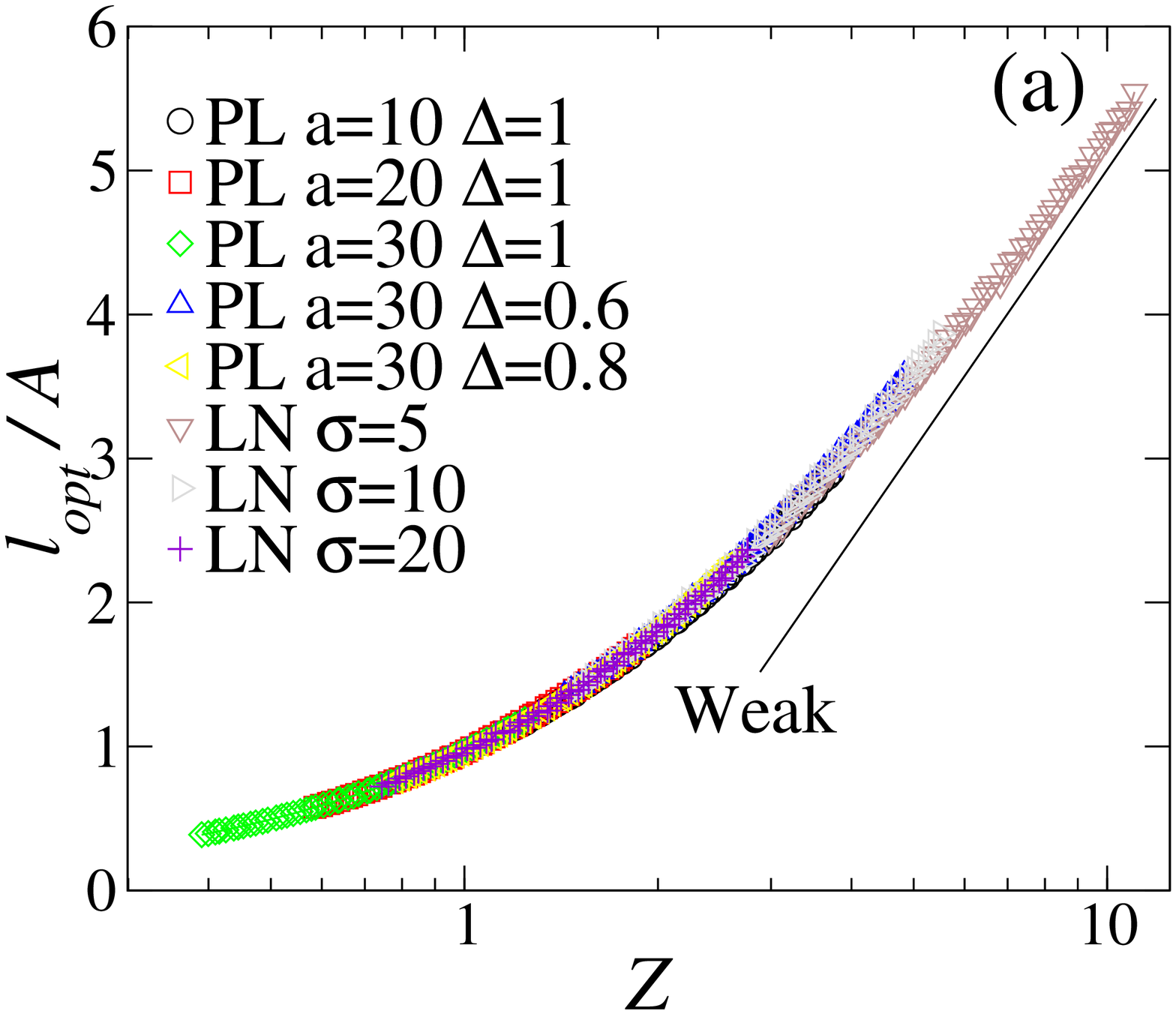}
\includegraphics[width=5.0cm,height=5.0cm,angle=0]{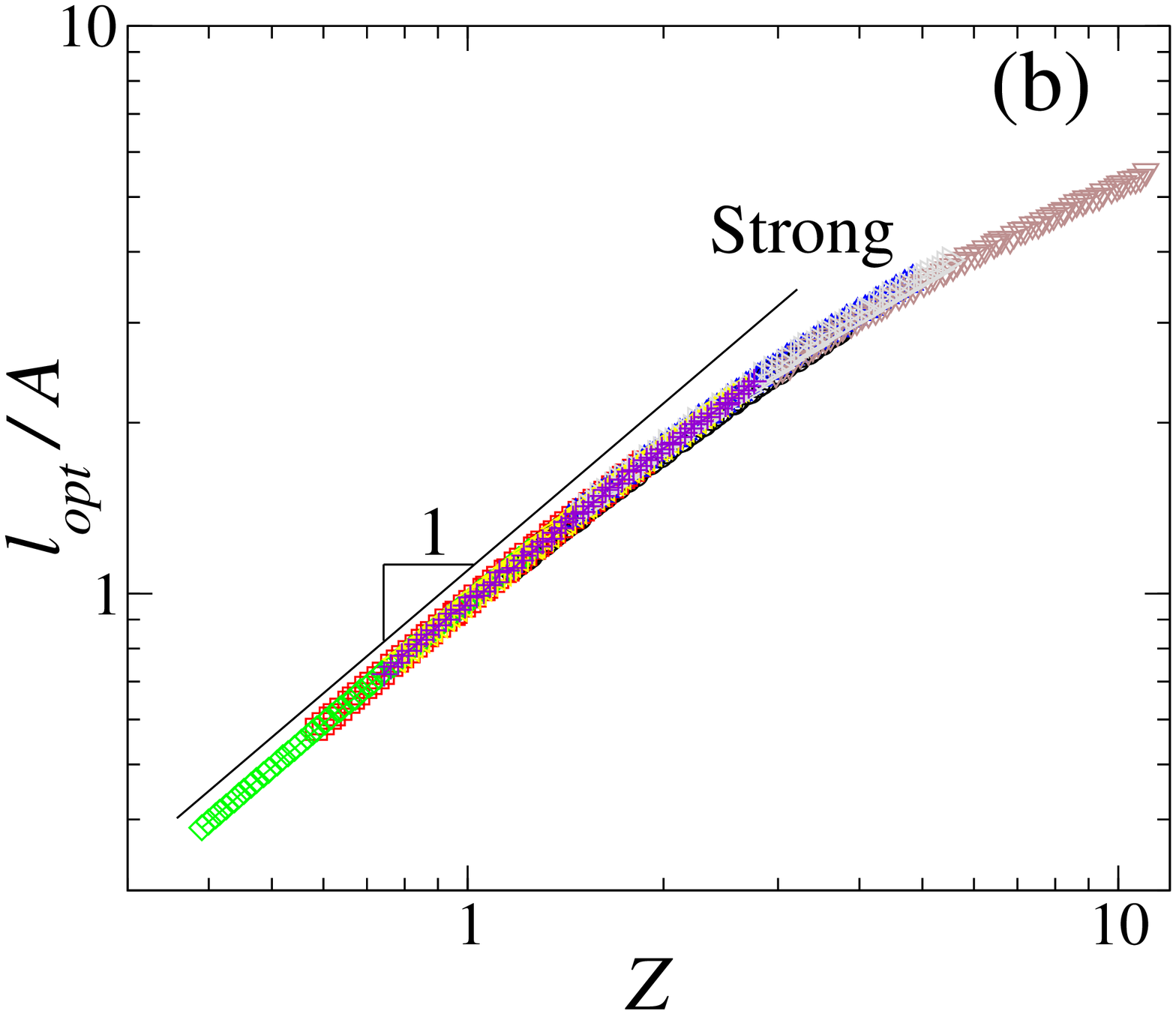}
\caption{The function $\ell_{opt}/A$ for ER networks after scaling, where $(a)$ is
  a linear-log plot and $(b)$ is a log-log plot. Distributions used are power
  law $x^a$ with $10\le a\le 30$ where $0\le x<1$, $x^a$ with $a=30$ and the
  range of $\Delta<x\le 1$ with $\Delta=0.6$ or $0.8$, and lognormal
  distribution with $10\le a \le 30$. The straight line in $(a)$
  indicates weak disorder and the straight line in $(b)$ indicates strong
  disorder (After \cite{yipingprl}).}
\label{graph3}
\end{center}
\end{figure}

Simulations for optimal paths on ER networks are shown in
Fig.~\ref{graph3}. Here we use the bombing
algorithm (See Section \ref{sec.alg} D) to determine the path length
$\ell_{\infty}$ in the strong disorder limit, which is related to
$N$ by $\ell_{\infty}\sim N^{\nu_{opt}}=N^{1/3}$~\cite{Brauns03}
(See Section \ref{sec.sd}). We see that for all disorder
distributions studied, $\ell_{opt}$ scales in the same universal way
with $Z \equiv \ell_{\infty}/A$. For $Z \gg 1$, $\ell_{opt}/A$ is
linear with $\mathrm{log}(\ell_{\infty}/A)$ as expected~(Fig. 3a).
For small $Z=\ell _\infty/A$ (Fig.~\ref{graph3} b), $\ell_{opt} \propto
\ell_{\infty}\sim N^{1/3}$, which is the strong disorder
behavior~\cite{Brauns03}. Thus, we see that when $N$ increases, a
crossover from strong to weak disorder occurs in the scaled
optimal paths $\ell_{opt}/A$ vs. $Z$. Again, the collapse of all
curves for different disorder distributions for ER networks
supports the general condition of Eq.~(\ref{eq:Z}).

Next we use Eq.~(\ref{eq:Z}) to analyze the other types of disorder given in
Table~\ref{table1} that do not have strong disorder behavior. For a uniform
distribution, $P(\tau)=1/a$ and we obtain $A=1$. The parameter $a$ cancels, so
$Z=L^{1/\nu}$ for lattices, and $Z=N^{1/3}$ for ER networks. Hence for any
value of $a$, $Z << 1$, and strong disorder behavior cannot occur for a
uniform distribution.

\begin{figure}[h ]
\begin{center}
\includegraphics[width=5.0cm,height=5.0cm,angle=0]{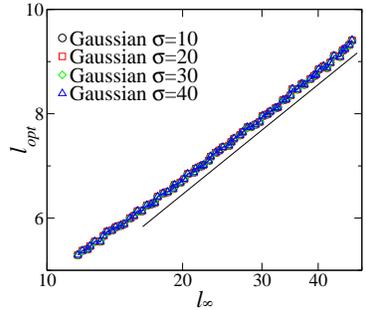}
\caption{The optimal path for Gaussian distribution of weights for ER
  networks. Note that these curves would collapse after scaling to the curves
  in Fig.~\ref{graph3} in the weak disorder tail of large $Z^{-1}$ (After \cite{yipingprl}).}
\label{graph4}
\end{center}
\end{figure}

Next we analyze the Gaussian distribution. We assume that all the weights
$\tau_j$ are positive and thus we consider only the positive regime of the
distribution. Using Eq.~(\ref{eq:A}) we obtain
\begin{equation}
  A=\frac{\sqrt{2\pi}p_c\mathrm{erf^{-1}}
  }{1-e^{-[\mathrm{erf^{-1}}(p_c)]^2}}.
\end{equation}
The disorder is controlled solely by $p_c$ which is related only to the type
of network, and $A$ cannot take on large values. Thus, also for the Gaussian
$P(\tau,a)$, all optimal paths are in the weak disorder regime.
Similar considerations lead to the same conclusion for the exponential
distribution where $A=\frac{-p_c\mathrm{ln}(1-p_c)}{p_c+(1-p_c)\mathrm{ln}(1-p_c)}$.
Simulation results for the Gaussian distribution shown in Fig.~\ref{graph4} display only
weak disorder (i.e. no weak-strong disorder crossover), thus supporting the
above conclusions.

To summarize, in this section we presented a criterion
for the inverse disorder strength $Z$ on the optimal path in weighted networks for
general distributions $P(\tau,a)$. We show an analytical expression,
Eq.~(\ref{eq:A}), which fully characterizes the behavior of the optimal path.
Simulation of several distributions support these analytical predictions. It
is plausible that the criterion of Eq.~(\ref{eq:A}) is valid also for other
physical properties in weighted networks --- such as conductivity and flow in
random resistor networks --- due to a recently-found close relation between
the optimal path and flow~\cite{strelniker,zhenhua}.

\section{Scaling of optimal-path-lengths distribution with finite disorder in complex networks}

\begin{figure}[h]
\includegraphics[width=6cm,height=8cm,angle=270]{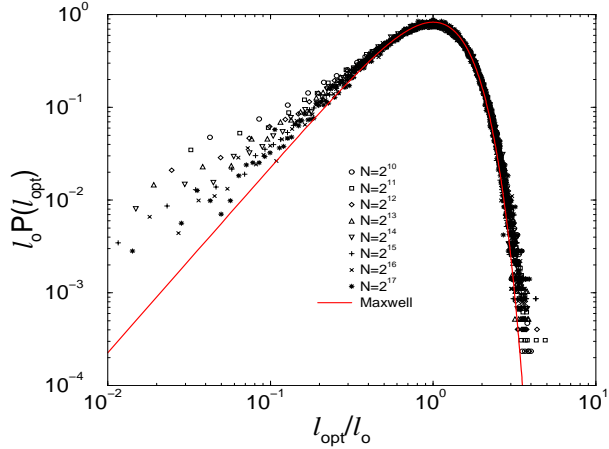}
\caption{Scaled curve for the probability distribution
  $P(\ell_{\mbox{\scriptsize opt}})$ of optimal path lengths for
  network sizes $N = 1024,2048,4096,8192,16384, 32768,65536$. The gray curve
  represents Maxwellian fit given by Eq.~(\ref{eq.s5}) (After \cite{Brauns203}). \label{fsantafe.2}}
\end{figure}

In this chapter we present further support~\cite{tomerdistributions} for the
general analytical results presented in section~\ref{sec.tws}B. The question
is how the different optimal paths in a network are distributed?  The
distribution of the optimal path lengths is especially important in
communication networks, in which the overall network performance depends on
the different path lengths between all nodes of the network, and not only the
average.  Ref.~\cite{Sameet04} studied the probability distribution
$P(\ell_{\mbox{\scriptsize opt}})$ of optimal path lengths in an ER network
in the SD limit. The scaled curve for $P(\ell_{\mbox{\scriptsize opt}})$ for
different network sizes is shown in Fig.~\ref{fsantafe.2} in a log-log plot.
We find that similarly to the behavior of self avoiding walks \cite{DeGennes}
there are two regimes in this distribution, the first one being
a power law $P(\ell_{\mbox{\scriptsize opt}}) \sim (\ell_{\mbox{\scriptsize
    opt}})^ g$ which is evident from the figure, with $g \approx
2$. The second regime is an exponential $P(\ell_{\mbox{\scriptsize
    opt}}) \sim e^{-C \ell_{\mbox{\scriptsize opt}}^{\delta}}$ where $C$ is a
constant and $\delta $ is around $2$. This leads us to the conjecture that
the distribution may have a Maxwellian functional form:
\begin{equation}\label{eq.s5}
P(\ell_{\mbox{\scriptsize opt}})=\frac{4\ell_{\mbox{\scriptsize opt}}^2
e^{-(\ell_{\mbox{\scriptsize opt}}/l_o)^2}}{\sqrt{\pi}l_o^3},
\end{equation}
Where $\ell_o=\sqrt{\pi}\langle\ell_{\mbox{\scriptsize
opt}}\rangle/2$ is the most probable value of
$\ell_{\mbox{\scriptsize opt}}$. The solid line in the figure is
the plot of this function and as seen it agrees with our numerical
results.

The exponents $g$ and $\delta$ can be obtained from the
following heuristic arguments. The right tail of the distribution
$P(\ell_{\mbox{\scriptsize opt}})$ is determined by the
distribution of the IIC size in the network of $N$ nodes. At
percolation threshold (Sec.~\ref{sec.tws}), $N$ nodes are divided into $N/2$ clusters,
obeying the power law distribution. However, the sum of all the
cluster sizes is equal to $N$, thus the distribution of the
largest cluster sizes must have a finite size exponential cutoff
$P(S)\sim \exp(-C S)$, as for the distribution of the segments of
an interval divided by random partitions. Since $S\sim
\ell_{\mbox{\scriptsize opt}}^{d_\ell}$, we have $\delta=d_\ell$.

To find the left tail distribution, we use the concept of MST. The
chemical diameter of the MST is $\ell_{\mbox{\scriptsize opt}}$
while its mass is $N \sim \ell_{\mbox{\scriptsize
opt}}^{1/\nu_{\mbox{\scriptsize opt}}}$. Due to self-similarity of
the MST the number of nodes $n(\ell)$ within a chemical distance
$\ell$ also scales as $n(\ell)\sim \ell^{1/\nu_{\mbox{\scriptsize
opt}}}$. Thus the probability density of the of the optimal path for
small values of $\ell$ scales as $d
n(\ell)/d\ell=\ell^{1/\nu_{\mbox{\scriptsize opt}}-1}$. Hence
$g=1/\nu_{\mbox{\scriptsize opt}}-1$. We expect that our
conjecture is valid also for SF networks. Using Eqs.
(\ref{eq:d_ell}) and (\ref{eq:nu_opt}) we have:
\begin{equation}
\delta =d_\ell=
\cases {
    \begin{array}{rll}
        2,  & \lambda>4, & {\rm ER} \\
        (\lambda-2)/(\lambda-3),  &  3<\lambda \leq 4 &
    \end{array}
    }\,,
\label{eq:delta}
\end{equation}
and
\begin{equation}
g =
\cases {
    \begin{array}{rll}
        2,  & \lambda>4, & {\rm ER} \\
        2/(\lambda-3),  &  3<\lambda \leq 4 &
    \end{array}
    }\,.
\label{eq:g}
\end{equation}

A recent work has studied the distribution form of shortest path lengths on
minimum spanning trees~\cite{santafe}, which corresponds to optimal paths on
networks with large variation in link weights ($a \rightarrow \infty$).

Using the scaling derived in Sect. \ref{sec.tws}, more precisely:
\begin{equation}
  \label{equ:weighted_length}
  \ell(a) \sim \ell_{\infty} F\left( \frac{\ell_{\infty}}{ap_c} \right),
\end{equation}
where $p_c$ is the percolation threshold and $\ell_{\infty} \sim
N^{\nu_{opt}}$ is the optimal path length for strong disorder ($a
\rightarrow \infty$). For Erd\H{o}s-R\'enyi (ER) graphs
$\nu_{opt}=1/3$.
We generalize these results and suggest that the distribution of the
optimal path lengths has the following scaling form:
\begin{equation}
  \label{equ:weighted_dist}
  P(\ell_{\rm opt},N,a) \sim \frac{1}{\ell_{\infty}} G \left( \frac{\ell_{\rm opt}}{\ell_{\infty}} , \frac{1}{p_c}\frac{\ell_{\infty}}{a} \right).
\end{equation}
The parameter $Z \equiv \frac{1}{p_c}\frac{\ell_{\infty}}{a}$, which is
equivalent to $Z$ in Eq. ~(\ref{eq:Z}), determines the functional form of the
distribution.  Relation~(\ref{equ:weighted_dist}) is supported by simulations
\cite{tomerdistributions} for both ER and SF graphs, including SF graphs with
$2<\lambda<3$, for which $p_c \rightarrow 0$ with system size
$N$~\cite{CEBH00}.

We simulate ER graphs with weights on the links for different values of graph
size $N$, control parameter $a$, and average degree $\langle k \rangle$
(which determines $p_c=1/ \langle k \rangle $). We then generate the shortest path tree (SPT)
using Dijkstra's algorithm (See Section~\ref{sec.dij}) from some randomly
chosen root node. Next, we calculate the probability distribution function of
the optimal path lengths for all nodes in the
graph~\cite{tomerdistributions}.

\begin{figure}[h]
\begin{center}
\includegraphics[width=6.0cm,height=6.0cm]{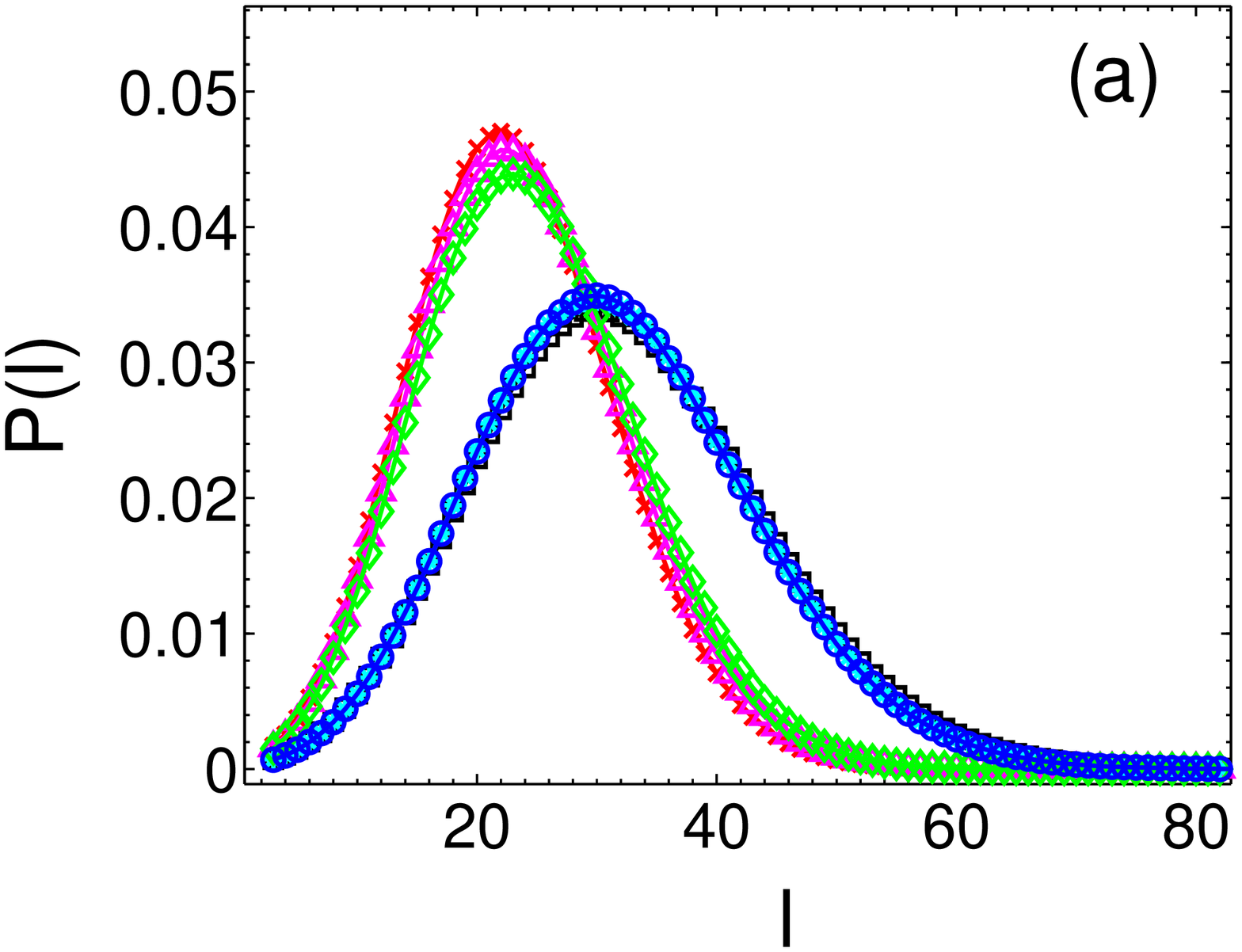}
\includegraphics[width=5.0cm,height=6.0cm]{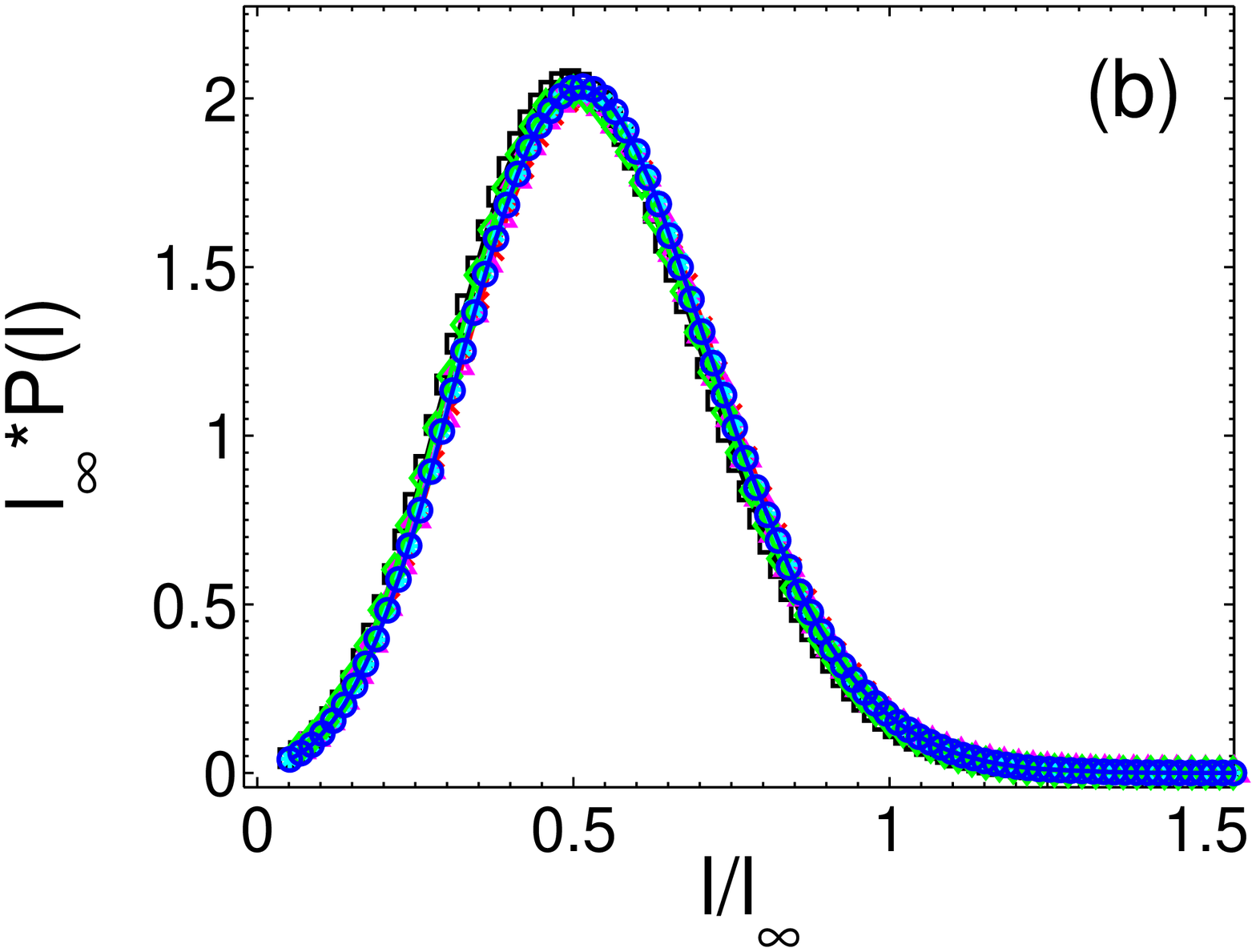}\\
\includegraphics[width=5.0cm,height=6.0cm]{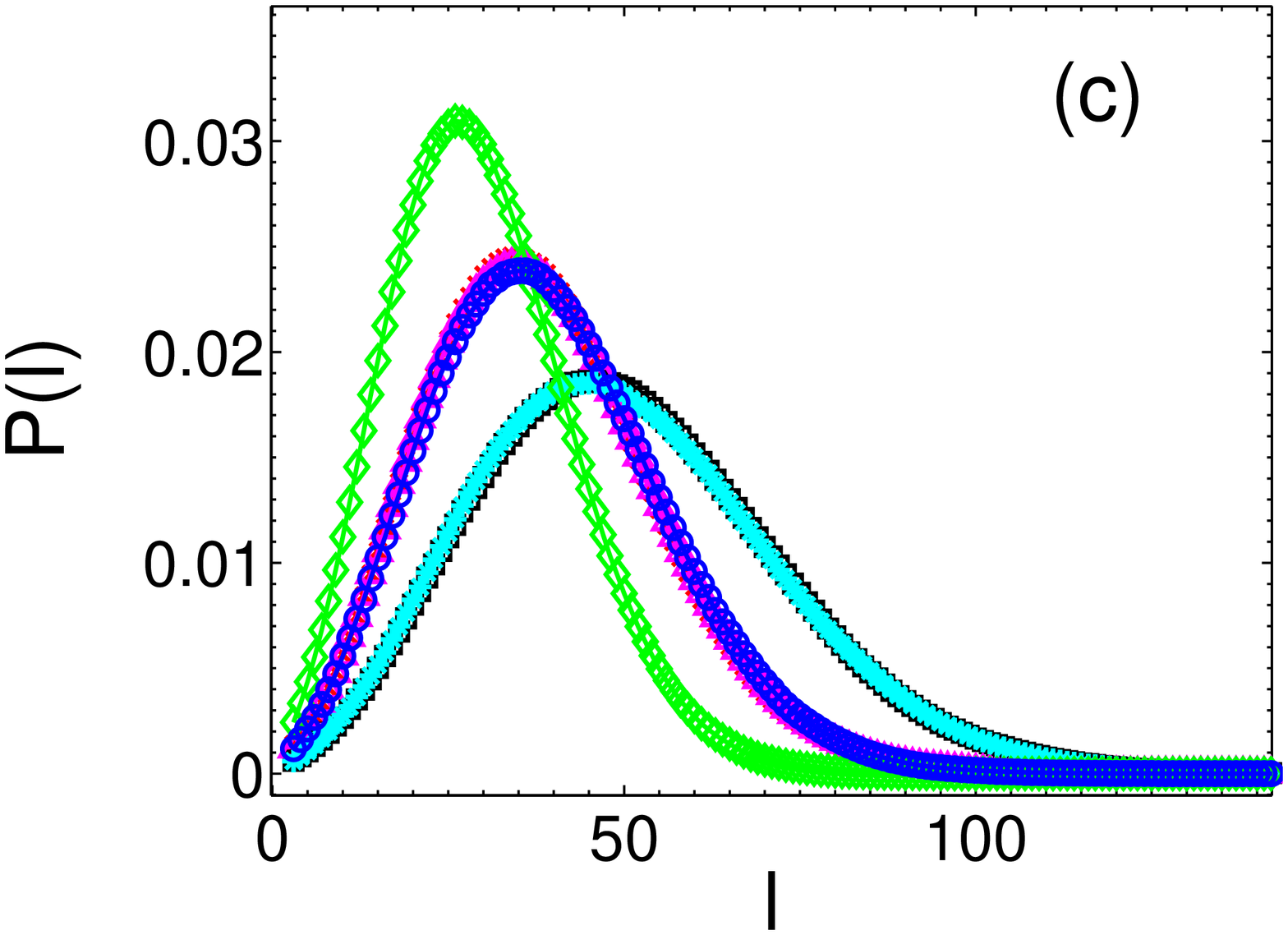}
\includegraphics[width=5.0cm,height=6.0cm]{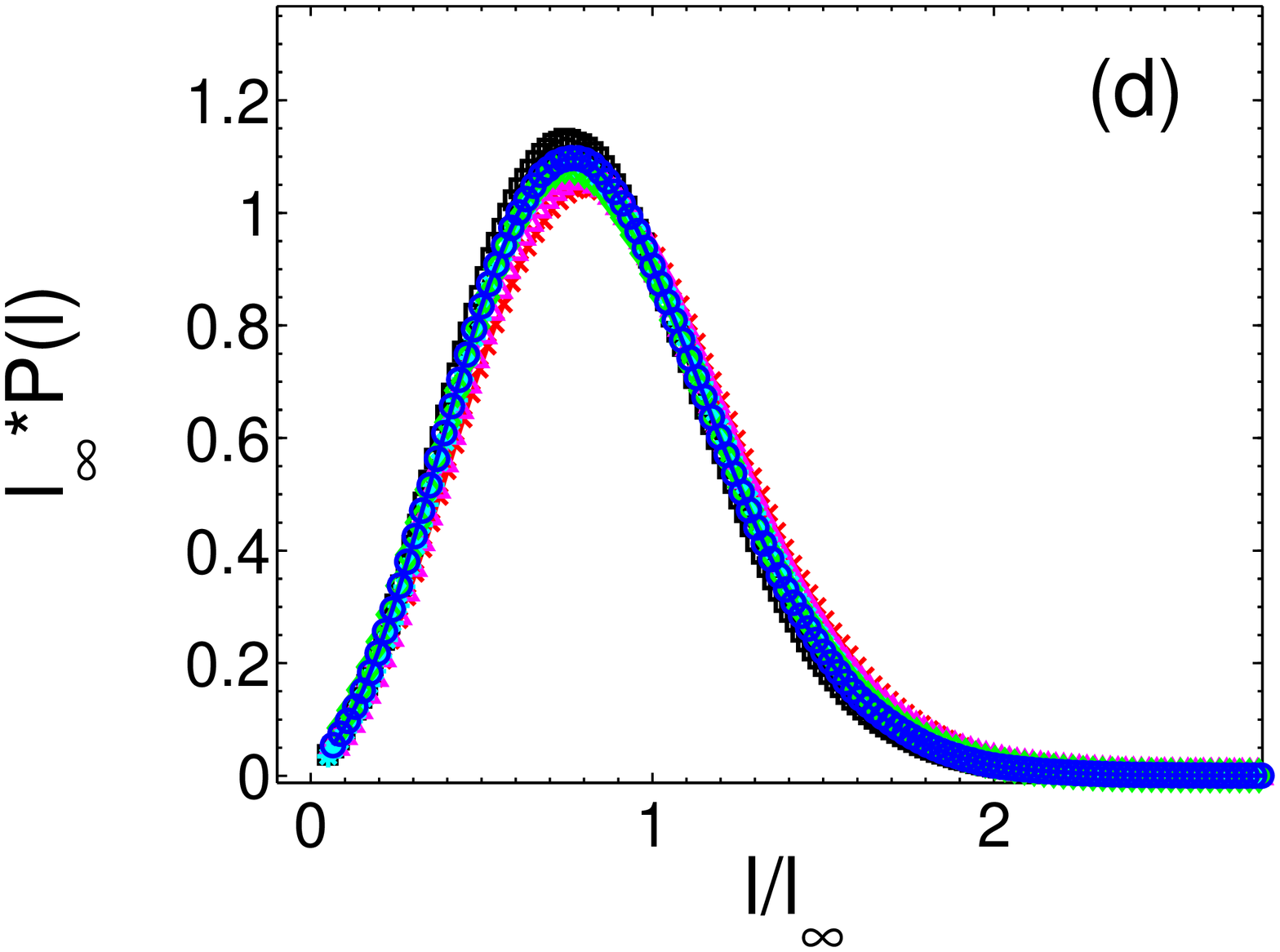}
\caption{Optimal path lengths
  distribution, $P(\ell_{\rm opt})$, for ER networks with (a,b) $Z \equiv
  \frac{1}{p_c}\frac{\ell_{\infty}}{a} = 10$ and (c,d) $Z =3$. (a)
  and (c) represent the unscaled distributions for $Z=10$ and
  $Z=3$ respectively, while (b) and (d) are the scaled
  distribution. Different symbols represent networks with
  different characteristics such as size $N=2000,4000,8000$ (which determines
  $\ell_{\infty} \sim N^{1/3}$), average degree $\langle k \rangle=3,5,8$ (which
  determines $p_c = 1/\langle k \rangle $), and disorder strength $a=\ell_\infty/(p_cZ)$
  Results were averaged
  over $1500$ realizations  (After \cite{tomerdistributions}).\label{fig:ERWeak}}
\end{center}
\end{figure}

In Fig.~\ref{fig:ERWeak} we plot $\ell_{\infty}P(\ell_{\rm opt},N,a)$ vs.
$\ell_{\rm opt}/\ell_{\infty}$ for different values of $N$, $a$, and $\langle k \rangle$. A
collapse of the curves is seen for all graphs with the same value of
$Z = \frac{1}{p_c}\frac{\ell_{\infty}}{a}$.

\begin{figure}
  \begin{center}
    \includegraphics[width=6.0cm,height=6.0cm]{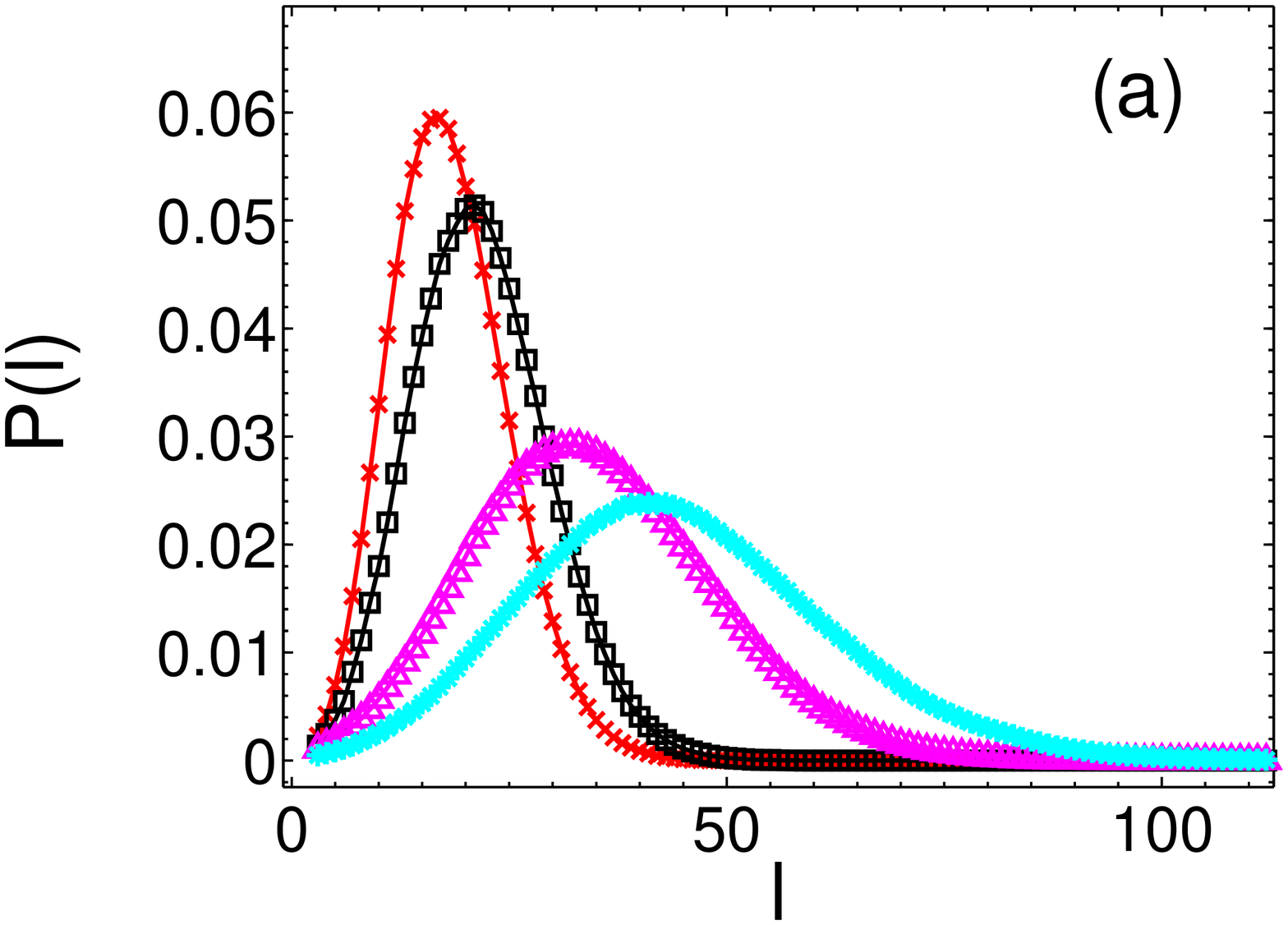}
    \includegraphics[width=6.0cm,height=6.0cm]{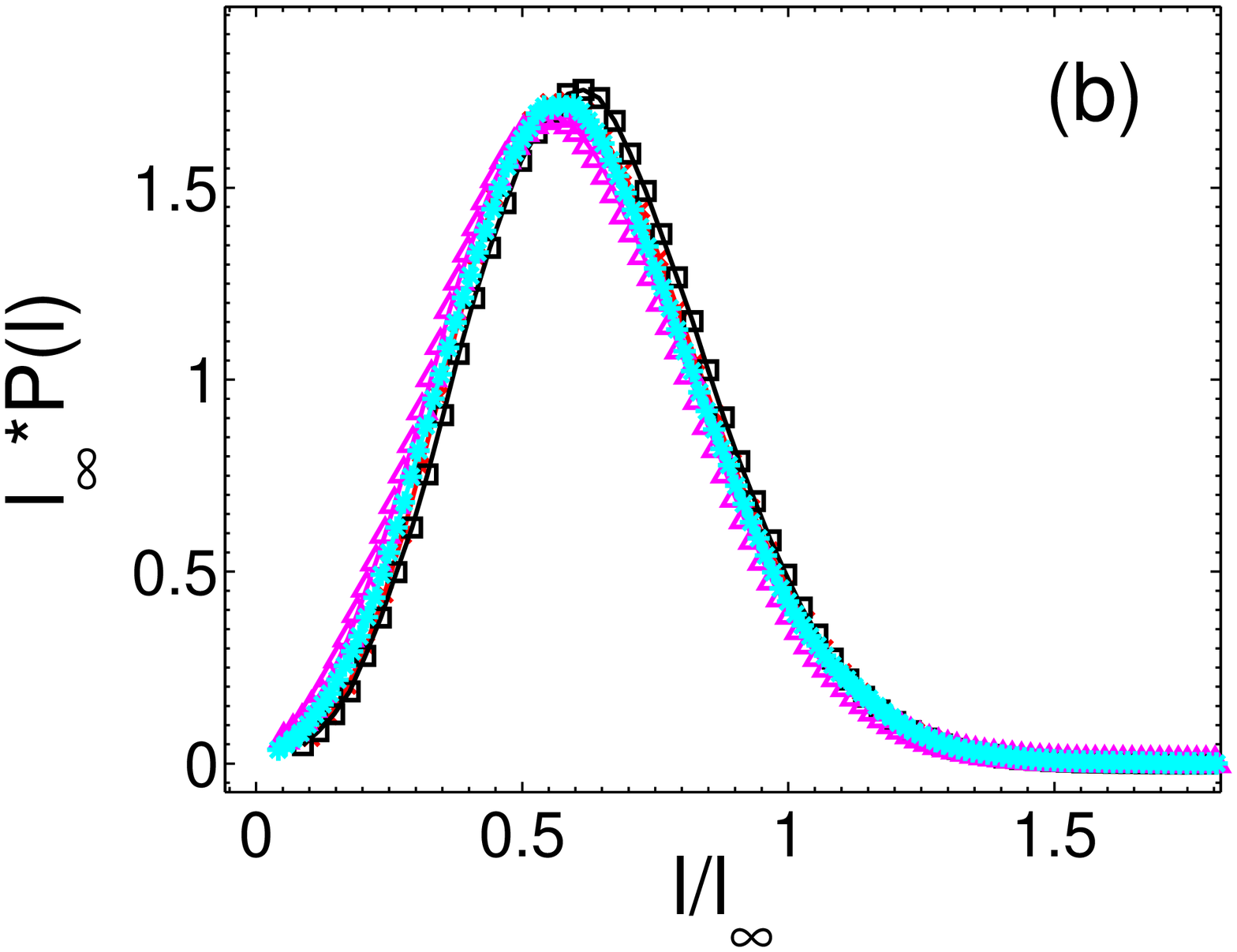}\\
    \includegraphics[width=6.0cm,height=6.0cm]{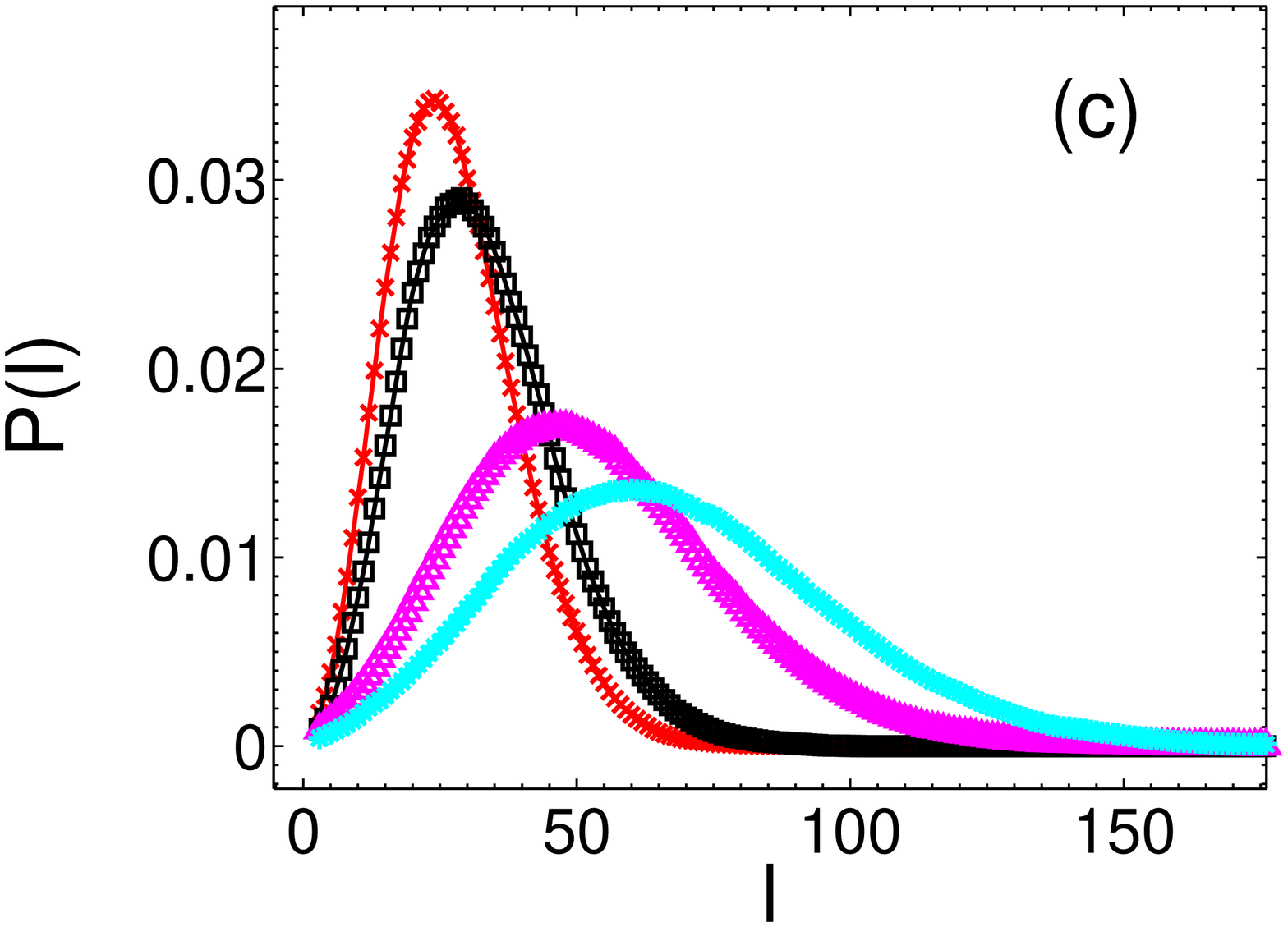}
    \includegraphics[width=6.0cm,height=6.0cm]{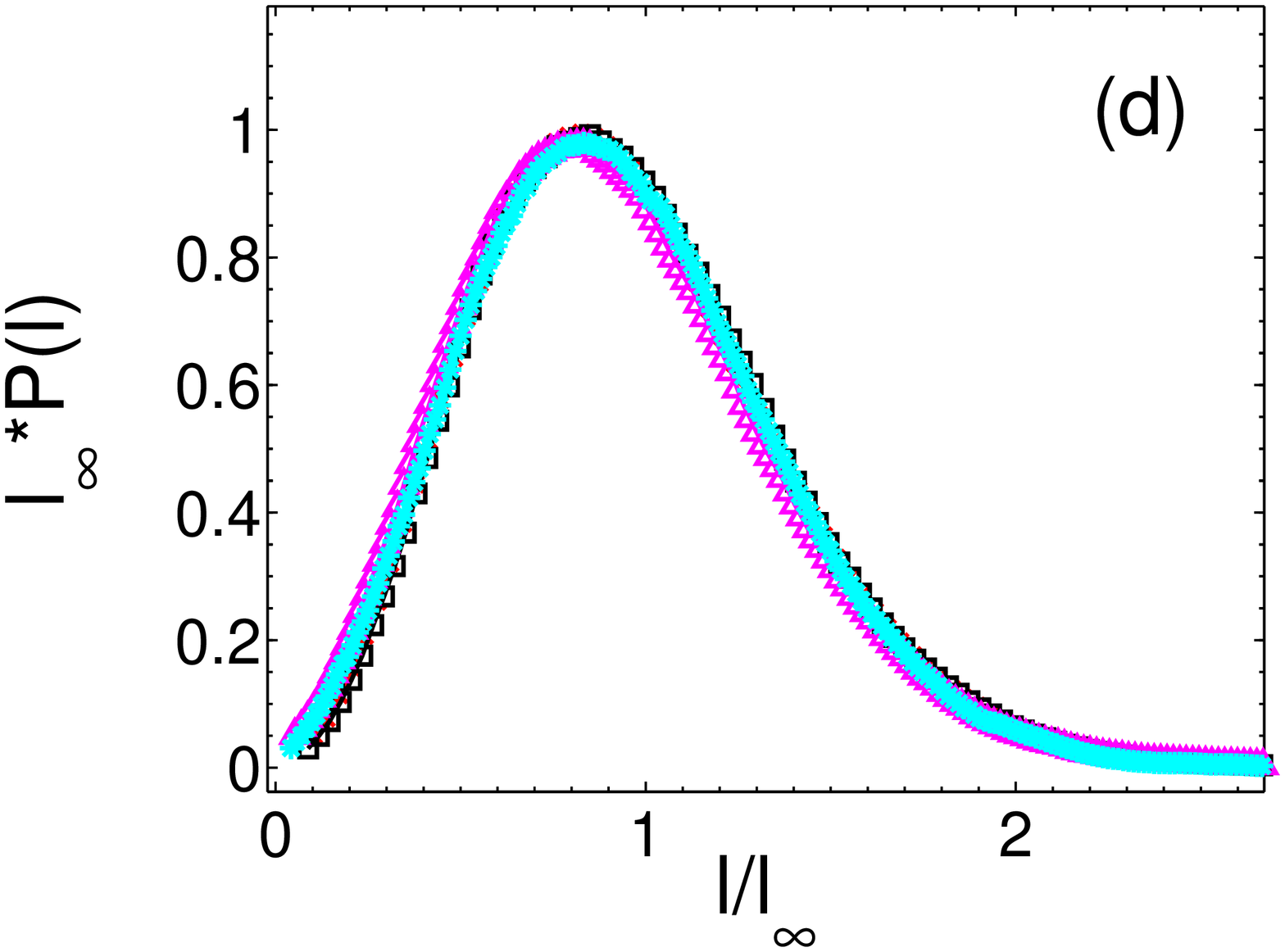}
    \caption{\label{fig:SFWeak} Optimal path lengths
      distribution, $P(l)$, for SF networks with (a,b) $Z \equiv
      \frac{1}{p_c}\frac{\ell_{\infty}}{a} = 10$ and (c,d) $Z =2$. (a)
      and (c) represent the unscaled distributions for $Z=10$ and
      $Z=2$ respectively, while (b) and (d) are the scaled
      distribution. Different symbols represent networks with
      different characteristics such as size $N=4000,8000$ (which determines
      $\ell_{\infty} \sim N^{\nu_{opt}}$), $\lambda=3.5,5$ and $m=2$ (which
      determine $p_c$), and disorder strength $a=\ell_\infty/(p_c Z)$. Results were averaged over $250$
      realizations  (After \cite{tomerdistributions}).}
     \end{center}
  \end{figure}

Figure~\ref{fig:SFWeak} shows similar plots for SF graphs -- with a degree
distribution of the form $P(k) \sim k^{-\lambda}$ and with a minimal degree
$m$. Scale-free graphs were generated according to the ``configuration
model'' or Molloy Reed algorithm (See
Section~\ref{sec.alg}A)~\cite{Molloy}~\footnote{Note that the minimal degree
  is $m=2$ thus ensuring that there exists an infinite cluster for any
  $\lambda$, and thus $0<p_c<1$. For the case of $m=1$ there is almost surely
  no infinite cluster for $\lambda > \lambda_c \approx 4$ (or for a slightly
  different model, $\lambda_c = 3.47875$), resulting in an effective
  percolation threshold $p_c = \frac{\av{k}}{\av{k(k-1)}} > 1$.}. A collapse
is obtained for different values of $N$, $a$, $\lambda$ and $m$, with
$\lambda>3$ and for the same values of $Z$.

\begin{figure}
  \begin{center}
  \includegraphics[width=6.0cm,height=6.0cm]{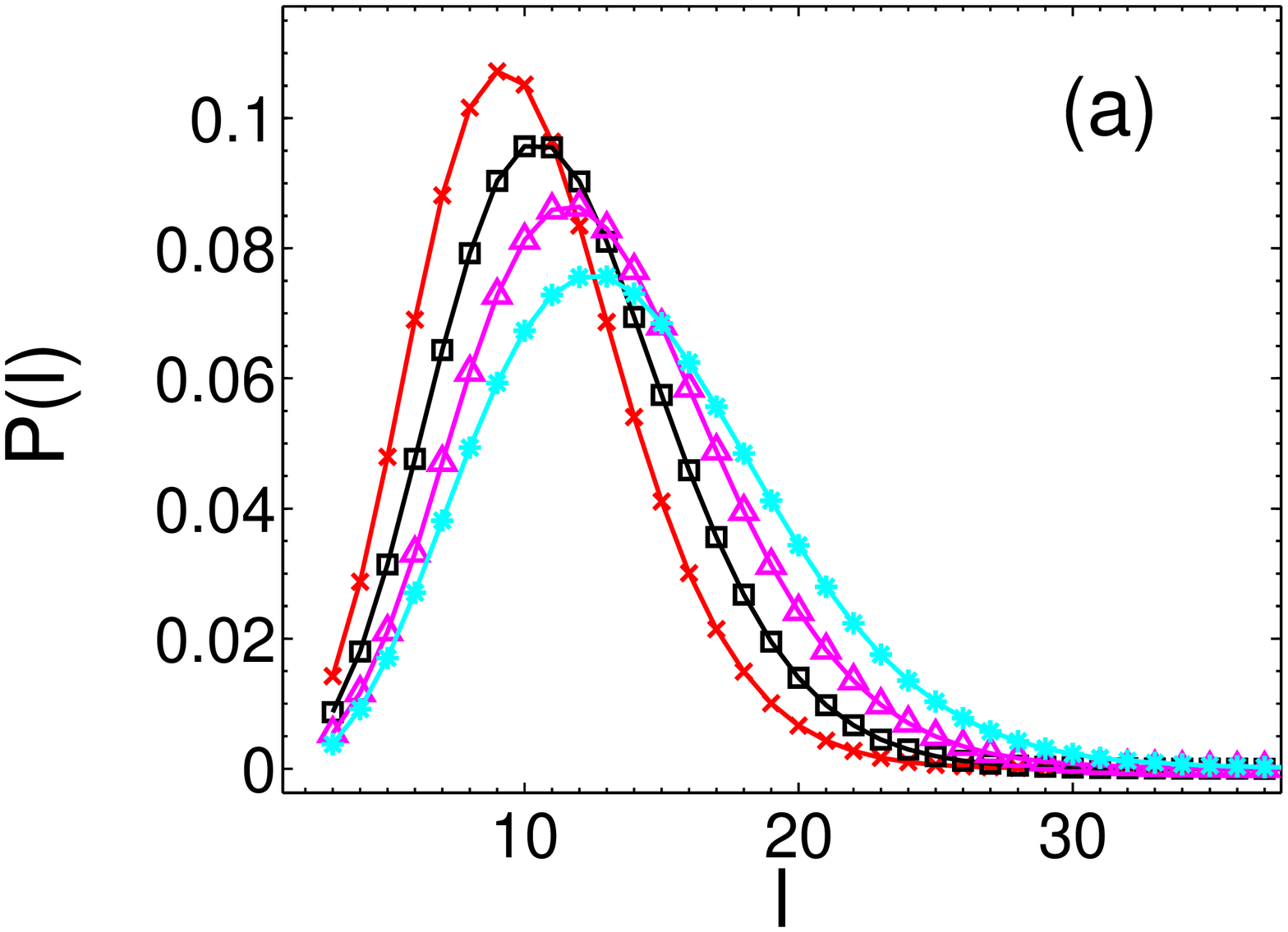}
  \includegraphics[width=6.0cm,height=6.0cm]{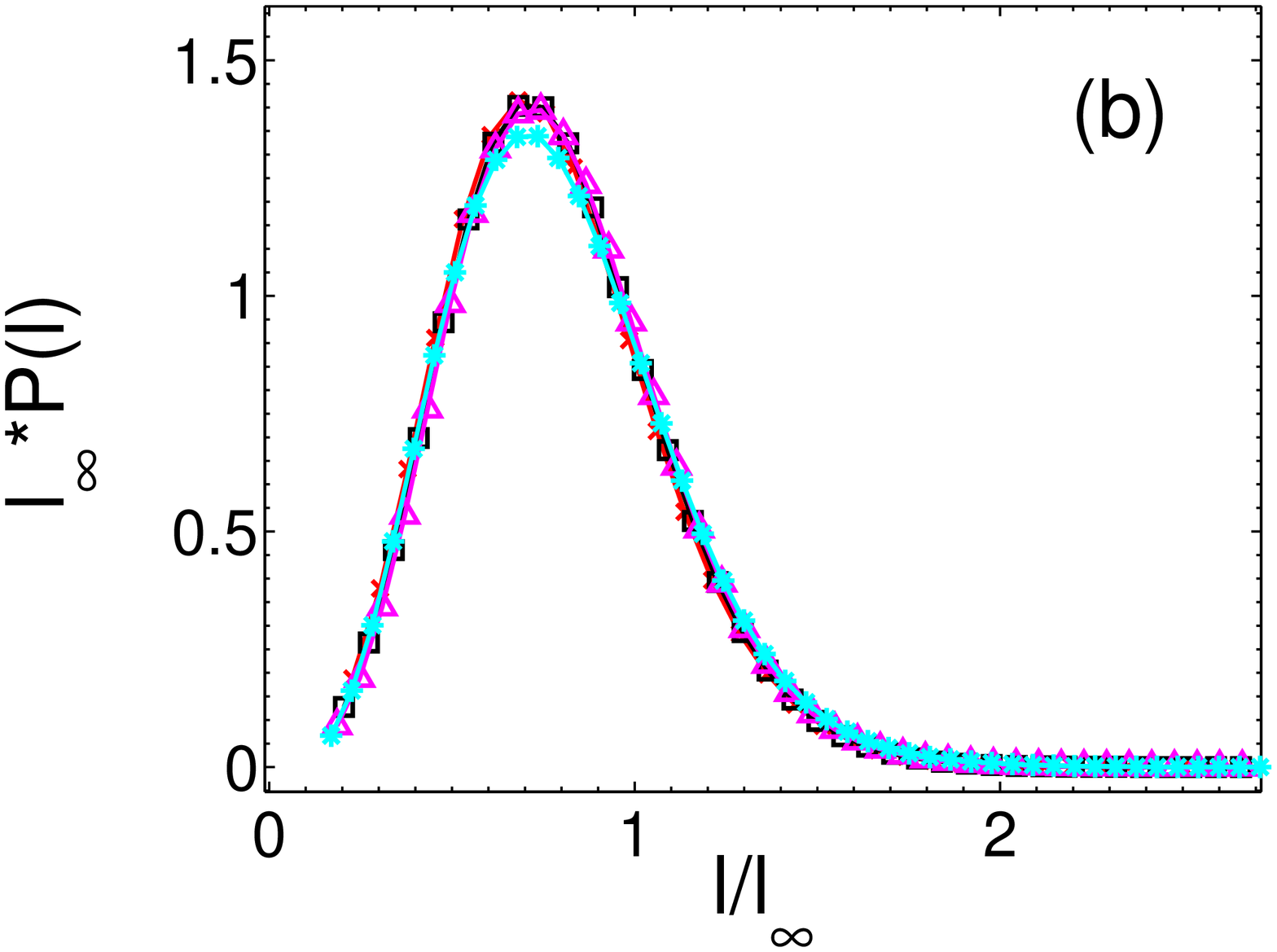}
    \caption{ Optimal path lengths
      distribution function for SF graphs with $\lambda=2.5$, $m=2$ and with
      $Z \equiv \frac{1}{p_c}\frac{\ell_{\infty}}{a} = 10$. (a)
      represents the unscaled distribution for $Z=10$ while (b) shows
      the scaled distribution. Different symbols represent graphs with
      different characteristics such as size $N=2000,4000,8000,1600$ (which determines
      $\ell_{\infty} \sim \log(N)$ and $p_c \sim N^{-1/3}$), and
      disorder strength $a=\ell_\infty/(p_c Z)$.
      Results were averaged over $1500$ realizations  (After \cite{tomerdistributions}).\label{fig:SF25Weak}}
    \end{center}
\end{figure}

  Next, we study SF networks with $2 < \lambda < 3$. In this regime the
  second moment of the degree distribution $\langle k^2 \rangle$ diverges,
  leading to several anomalous
  properties~\cite{CEBH00,Cohen3,newman-callaway-2000:networks_robustness}.
  For example: the percolation threshold approaches zero with system size according to Eq.~(\ref{eq:cohen}):
  $p_c \sim N^{-\frac{3-\lambda}{\lambda-1}} \rightarrow 0$, and the optimal
  path length $\ell_{\infty}$ in SD was found numerically to scale
  logarithmically with $N$ compared to polynomially, found in $\lambda
  >3$~\cite{Brauns03}.  Nevertheless, it seems from Fig.~\ref{fig:SF25Weak}
  that the optimal paths probability
  distribution for SF networks with $2 < \lambda < 3$ exhibits similar
  collapse for different values of $N$ and $a$ for the same $Z$ (although its
  functional form is different compared to the $\lambda>3$
  case)~\cite{tomerdistributions}.


We present evidence that the optimal path is related to
percolation~\cite{Sameet04}. The numerical results suggest
that for a finite disorder parameter $a$, the optimal path (on
average) follows the percolation cluster in the network (i.e.,
links with weight below $p_c$) up to a typical ``characteristic
length'' $\xi = a p_c$, before deviating and making a ``shortcut''
(i.e. crossing a link with weight above $p_c$). For length scales
below $\xi$ the optimal path behaves as in strong disorder and its
length is relatively long. The shortcuts have an effect of
shortening the optimal path length from a polynomial to
logarithmic form according to the universal function $F(u)$
(Eq.~(\ref{equ:weighted_length})).
\omitit{~\footnote{This is somewhat
    analogous to the small world model, see
    e.g.~\cite{barthelmy-amaral-1999:small_world_crossover}.}}
Thus, the optimal path for finite $a$ can be viewed as consisting of
``blobs'' of size $\xi$ in which strong disorder persists. These blobs
are interconnected by shortcuts, which result in the total path being
in weak disorder.

\begin{figure}[h]
     \begin{center}
    \includegraphics[width=6.0cm,height=6.0cm]{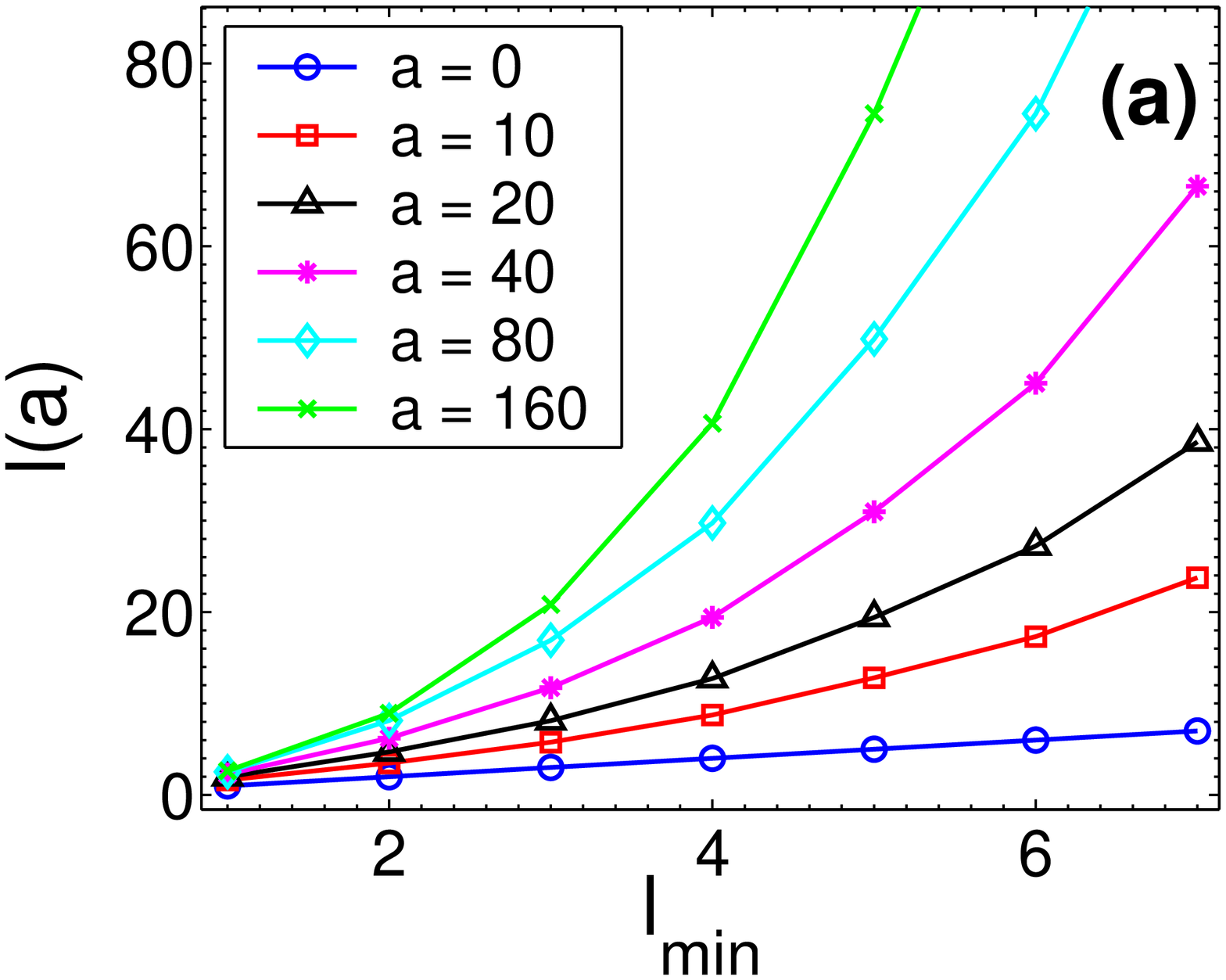}
    \includegraphics[width=6.0cm,height=6.0cm]{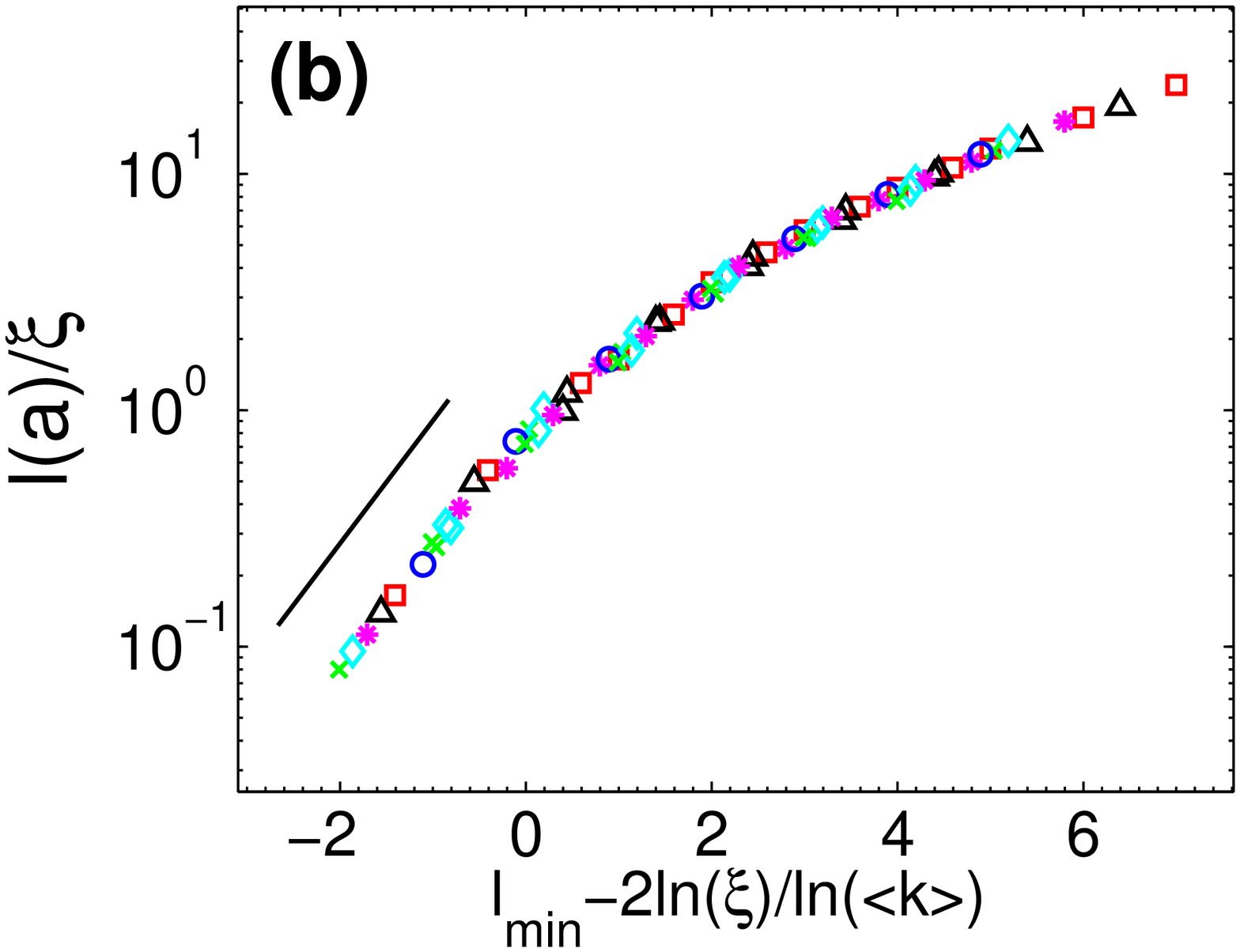}
    \caption{\label{fig:TransitionSameGraph} Transition
      between different scaling regimes for the optimal path length
      $l(a)$ inside an ER graph with $N=128,000$ nodes and
      $\av{k}=10$. (a) shows the unscaled and (b) shows the scaled
      length of the optimal path $l(a)$ averaged over all nodes with
      same value of $\ell_{min}$. Different symbols represent different
      values of the disorder strength $a$.
      Fig. (b) shows that for length scales $\ell(a)$ smaller than the
      ``characteristic length'', $\xi=ap_c$, $l(a)$ grows
      exponentially relative to the shortest hopcount path $\ell_{min}$
      (see solid line). This is consistent with $l(a) \sim N^{1/3}$
      and $\ell_{min} \sim \log{N}$ inside the range of size $\xi = a
      p_c$. For length scales above $\xi$ both quantities scale as
      $\log{N}$.  (After \cite{tomerdistributions}). }
    \end{center}
  \end{figure}

We next present direct simulations supporting this argument
\cite{tomerdistributions}. We calculate the optimal path length $\ell(a)$
inside a single network of size $N$, for a given $a$, and find
(Fig.~\ref{fig:TransitionSameGraph}) that it scales differently below and
above the characteristic length $\xi = ap_c$.  For each node in the graph we
find $\ell_{min}$, which is the number of links (``hopcounts'') along the
shortest path from the root to this node \textit{without regarding the weight
of the link}.

In Fig.~\ref{fig:TransitionSameGraph} we plot the length of the optimal
path $\ell(a)$, averaged over all nodes with the same value of $\ell_{min}$ for
different values of $a$. The figure strongly suggests that $l(a) \sim
\exp(\ell_{min})$ for length scales below the characteristic length $\xi = a
p_c$ (see the linear regime in Fig.~\ref{fig:TransitionSameGraph}b), while for
large length scales $\ell(a) \sim \ell_{min}$.

For length scales smaller
  than $\xi$ we have $\ell_{opt}=A N^{1/3}$ and $\ell_{min} = B \ln{N}$,
  where $A$ and $B$ are constants.  Thus $N = \exp{(\ell_{min}/B)}$ and
  $\ell_{opt} = A \exp{(l_{min}/3B)}$.  Consequently, we expect that:
  $\frac{\ell_{opt}}{\xi} = \frac{A \exp{(\ell_{min}/3B)}}{\xi} = A
  \exp{[(l_{min}-3B \ln{\xi})/3B]}$.  We find the best scaling in
  Fig.~\ref{fig:TransitionSameGraph} for $B =
  \frac{2}{3 \ln{\av{k}}}$.

This is consistent with our hypothesis that below the characteristic
length ($\xi = a p_c$) $\ell_{min} \sim \log{N}$ and $l(a) \sim N^{1/3}$,
while $\ell_{min} \sim \log{N}$ and $l(a) \sim \log{N}$ above.

  In order to better understand why the distributions of $\ell_{opt}$ depend on
  $Z$ according to Eq.~(\ref{equ:weighted_dist}), we suggest the following
  argument.  The optimal path for $a \rightarrow \infty$, was shown to be
  proportional to $N^{1/3}$ for ER graphs and $N^{(\lambda - 3)/(\lambda -
    1)}$ for SF graphs with $3<\lambda<4$~\cite{Brauns03}.  For finite $a$
  the number of shortcuts, or number of blobs, is $Z =
  \frac{\ell_{\infty}}{\xi} = \frac{\ell_{\infty}}{ap_c}$.  The deviation of
  the optimal path length for finite $a$ from the case of $a \rightarrow
  \infty$ is a function of the number of shortcuts.  These results explain
  why the parameter $Z \equiv \frac{\ell_{\infty}}{ap_c}$ determines the
  functional form of the distribution function of the optimal paths (see also
  section~\ref{sec.tws}B) .

To summarize, we have shown that the optimal path length distribution
in weighted random graphs has a universal scaling form according to
Eq.~(\ref{equ:weighted_dist}). We explain this behavior and
demonstrate the transition between polynomial and logarithmic behavior
of the average optimal path in a single graph.
Our results are consistent with results found for finite dimensional
systems~\cite{Porto99,zhenhua,strelniker}: In finite dimension the parameter
controlling the transition is $Z=\frac{L^{1/\nu}}{ap_c}$, where $L$ is the
system length and $\nu$ is the correlation length critical exponent as in Eq.(\ref{eq:Zlat}). This is
because only the ``red bonds'' - bonds that if cut would disconnect the
percolation cluster~\cite{coniglio-1982:cluster_structure} - control the
transition (see also section~\ref{sec.tws}B).

\section{Scale-Free Networks Emerging from Weighted Random Graphs}
In this section we introduce a simple process that generates random
scale-free networks with $\lambda=2.5$ from weighted Erd\"{o}s-R\'enyi
graphs~\cite{Tomer_grey}.  We further show that the
minimum spanning tree (MST) on an Erd\"{o}s-R\'enyi graph is related
to this network, and is composed of percolation clusters, which we
regard as ``super nodes'', interconnected by a scale-free tree.
We will see that due to optimization this scale-free tree is dominated
by links having high weights --- significantly higher than the
percolation threshold $p_c$. Hence, the MST naturally distinguishes
between links below and above the percolation threshold, leading to a
scale-free ``supernode network''. Our results may explain the origin
of scale-free degree distribution in some real world networks.

\begin{figure}[h]
   \begin{center}
    \includegraphics[width=9.5cm,height=9.5cm]{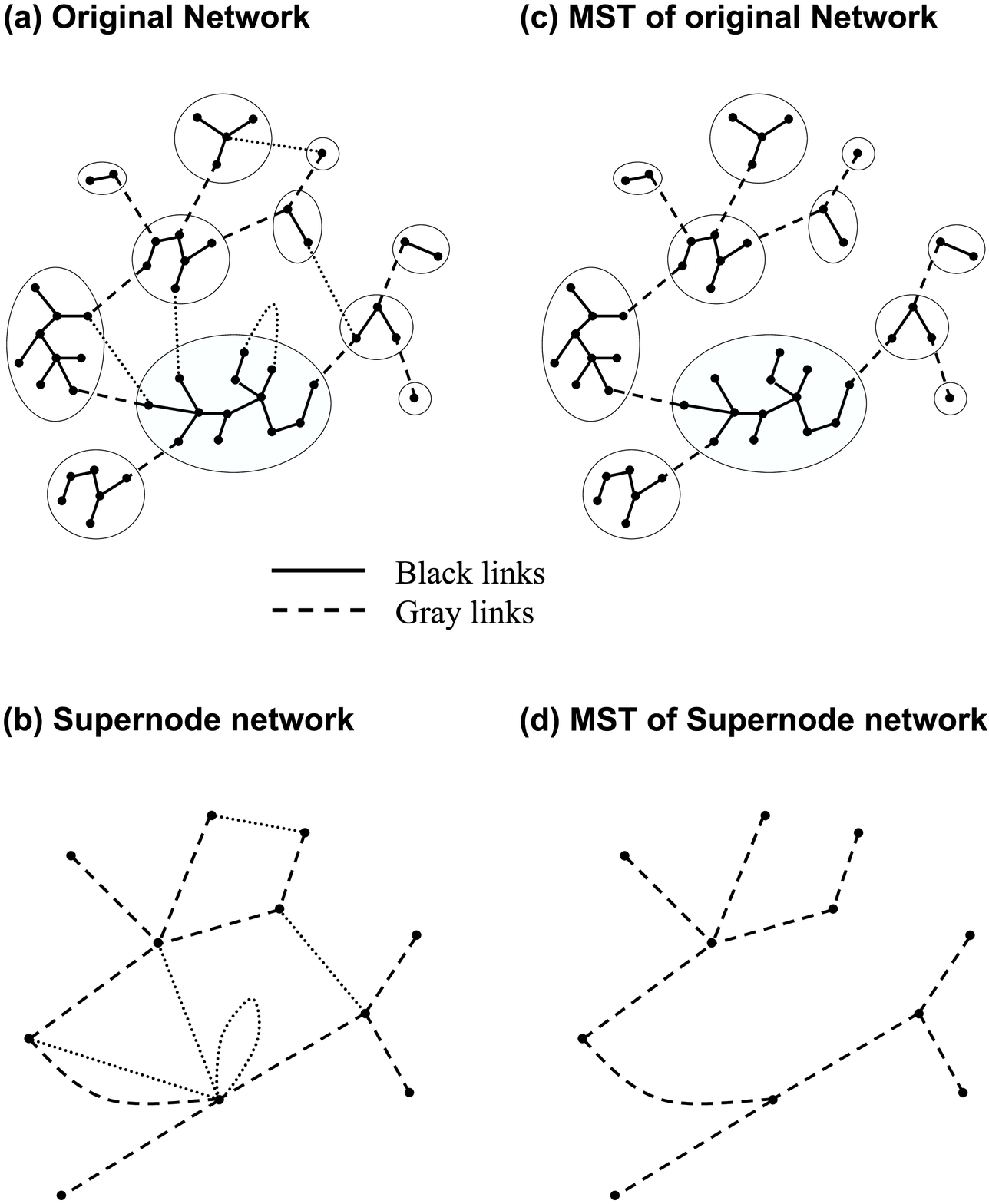}
    \caption{\label{fig:gray_network_sketch} Sketch of the
      ``supernode network''. (a) The original ER network, partitioned
      into percolation clusters whose sizes $s$ are power-law
      distributed, with $n_s \sim s^{-\tau}$ where $\tau = 2.5$ for ER
      graphs. The ``black'' links are the links with weights below
      $p_c$, the ``dotted'' links are the links that are removed by
      the bombing algorithm, and the ``gray'' links are the links
      whose removal will disconnect the network (and therefore are not
      removed even though their weight is above $p_c$).
      (b) The ``supernode network'': the nodes are the clusters in the
      original network and the links are the links connecting nodes in
      different clusters (i.e., ``dotted'' and ``gray'' links). The
      supernode network is scale-free with $P(k)\sim k^{-\lambda}$ and
      $\lambda=2.5$. Notice the existence of self loops and of double
      connections between the same two supernodes.
      (c) The minimum spanning tree (MST), composed of black and gray
      links only.
      (d) The MST of the supernode network (``gray tree''), which is
      obtained by bombing the supernode network (thereby removing the
      ``dotted'' links), or equivalently, by merging the clusters in
      the MST to supernodes. The gray tree is scale-free, with
      $\lambda=2.5$ (After \cite{Tomer_grey}).}
   \end{center}
  \end{figure}

Consider an Erd\"{o}s-R\'enyi (ER) graph with $N$ nodes and an average degree
$\av{k}$, thus having a total of $N \langle k \rangle /2$ links. To each link
we assign a weight chosen randomly and uniformly from the range $[0,1]$. We
define black links to be those links with weights below a threshold
$p_c=1/\langle k\rangle$. Two nodes belong to the same cluster if they are
connected by black links [Fig.~\ref{fig:gray_network_sketch}(a)].

\begin{figure}
  \begin{center}
    \includegraphics[width=6.0cm,height=6.0cm]{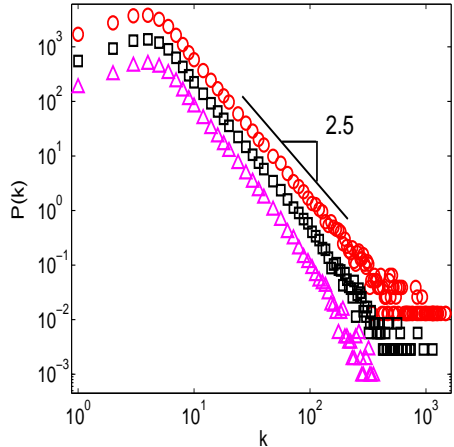}
    \caption{\label{fig:graynet-degrees-layers} The degree distribution of
      the supernode network of Fig.~1(b), where the supernodes are the
      percolation clusters, and the links are the links with weights larger
      than $p_c$ ($\bigcirc$). The distribution exhibits a scale-free tail
      with $\lambda\approx 2.5$.  If we choose a threshold less than $p_c$,
      we obtain the same power law degree distribution with an exponential
      cutoff.  The different symbols represent slightly different threshold
      values: $p_c-0.03$ ($\Box$) and $p_c-0.05$ ($\bigtriangleup$).  The
      original ER network has $N=50,000$ and $\langle k \rangle =5$.  Note
      that for $k\approx\av{k}$ the degree distribution has a maximum (After \cite{Tomer_grey}).}
\end{center}
  \end{figure}

From Section~\ref{sec.tws} follows that the number of clusters of
$s$ nodes scales as a power law, $n_s \sim s^{-\tau}$, with
$\tau=2.5$ for ER networks.
Next, we merge all nodes inside each cluster into a single ``supernode''. We
define a new ``supernode network'' [Fig.~\ref{fig:gray_network_sketch}(b)] of
$N_{\rm sn}$ supernodes \cite{Sameet04}. The links between two supernodes
[see Figs.~\ref{fig:gray_network_sketch}(a) and
\ref{fig:gray_network_sketch}(b)] have weights {\em larger\/} than $p_c$. The
degree distribution $P(k)$ of the supernode network can be obtained as
follows. Every node in a supernode has the same (finite) probability to be
connected to a node outside the supernode. Thus, we assume that the degree
$k$ of each supernode is proportional to the cluster size $s$, which obeys
$n_s\sim s^{-2.5}$.  Hence $P(k)\sim k^{-\lambda}$, with $\lambda=2.5$, as
supported by simulations shown in Fig.~\ref{fig:graynet-degrees-layers}.
Furthermore, we also see that if the threshold for obtaining the clusters
which are merged into supernodes is changed slightly, the degree distribution
still remains scale free with $\lambda=2.5$, but with an exponential cutoff.
This is an indication of the fact that there are still supernodes of high
degree which are connected to many other (small) supernodes by links with
weights significantly higher than $p_c$; if this was not the case, a small
change in the threshold would cause many clusters to merge and destroy the
power law in the supernode network degree distribution.

We next show that the MST on an ER graph is
related to the supernode network, and therefore also exhibits
scale-free properties.
In the MST each path between two sites on the MST is the optimal
path in the ``strong disorder'' limit \cite{Cieplak,Dob}, meaning
that along this path the maximum {\em barrier} (weight) is the
smallest possible \cite{Dob,Brauns03,Sameet04}.

Here we use the bombing algorithm (See Section~\ref{sec.alg}D ). If
the removal of a link disconnects the graph, we restore the link
and mark it ``gray'' ; otherwise the link [shown dotted in
Fig.~\ref{fig:gray_network_sketch}(a)] is removed. The links that
are not bombed are marked as ``black''. In the bombing algorithm,
only links that close a loop can be removed. Because below
criticality loops are negligible \cite{ER59,Albert02} for ER networks ($d\to\infty$), bombing does not modify the
percolation clusters
--- where the links are black and have weights below $p_c$.
Thus, bombing modifies only links {\em outside} the clusters, so
actually it is only the links of the {\em supernode network} that are
bombed.  Hence the MST resulting from bombing is composed of
percolation clusters (composed of black links) and connected by gray links
[Fig.~\ref{fig:gray_network_sketch}(c)].

\begin{figure}
\begin{center}
    \includegraphics[width=6.0cm,height=6.0cm]{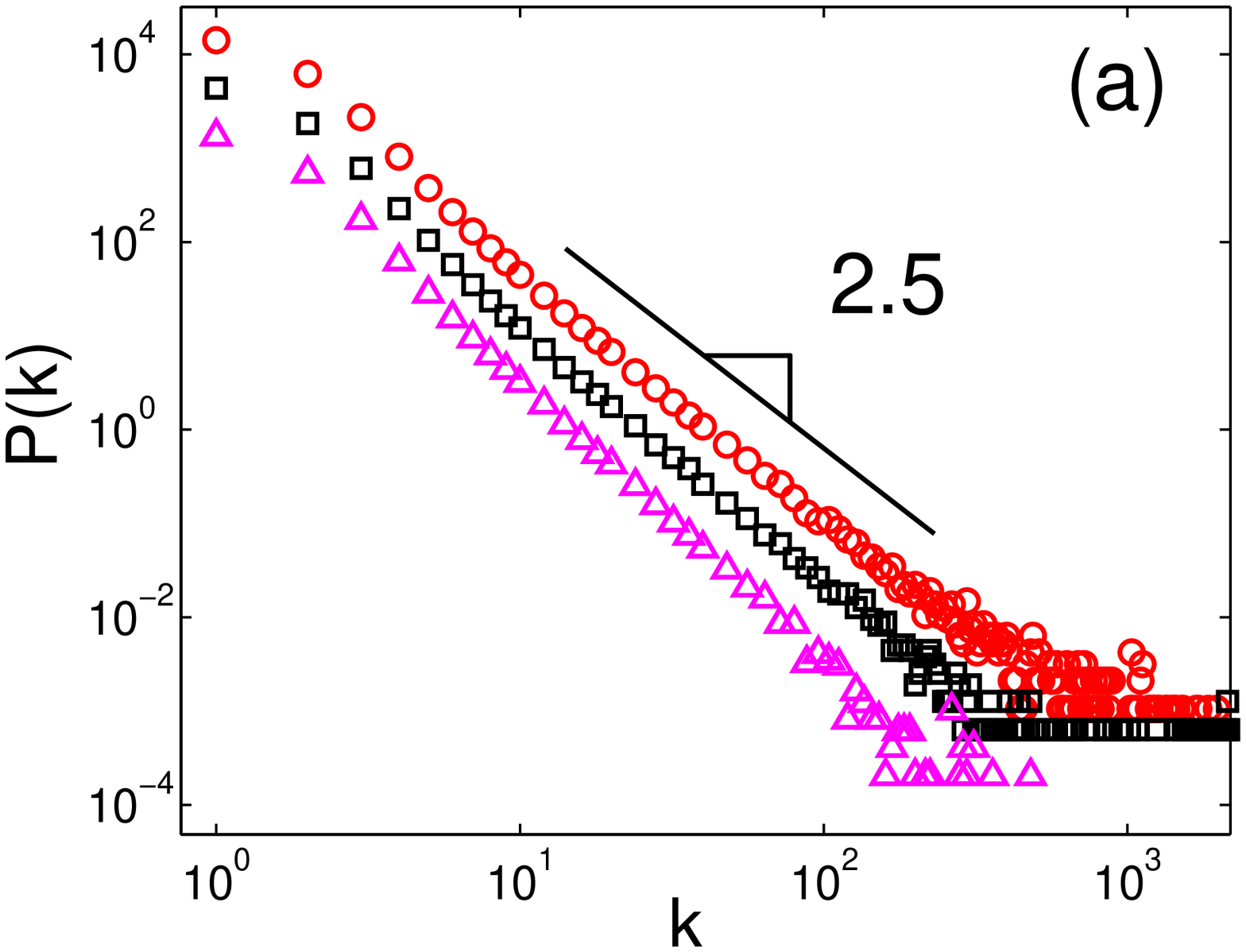}
    \includegraphics[width=6.0cm,height=6.0cm]{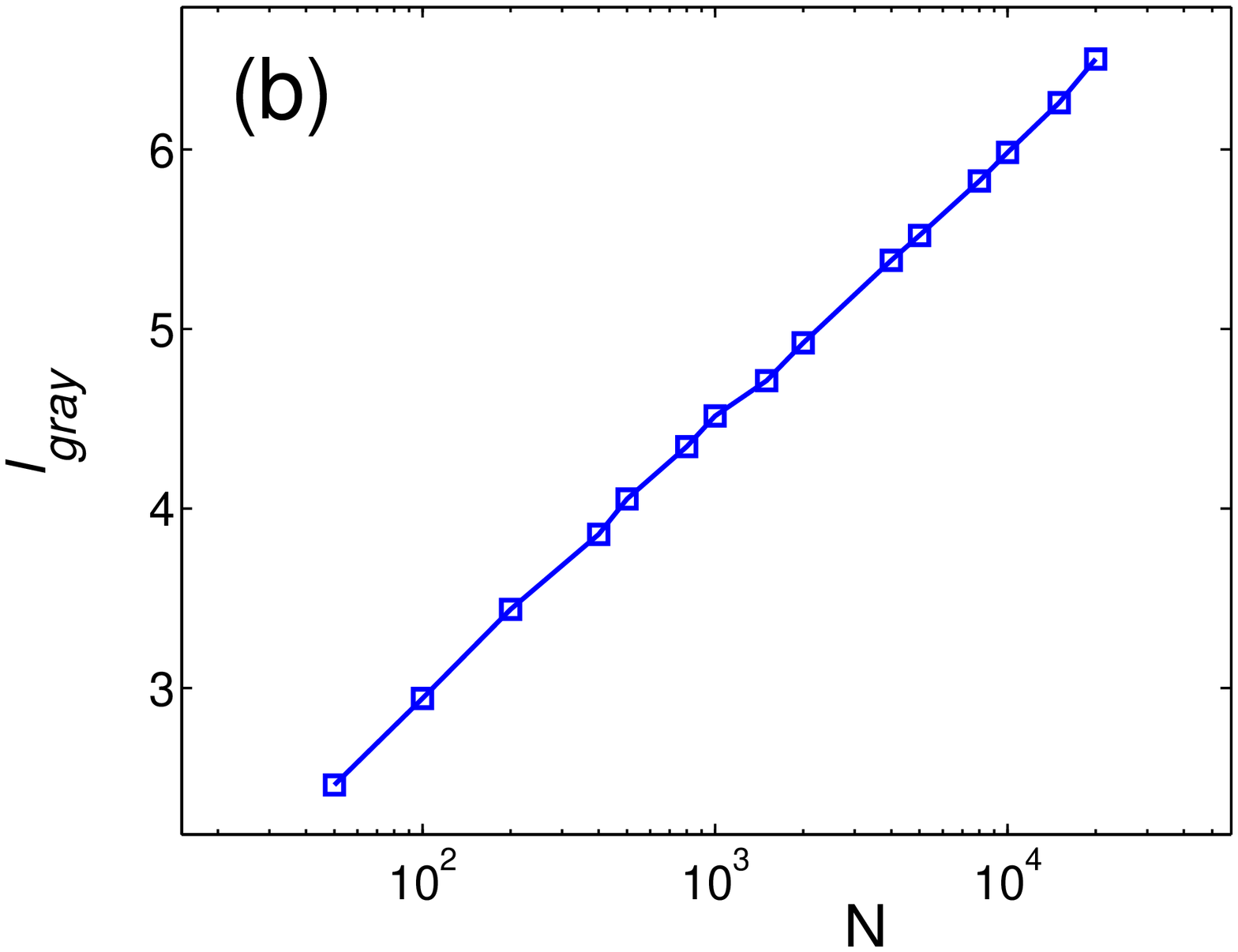}
    \caption{\label{fig:tree_degrees_radius} (a) The
      degree distribution of the ``gray tree'' (the MST of the
      supernode network, shown in
      Fig.~\ref{fig:gray_network_sketch}(d)), in which the supernodes
      are percolation clusters and the links are the gray links.
      Different symbols represent different threshold values: $p_c$
      ($\bigcirc$), $p_c+0.01$ ($\Box$) and $p_c+0.02$
      ($\bigtriangleup$). The distribution exhibits a scale-free tail
      with $\lambda\approx 2.5$, and is relatively insensitive to
      changes in $p_c$.
    (b) The average path length $\ell_{\rm gray}$ on a the gray tree as
      a function of original network size. It is seen that $\ell_{\rm
      gray}\sim\log N_{\rm sn}\sim\log N$ (After \cite{Tomer_grey}).\omitit{The same behavior can be seen for
      MST's on random scale-free graphs with $\lambda=2.5$ if we allow
      for double connections}}
\end{center}
  \end{figure}

From the MST of Fig.~\ref{fig:gray_network_sketch}(c) we now
generate a new tree, the MST of the supernode network, which we
call the ``gray tree'', whose nodes are the supernodes and whose
links are the gray links connecting them [see
Fig.~\ref{fig:gray_network_sketch}(d)]. Note that bombing the
original ER network to obtain the MST of
Fig.~\ref{fig:gray_network_sketch}(c) is equivalent to bombing the
supernode network of Fig.~\ref{fig:gray_network_sketch}(b) to
obtain the gray tree, because the links inside the clusters are
not bombed.
%
We find [Fig~\ref{fig:tree_degrees_radius}(a)] that the gray tree has
also a scale-free degree distribution $P(k)$, with $\lambda=2.5$---the
same as the supernode network \cite{text3}.
We also find [Fig.~\ref{fig:tree_degrees_radius}(b)] the average path
length $\ell_{\rm gray}$ scales as $\ell_{\rm gray} \sim \log N_{\rm
  sn} \sim \log N$
\cite{Sameet04,text4}.
Note that even though the gray tree is scale-free, it is not
ultra-small~\cite{Cohen3}, since the length does not scale as $\log\log N$.

Next we show that the bombing optimization, which leads to the
MST, yields a significant separation between the weights of the
links inside the supernodes and the links connecting the
supernodes. As explained above, the MST is optimal in two senses:
(i) the total weight of all links is minimal (ii) any path between
any two nodes on the MST will encounter the smallest maximal
\textit{barrier} (weight) between these nodes. The last property
is common to many physical systems (e.g. the protein folding
network - see below).  Accordingly, we study the weights
encountered when traveling along a typical path on the MST.

\begin{figure}
  \begin{center}
    \includegraphics[width=6.0cm,height=6.0cm]{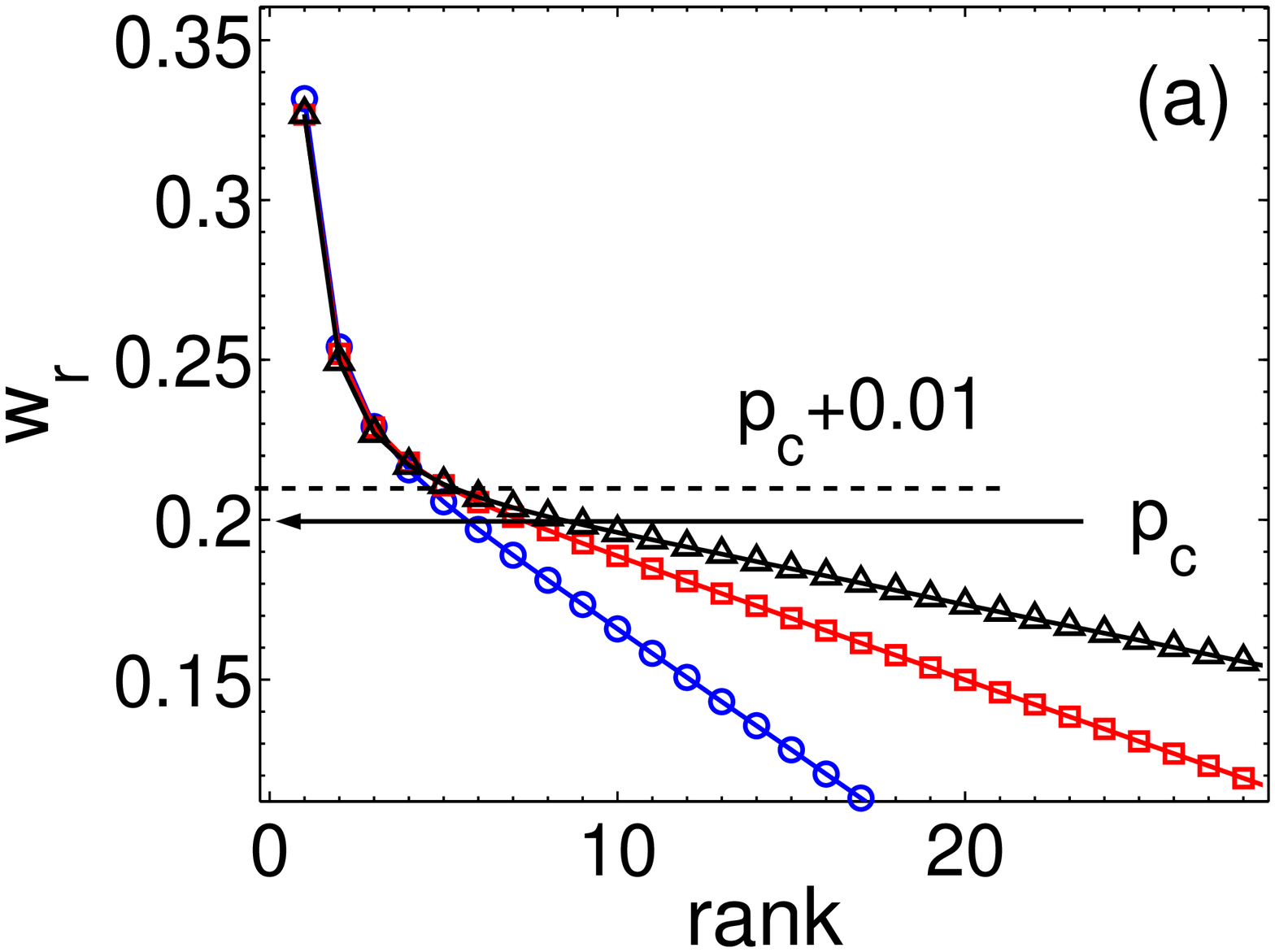}
    \includegraphics[width=6.0cm,height=6.0cm]{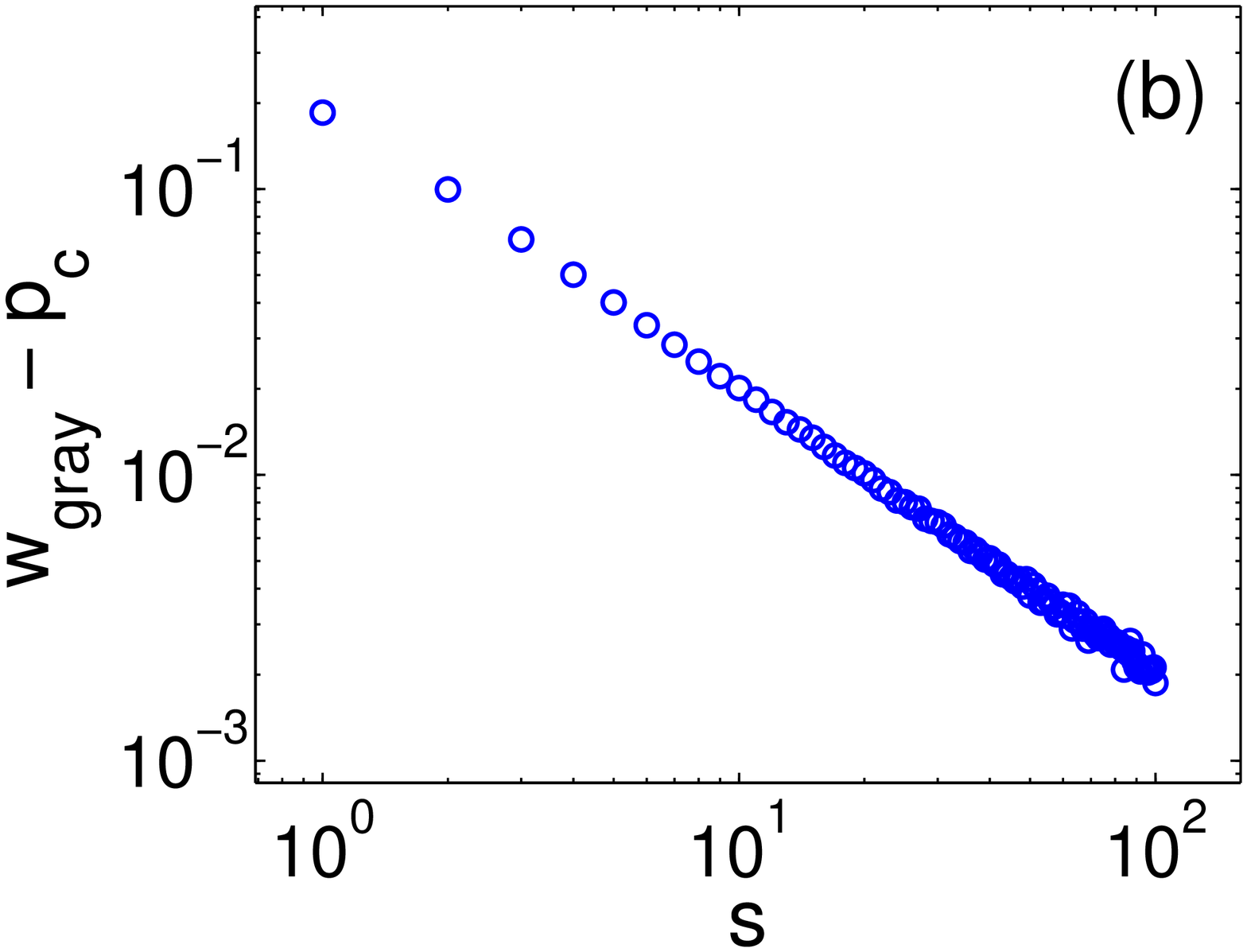}
    \caption{\label{fig:weights_optimal_path} (a) The average
      weights $w_r$ along the optimal path of an ER graph with
      $\av{k}=5$, sorted according to their rank. Different symbols
      represent different system sizes: $N = 2000$ ($\bigcirc$), $N =
      8000$ ($\Box$) and $N = 32000$ ($\bigtriangleup$). Below
      $p_c=0.2$, the weights are uniformly distributed, while weights
      above $p_c$ are significantly higher and independent of $N$. (b)
      Cluster size vs. the minimal gray link emerging from each
      cluster, for ER graphs with $\av{k}=5$ and $N=10000$. Small
      clusters are associated with higher weights because they have a
      small number of exits and thus cannot be optimized (After \cite{Tomer_grey}).}
  \end{center}
\end{figure}

We consider all pairs of nodes in the original MST of $N$ nodes
[Fig.~\ref{fig:gray_network_sketch}(c)] and calculate the typical path length $\ell_{\rm typ}$,
which is the most probable path length on the MST. For each path of
length $\ell_{\rm typ}$ we rank the weights on its links in descending
order. For the largest weights (``rank 1 links''), we calculate the
average weight $w_{r=1}$ over all paths. Similarly, for the next
largest weights (``rank 2 links'') we find the average $w_{r=2}$
over all paths, and so on up to $r=\ell_{\rm typ}$.
Fig.~\ref{fig:weights_optimal_path}(a) shows $w_r$ as a function
of rank $r$ for three different network sizes $N=2000$, $8000$, and
$32000$.  It can be seen that weights below $p_c$ (black links inside
the supernodes) are uniformly distributed and approach one another as
$N$ increases.  As opposed to this, weights above $p_c$ (``gray
links'') are {\it not\/} uniformly distributed, due to the bombing
algorithm, and are independent of $N$.
Actually, weights above $p_c$ encountered along the optimal path (such
as the largest weights $w_1$, $w_2$ and $w_3$) are significantly
higher than those below $p_c$.
Fig.~\ref{fig:weights_optimal_path}(b) shows that the links with the
highest weights on the MST can be associated with gray links from very
small clusters [Figs.~\ref{fig:gray_network_sketch}(a) and \ref{fig:gray_network_sketch}(c)] (similar results have been
obtained along the optimal path).

As mentioned earlier, this property is present also in the original supernode
network and hence the change in the threshold used to obtain the supernodes
does not destroy the power law degree distribution but only introduces an
exponential cutoff.  We thereby obtain a scale-free supernode network with
$\lambda=2.5$, which is not very sensitive to the precise value of the
threshold used for defining the supernodes.
For example, the scale-free degree distribution shown in
Fig.~\ref{fig:tree_degrees_radius}(a) for a threshold of $p_c+0.01$
corresponds to having only four largest weights on the optimal paths
[see Fig.~\ref{fig:weights_optimal_path}(a)]. However, even for
$p_c+0.02$ the degree distribution is well approximated by a
scale-free distribution with $\lambda=2.5$ [see
Fig.~\ref{fig:tree_degrees_radius}(a)]. This means that mainly very
small clusters, connected with high-weight links to large clusters,
dominate the scale-free distribution $P(k)$ of the MST of the
supernode network (gray tree).
Hence, the bombing optimization process on an ER graph causes a
significant separation between links below and above $p_c$ to emerge
{\it spontaneously\/} in the system, and by merging nodes connected
with links of low weights, a scale-free network can arise.

The process described above may be related to the evolution of some real
world networks. Consider a homogeneous network with many components whose
average degree $\langle k\rangle$ is well defined.  Suppose that the links
between the components have different weights, and that some optimization
process separates the network into nodes which are well connected (i.e.,
connected by links with low weights) and nodes connected by links having much
higher weights. If the well-connected components merge into a single node,
this results in a new heterogeneous supernode network with scale free degree
distribution.
\omitit{and that some {\em optimization} process eventually unites
  components that are strongly linked.  This results in a new
  heterogeneous network with components that vary in size, and thus in
  number of outgoing connections.}

An example of a real world network whose evolution may be related to
this model is the protein folding network, which was found to be
scale-free with $\lambda \approx 2.3$
\cite{rao-caflisch-2004:protein_folding}. The nodes are the possible
physical configurations of the system and the links between them
describe the possible transitions between the different
configurations. We assume that this network is {\em optimal\/} because
the system chooses the path with the smallest energy {\em barrier\/}
from all possible trajectories in phase space. It is possible that the
scale-free distribution evolves through a similar procedure as
described above for random graphs: adjacent configurations with close
energies (nodes in the same cluster) cannot be distinguished and are
regarded as a single supernode, while configurations (clusters) with
high barriers between them belong to different supernodes.

A second example is computer networks. Strongly interacting computers
(such as computers belonging to researchers from the same company or
research institution) are likely to converge into a single domain, and
thus domains with various sizes and connectivities are formed.  This
network might be also optimal, because packets destined to an external
domain are presumably routed through the router which has the best
connection to the target domain.

To summarize, we have seen that any weighted random network hides an
inherent scale-free ``supernode network'' \cite{text5}.
We showed that the minimum spanning tree, generated by the bombing
algorithm, is composed of percolation clusters connected by a scale-free
tree of ``gray'' links. Most of the gray links connect small clusters to
large ones, thus having weights well above the percolation threshold
that do not change with the original size of the network.
Thus the optimization in the process of building the MST distinguishes
between links with weights below and above the threshold, leading to a
spontaneous emergence of a scale-free ``supernode network''.
We raise the possibility that in some naturally optimal real-world
networks, nodes connected well merge into one single node, and thus a
scale-free network emerges.

\section{Partition of the minimum spanning tree into superhighways and roads}
\label{seq.shw}

The centrality, $C$, quantifies the ``importance'' of a node for transport in
the network. Moreover, identifying the nodes with high $C$ enables, as shown
below, to improve their transport capacity and thus improve the global
transport in the network. Several definitions of centrality exists. Here we
deal with the ``betweeness centrality'' which is defined as the relative
number of shortest path in the network passing through a node (or a link).
The probability density function (pdf) of $C$ was studied on the MST for both
scale-free (SF)~\cite{Barabasi_sf} and Erd\H{o}s-R\'{e}nyi
(ER)~\cite{ER59,ER60} networks and found to satisfy a power law,
\begin{equation}
  {\cal P}_{\rm MST}(C) \sim C^{-\delta_{\rm MST}}
\end{equation}
with $\delta_{\rm MST}$ close to $2$~\cite{Goh_centrality, Kim_bc}.  An
important question is weather there are substructure of the MST which are
more central and play a major role on the
transport. Reference~\cite{highway_prl} shows that a sub-network of the MST,
the infinite incipient percolation cluster (IIC) has a significantly higher
average $C$ than the entire MST~\cite{FN_IIC_isIn_MST} --- i.e., the set of
nodes inside the IIC are typically used by transport paths more often than
other nodes in the MST~\cite{highway_prl}. --- In this sense the IIC can be
viewed as a set of {\it superhighways} (SHW) in the MST. The nodes on the MST
which are not in the IIC are called {\it roads}, due to their analogy with
roads of less traffic (usually used by local residents). We
demonstrate the impact of this finding by showing that improving the capacity
of the superhighways (IIC) is significantly a better strategy to enhance
global transport compared to improving the same number of links with the
highest $C$ in the MST, although they have higher $C$~\cite{highway_prl}.
This counterintuitive result shows the advantage of identifying the IIC
subsystem, which is very small compared to the full
network~\cite{FN_iic_mst_mass_ratio}. These results are based on extensive
numerical studies for centrality of the IIC, and comparison with the
centrality of the entire MST~\cite{highway_prl,Newman_centrality}, as described below.
ER and SF network of size $N$ are generated by the methods explained in
section~\ref{sec.draw_net}. Multiple connections between two nodes and
self-loops in a single node are disallowed. To construct a {\it weighted}
network, a weight $w_i$ is assigned to each link from a uniform distribution
between $0$ and $1$. The MST is obtained from the weighted network using
Prim's algorithm~\cite{network_flow_book} (see Section~\ref{sec.mst}). Once
the MST is built, one can compute the value of $C$ of each node by counting the
number of paths between all possible pairs passing through that node and
normalize $C$ by the total number of pairs in the MST, $N(N-1)/2$, which
ensures that $C$ is between 0 and 1~\cite{FN_sd_mst}. The IIC of ER and SF
networks is simulated as explained in Section~\ref{sec.iic}.

\begin{figure}
  \includegraphics[width=\textwidth]{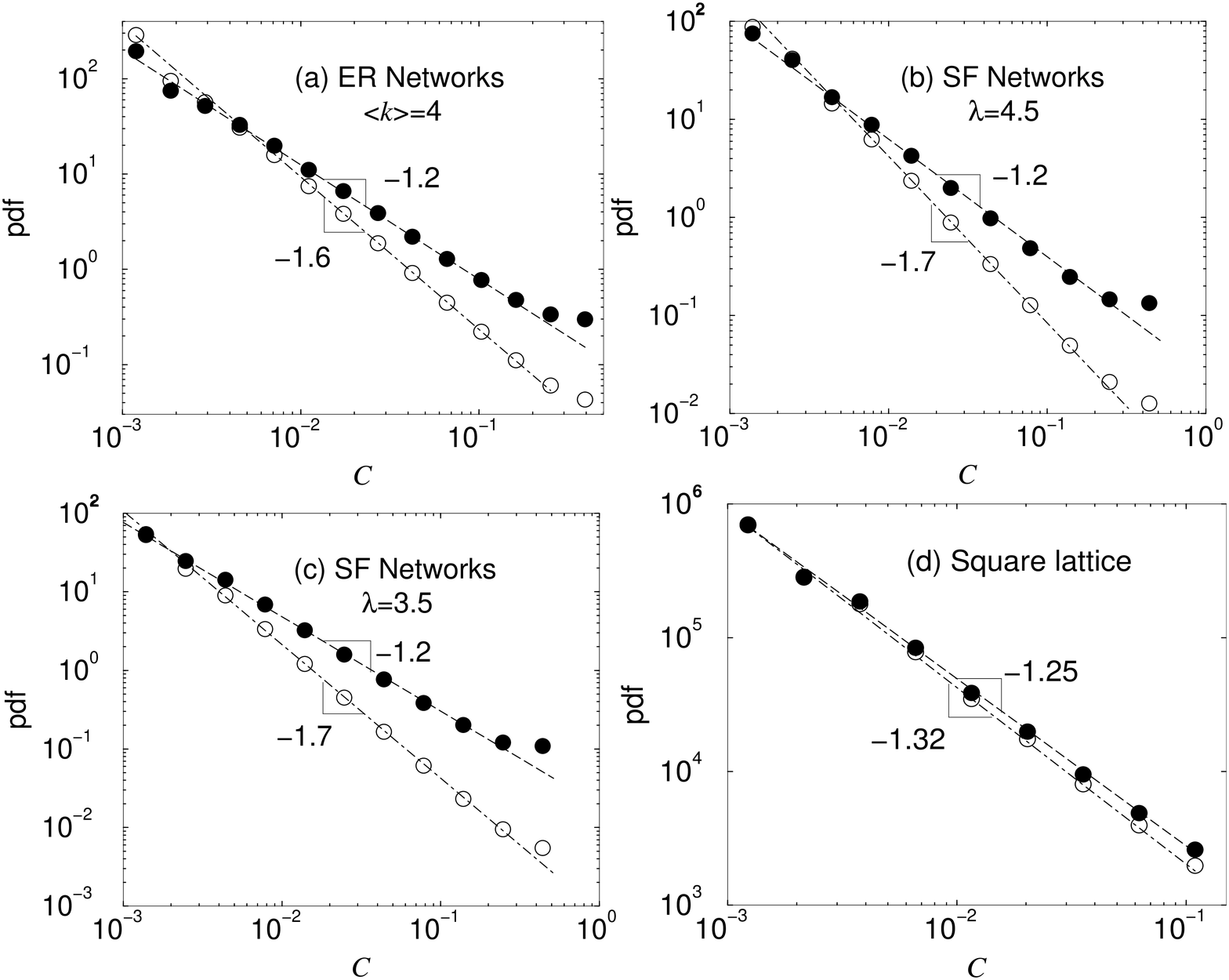}
  \caption{The pdf of the centrality $C$ of nodes for (a) ER graph with $\langle k
    \rangle = 4$, (b) SF with $\lambda = 4.5$, (c) SF with $\lambda = 3.5$
    and (d) $90 \times 90$ square lattice. For ER and SF, $N = 8192$ and for
    the square lattice $N = 8100$ . We analyze $10^4$ realizations. For each
    graph, the full circles show ${\cal P}_{\rm IIC}({\rm C})$; the
    empty circles show ${\cal P}_{\rm MST}({\rm C})$ (After \cite{highway_prl}).}
  \label{graph_Pcent_node}
\end{figure}

To quantitatively study the centrality of the nodes in the IIC, we calculate
the pdf, ${\cal P}_{\rm IIC}(C)$ of $C$.  Figure~\ref{graph_Pcent_node} shows
 that for all three cases studied, ER, SF and square lattice
networks, ${\cal P}_{\rm IIC}(C)$ for nodes satisfies a power law
\begin{equation}
  {\cal P}_{\rm IIC}(C) \sim C^{-\delta_{\rm IIC}},
  \label{BC_IIC_scaling}
\end{equation}
where
\begin{equation}
  \delta_{\rm IIC} \approx \left\{ \begin{array}{ll}
    1.2  & {\rm [ER, SF] } \\
    1.25 & {\rm [square~lattice]}
  \end{array}\right..
\end{equation}

Moreover, from Fig.~\ref{graph_Pcent_node}, it is seen that $\delta_{\rm
  IIC} < \delta_{\rm MST}$, implying a larger probability to find a larger
value of $C$ in the IIC compared to the entire MST. The values for
$\delta_{\rm MST}$ are consistent with those found in
Ref.~\cite{Goh_centrality}. Similar results for the centrality of the links
were obtained. The results thus show that the IIC is like a network of {\it
  superhighways} inside the MST. When we analyze centrality of the entire
MST, the effect of the high $C$ of the IIC is not seen since the IIC is only
a tiny fraction of the MST. Some results are summarized in
Table~\ref{table_para}.

The values of $\delta_{\rm MST}$ and $\delta_{\rm IIC}$ can be understood
from the following scaling arguments, based self-similarity properties of the
MST and the IIC. Similar arguments are used in \cite{stauffer,BH96} to
express exponent $\tau_s$ describing the cluster size distribution at
percolation threshold on a lattice in terms of the cluster fractal dimension
and the dimension of the lattice.  Indeed, the majority of nodes are
connected through the superhighway links, whose centrality is proportional to
$N^2$. The number of these links for the entire network scales as
$\ell_\infty(N) \sim N^{1/\nu_{opt}}$ \cite{Brauns03,zhenhua}. Thus small
regions of the MST of chemical diameter $\ell$ consists of $n^{1/\nu_{\rm
opt}}$ nodes. These regions have length $\ell_\infty(n) \sim n^{1/\nu_{\rm
opt}}$. These roads connect the nodes in this region with the rest of the
nodes of the MST, and thus their centrality is at least $nN$. The total number of such
links in all the regions of size $n$ is $m(n)=\ell_\infty(n)N/n\sim
n^{\nu{_{\rm opt}-1}}$. This is the number of links with centrality larger
than $nN$. Thus number of links with centrality exactly $nN$ is
$m(n)-m(n+1)\sim n^{\nu{_{\rm opt}-2}}=n^{-\delta_{\rm MST}}$. Using Eq.
(\ref{eq:nu_opt}) we have:
\begin{equation}
\delta_{\rm MST} =
\cases {
    \begin{array}{rll}
        5/3,  & \lambda>4, & {\rm ER} \\
        (\lambda+1)/(\lambda-1),  &  3<\lambda \leq 4 &
    \end{array}
    }\,.
\label{eq:delta_MST}
\end{equation}
Similar arguments lead to the centrality distribution of the nodes on the
IIC. The small regions of chemical diameter $\ell$ have centrality larger or
equal to $Nn(\ell)$. The number of links in the IIC belonging to these
regions is $s(\ell)=\ell^{d_\ell}$ \cite{Cohen}. The total number of such
regions is $S/s(\ell)$, thus the total number of the links of the IIC with
centrality larger than $Nn(\ell)$ is $m[n(\ell)]=\ell/s(\ell)\sim
\ell^{1-d_{\ell}}=n^{(1-d_{\ell})\;\nu_{\rm opt}}$. Accordingly
\begin{equation}
\delta_{\rm IIC} =1+(d_{\ell}-1)\nu_{\rm opt}=
\cases {
    \begin{array}{rll}
        4/3,  & \lambda>4, & {\rm ER} \\
        \lambda/(\lambda-1),  &  3<\lambda \leq 4 &
    \end{array}
    }\,.
\label{eq:delta_IIC}
\end{equation}
The values predicted by Eqs. (\ref{eq:delta_MST}) and (\ref{eq:delta_IIC}) are in good agreement with
the simulation results presented in Table~\ref{table_para}.

To further demonstrate the significance of the IIC, we compute the average  $\langle {\rm C}
\rangle$ for each realization of the network over all nodes. Fig.~\ref{graph_ave_bc_node},
shows the histograms of $\langle {\rm C} \rangle$ for both the IIC and for
the other nodes on the MST. We see that the nodes on the IIC have
significantly larger $\langle {\rm C}
\rangle$ compared to the other nodes of the MST.

\begin{table}[!h]
  \begin{tabular}{|c|c|c|c|c|}
    \hline                       & ER  & SF ($\lambda = 4.5$) & SF ($\lambda = 3.5$) & square lattice \\
    \hline $\delta_{\rm IIC}$    & 1.2 & 1.2 & 1.2 & 1.25 \\
    \hline $\delta_{\rm MST}$    & 1.6 & 1.7 & 1.7 & 1.32 \\
    \hline $\nu_{\rm opt}$       & 1/3 & 1/3 & 0.2 & 0.61 \\
    \hline $\langle u \rangle$   & $0.29$ & $0.20$ & $0.13$ & $0.64$ \\
    \hline
  \end{tabular}
  \caption{Results for the IIC and the MST (After~\cite{highway_prl}).}
  \label{table_para}
\end{table}

\begin{figure}[h]
  \includegraphics[width=0.5\textwidth]{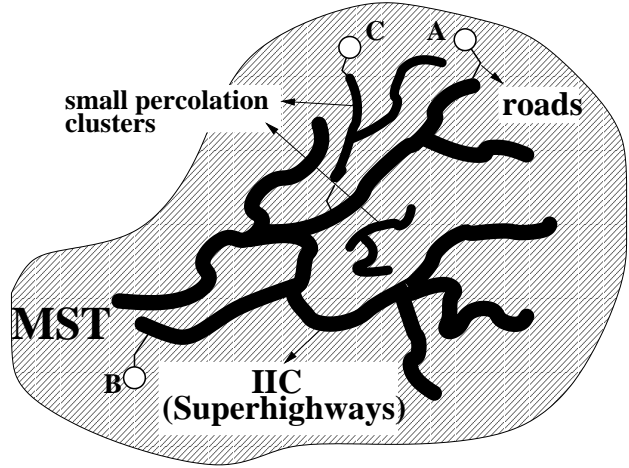}
  \caption{Schematic graph of the network of connected superhighways (heavy
    lines) inside the MST (shaded). A, B and C are examples of possible entry
    and exit nodes, which connect to the network of superhighways by
    ``roads'' (thin lines). The middle size lines indicate other percolation
    clusters with much smaller size compared to the IIC  (After \cite{highway_prl}).}
  \label{graph_schematic_of_hw}
\end{figure}

\begin{figure}[h]
  \includegraphics[width=10.0cm,height=10.0cm]{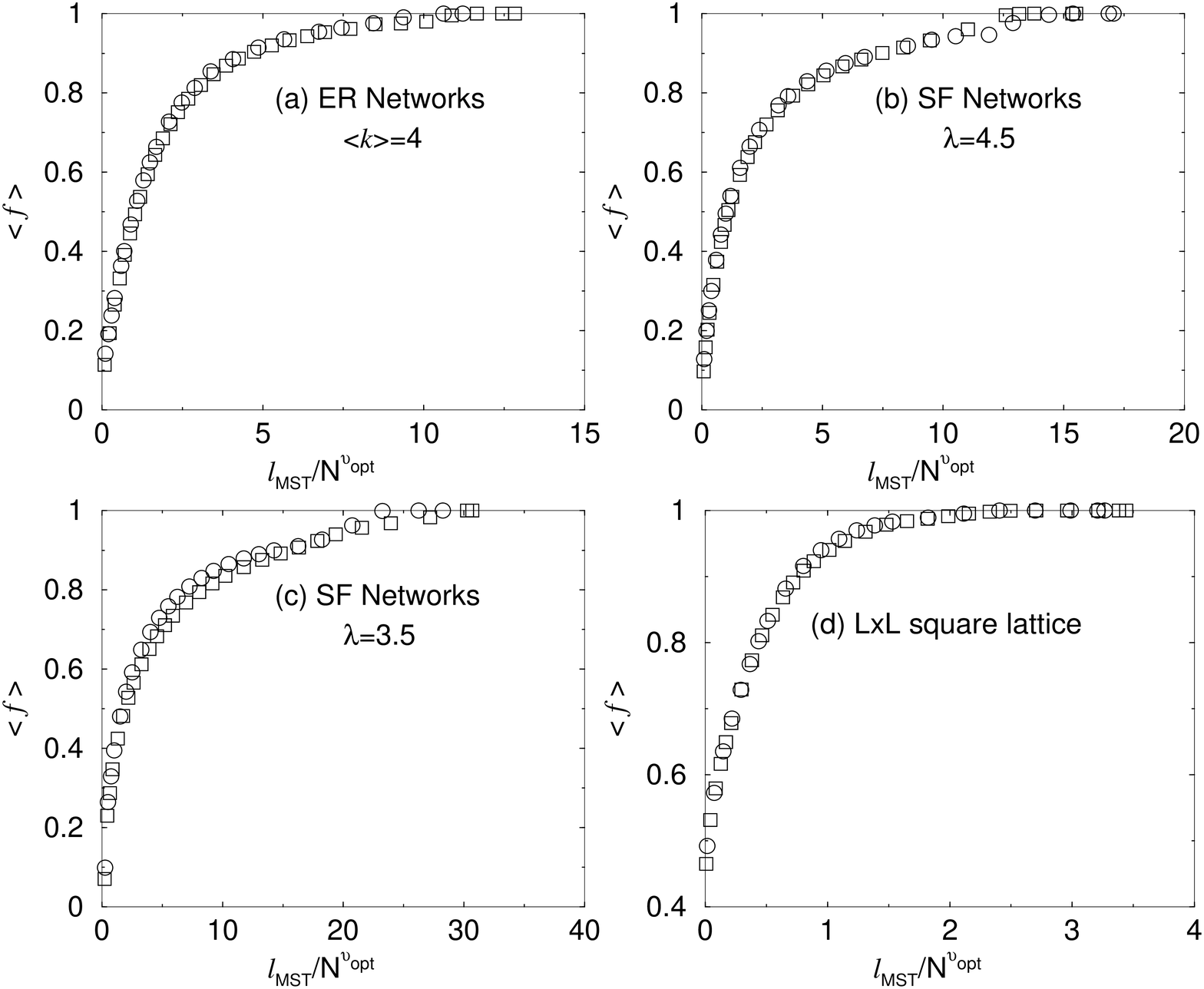}
  \caption{The average fraction, $\langle f \rangle$, of pairs using the SHW,
    as a function of $\ell_{\rm MST}$, the distance between the pair on the
    {\rm MST}. (a) ER graph with $\langle k \rangle = 4$, (b) SF with
    $\lambda = 4.5$, (c) SF with $\lambda = 3.5$ and (d) square lattice. For
    ER and SF: ($\bigcirc$)$N = 1024$ and ($\Box$)$N = 2048$ with $10^4$
    realizations. For square lattice: ($\bigcirc$)$N = 1024$ and ($\Box$)$N =
    2500$ with $10^3$ realizations. The $x$ axis is rescaled by $N^{\nu_{\rm
        opt}}$, where $\nu_{\rm opt} = 1/3$ for ER and for SF with $\lambda >
    4$, and $\nu_{\rm opt} = (\lambda - 3)/\lambda -1)$ for SF networks with
    $3 < \lambda < 4$~\cite{Brauns03}. For the $L \times L$ square lattice,
    $\ell_{\rm MST} \sim L^{d_{\rm opt}}$ with $d_{\rm opt} = 1.22$ and since
    $L^2 = N$, $\nu_{\rm opt} = d_{\rm opt} / 2 \approx 0.61$~\cite{Cieplak,
      Porto99}  (After \cite{highway_prl}).}
  \label{graph_ave_frac}
\end{figure}

\begin{figure}[h]
  \includegraphics[width=10.0cm,height=10.0cm]{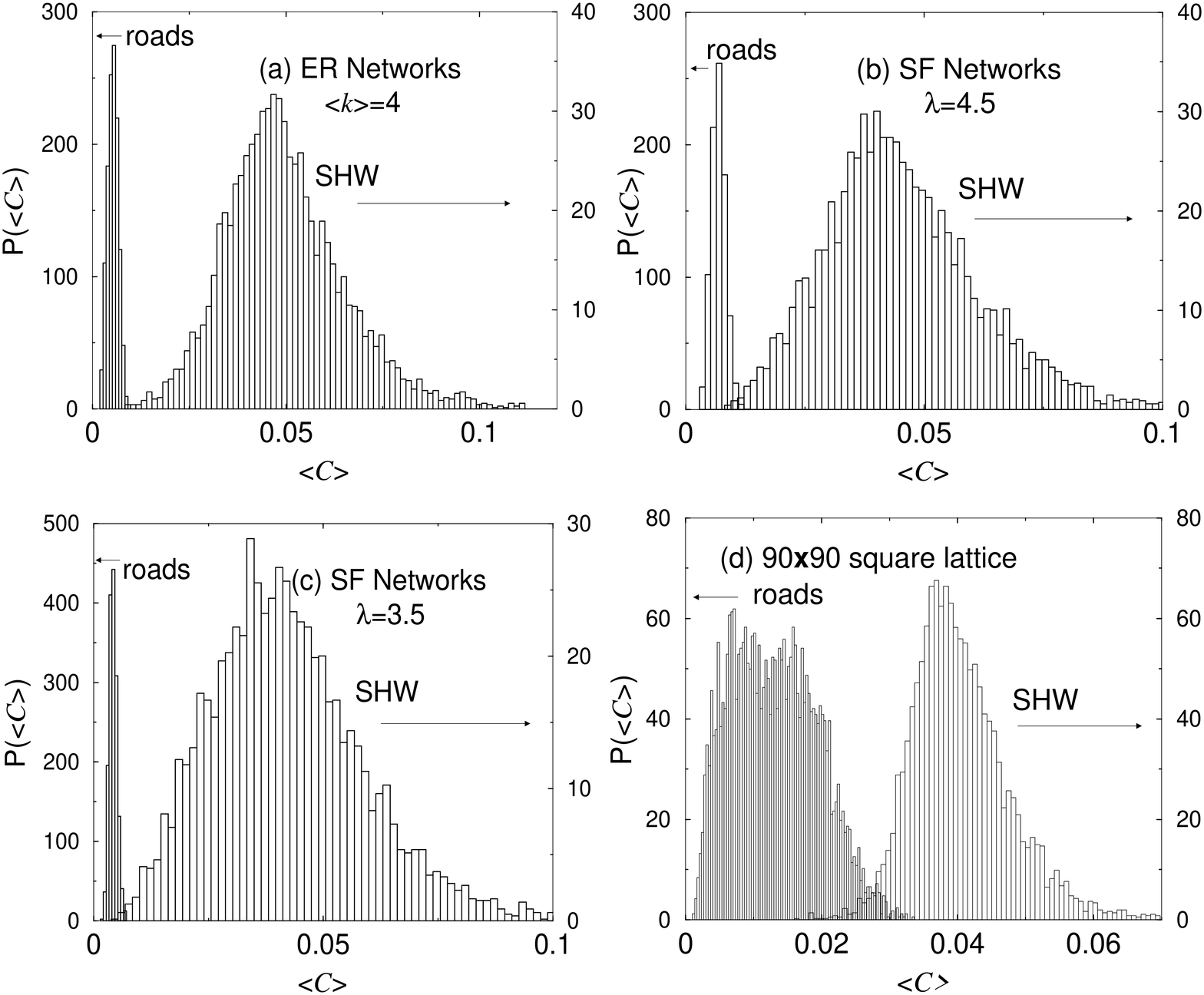}
  \caption{The normalized pdf for superhighway and roads of $\langle {\rm C}
    \rangle$, the $C$ averaged over all nodes in one realization. (a) ER
    network, (b) SF network with $\lambda = 4.5$, (c) SF network with
    $\lambda = 3.5$ and (d) square lattice network. To make each histogram,
    we analyze 1000 network configurations  (After \cite{highway_prl}).}
  \label{graph_ave_bc_node}
\end{figure}

Figure~\ref{graph_schematic_of_hw} shows a schematic plot of the SHW inside
the MST and demonstrates its use by the path between pairs of nodes. The MST
is a ``skeleton'' subset of links inside the network, which plays a key role
in transport between the nodes. However, the IIC in the MST is like the
``spine in the skeleton'', which plays the role of the superhighways inside a
road transportation system. To illustrate our result a car can drive from the
entry node ${\rm A}$ on roads until it reaches a superhighway, and finds the
exit which is closest to node ${\rm B}$. Thus those nodes which are far from
each other in the MST use the IIC superhighways more than those nodes which
are close to each other. In order to demonstrate this, we compute $f$, the
average fraction of pairs of nodes using by the shortest paths the {\rm IIC},
as a function of $\ell_{\rm MST}$, the distance between a pair of nodes on
the MST (Fig.~\ref{graph_ave_frac}). We see that $f$ increases and approaches
one as $\ell_{\rm MST}$ grows. We also show that $f$ scales as $\ell_{\rm
MST} / N^{\nu_{\rm opt}}$ for different system sizes, where $\nu_{\rm opt}$
is the percolation connectedness exponent~\cite{Brauns03,zhenhua}.

\begin{figure}
  \includegraphics[width=\textwidth]{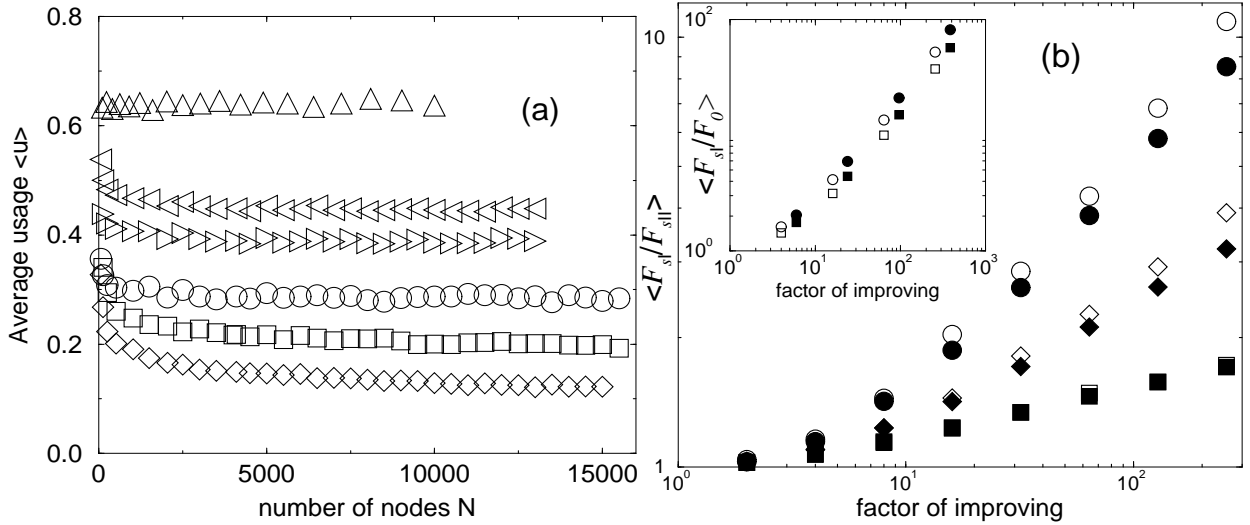}
  \caption{(a) The average usage $\langle u \rangle \equiv \langle \ell_{\rm
    IIC} / \ell_{\rm MST} \rangle$ for different networks, as a function of
    the number of nodes $N$. $\bigcirc$ (ER with $\langle k \rangle = 4$),
    $\Box$ (SF with $\lambda = 4.5$), $\Diamond$ (SF with $\lambda = 3.5$),
    $\bigtriangleup$ ($L \times L$ square lattice). The symbols ($\rhd$) and
    ($\lhd$) represent the average usage for ER with $\langle k \rangle = 4$
    when the two largest percolation clusters and the three largest
    percolation clusters are taken into account, respectively. (b) The ratio
    between the flow using strategy I, $F_{s\rm I}$,
    and that using strategy II, $F_{s\rm II}$, as a function of the
    factor of improving conductivity/capacity. The inset is
    the ratio between the flow using strategy I and the flow in the original
    network, $F_{\rm 0}$. The data are all for ER networks
    with $N = 2048$, $\langle k \rangle = 4$ and $n = 50$($\bigcirc$), $n =
    250$($\Diamond$) and $n = 500$($\Box$). The unfilled symbols are for
    current flow and the filled symbols are for maximum flow  (After \cite{highway_prl}).}
  \label{graph_ratio_and_flow}
\end{figure}

The next question is how much the IIC is used in transport on the MST? We
define the {\it IIC superhighway usage},
\begin{equation}
  u \equiv \frac{\ell_{\rm IIC}}{\ell_{\rm MST}},
\end{equation}
where $\ell_{\rm IIC}$ is the number of links in a given path of length
$\ell_{\rm MST}$ belonging to the {\rm IIC} superhighways. The average usage
$\langle u \rangle$ quantifies what fraction of nodes/links of the IIC is
used by the transport between all pairs of nodes.
In Fig.~\ref{graph_ratio_and_flow}(a), we show
$\langle u \rangle$ as a function of the system size $N$. Our results suggest
that $\langle u \rangle$ approaches a constant value and becomes independent
of $N$ for large $N$. This is surprising since the average value of the ratio
between the number of nodes on the IIC and on the MST, $\langle N_{\rm IIC} /
N_{\rm MST} \rangle$, approaches zero as $N \to
\infty$~\cite{FN_iic_mst_mass_ratio}, showing that although the IIC contains
only a tiny fraction of the nodes in the entire network, its usage for the
transport in the entire network is constant. We find that $\langle u \rangle
\approx 0.3$ for ER networks, $\langle u \rangle \approx 0.2$ for SF networks
with $\lambda = 4.5$, and $\langle u \rangle \approx 0.64$ for the square
lattice. The reason why $\langle u \rangle$ is not close to $1.0$ is that in
addition to the IIC, the optimal path passes also through other percolation
clusters, such as the second largest and the third largest percolation
clusters. In Fig.~\ref{graph_ratio_and_flow}, we also show for ER networks,
the average usage of the two largest and the three largest percolation
clusters for a path on the MST and we see that the average usage increases
significantly and is also independent of $N$. However, the number of clusters
used by a path on MST is relatively small and proportional to $\ln
N$~\cite{Sameet04}, suggesting that the path on the MST uses only few
percolation clusters and few jumps between them (of order $\ln N$)
when traveling from an entry node to an exit node on the network. When $N \to \infty$
the average usage of all percolation clusters should approach $1$.

Can we use the above results to improve the transport in networks? It is
clear that by improving the capacity or conductivity of the highest $C$ links
one can improve the transport (see Fig.~\ref{graph_ratio_and_flow}(b)
inset). We hypothesize that improving the IIC links (strategy I), which
represent the superhighways is more effective than improving the same number
of links with the highest $C$ in the MST (strategy II), although have
higher centrality~\cite{FN_highest_bc_IIC}. To test the hypothesis, we study
two transport problems: (i) current flow in random resistor networks, where
each link of the network represents a resistor and (ii) the maximum flow
problem well known in computer science~\cite{Cormen90}. We assign to
each link of the network a resistance/capacity, $e^{ax}$, where $x$ is an
uniform random number between 0 and 1, with $a = 40$. The value of $a$ is
chosen such as to have a broad distribution of disorder so that the MST
carries most of the flow~\cite{zhenhua, Sameet04}. We randomly choose $n$
pairs of nodes as sources and other $n$ nodes as sinks and compute the flow
between them. We compare the transport by improving the conductance/capacity
of all links on the IIC (strategy I) with that by improving the same number
of links with the highest $C$ in the MST (strategy II). Since the
two sets are not the same and therefore higher centrality links will be
improved in II~\cite{FN_highest_bc_IIC}, it is tempting to suggest that the
better strategy to improve the global flow is  strategy II. However, here we
demonstrate using ER networks as an example that counterintuitively strategy
I is better. We also find similar advantage of strategy I compared to
strategy II for SF networks with $\lambda = 3.5$.

In Fig.~\ref{graph_ratio_and_flow}(b), we compute the ratio between the flow
using strategy I ($F_{s\rm I}$) and the flow using strategy II ($F_{s\rm
II}$) as a function of the factor of improving conductivity/capacity of the
links. The figure clearly shows that strategy I is better than strategy
II. Since the number of links in the IIC is relatively very small comparing
to the number of links in the whole network~\cite{FN_iic_mst_mass_ratio}, it
could proven to be a very efficient strategy.

In summary, we find that the centrality of the IIC for transport in networks
is significantly larger than the centrality of the other nodes in the
MST. Thus the IIC is a key component for transport in the MST. We demonstrate
that improving the capacity/conductance of the links in the IIC is useful
strategy to improve transport.

\section{Summary}

We reviewed recent studies on the scaling of the average optimal path length
$\ell_{\mbox{\scriptsize opt}}$ in a disordered network.  There are two
scaling regimes of $\ell_{\mbox{\scriptsize opt}}$ corresponding to the
regimes of weak and strong disorder. For ER networks and SF networks with
$\lambda > 4$, $\ell_{\mbox{\scriptsize opt}} \sim \ln N$ in the weak
disorder regime while $\ell_{\mbox{\scriptsize opt}} \sim N^{1/3}$ in the
strong disorder regime.  For SF networks with $3 < \lambda < 4$,
$\ell_{\mbox{\scriptsize opt}} \sim \ln N$ in the weak disorder regime while
$\ell_{\mbox{\scriptsize opt}} \sim N^{\frac{\lambda - 3}{\lambda -1 }}$ in
the strong disorder regime. For SF networks with $2 < \lambda < 3$,
$\ell_{\mbox{\scriptsize opt}} \sim \ln N $ in the weak disorder regime while
$\ell_{\mbox{\scriptsize opt}} \sim \ln^{\lambda -1 } N$ in the strong
disorder regime.  The scaling behavior of $\ell_{opt}$ in the strong disorder
regime for ER and SF networks with $\lambda > 3$ is obtained analytically
using percolation theory \cite{exp1}.  For exponential disorder, for both ER
random networks and SF networks we obtain a scaling function for the
crossover from weak disorder characteristics to strong disorder
characteristics. We show that the crossover occurs when the min-max path
reaches a crossover length $\ell^*(a)$ and $\ell^*(a)\sim a$. Equivalently,
the crossover occurs when the network size $N$ reaches a crossover size
$N^*(a)$, where $N^*(a)\sim a^3$ for ER networks and for SF networks with
$\lambda \ge 4$ and $N^*(a)\sim a^{\frac{\lambda - 1}{\lambda -3}}$ for SF
networks with $3 < \lambda < 4$.

We also have shown that the optimal path length distribution in weighted
random graphs has a universal scaling form according to
Eq.~(\ref{equ:weighted_dist}). We explain this behavior and demonstrate the
transition between polynomial to logarithmic behavior of the average optimal
path in a single graph.

Our results are consistent with results found for finite dimensional
systems~\cite{Porto99, zhenhua,
  strelniker-havlin-berkovits-frydman-2005:Tomography, perlsman}: In finite
dimension the parameter controlling the transition is
$\frac{L^{1/\nu}}{ap_c}$, where $L$ is the system length and $\nu$ is the
correlation length critical exponent. This is because only the ``red
bonds'' - bonds that if cut would disconnect the percolation
cluster~\cite{coniglio-1982:cluster_structure} - control the transition.

We also show that any weighted random network hides an
inherent scale-free ``supernode network'' \cite{text5}.
We showed that the minimum spanning tree, generated by the bombing
algorithm, is composed of percolation clusters connected by a scale-free
tree of ``gray'' links. Most of the gray links connect small clusters to
large ones, thus having weights well above the percolation threshold
that do not change with the size of the network.
Thus the optimization in the process of building the MST distinguishes
between links with weights below and above the threshold, leading to a
spontaneous emergence of a scale-free ``supernode network'' with $\lambda =
2.5$.
We raise the possibility that in some naturally optimal real-world
networks, nodes connected well merge into one single node, and thus a
scale-free network emerges.

The centrality in networks for transport on the MST is studied.  We found that
the centrality of the nodes in the IIC is significantly larger than the
centrality of the other nodes in the MST.  The analytical estimation for the
exponents of the centrality distribution for both the MST and the IIC are
provided. Thus the IIC is a key component for transport in the MST. As a
result of this finding, we demonstrated that improving the
capacity/conductance of the links in the IIC is a useful strategy to improve
transport which is a better strategy compare to improving the same number of
links with the highest centrality in the MST.  This is probably due to the
global nature of transport which prefer global improvement of the
superhighways rather than local improvement of high centrality links.

\subsubsection*{Acknowledgments}

We thank ONR, Israel Science Foundation, European NEST project DYSONET,
FONCyt (PICT-O 2004/370) and Israeli Center for Complexity Science for
financial support.


\begin{thebibliography}{99}
\bibitem{spt} In the literature of computer science it is called
``Shortest Path Tree''
\bibitem{exp1} The results for SF networks with $2 < \lambda < 3$
  have been obtained numerically and a theoretical explanation for these
  results is still pending.
\bibitem{exp_1} This is a consequence of the fact that for the
original network the clusters at percolation have sizes $s$
distributed as $P(s) \sim s^{-\tau}$ \cite{Cohen02}, (with $\tau =
2.5$ for ER networks and for SF networks with $\lambda \ge 4$, and
$\tau = (2\lambda-3)/(\lambda-2)$ for SF networks with $3 <
\lambda < 4$) and each node within this cluster has a non-zero
probability of connecting to a node outside the cluster.
\bibitem{FN_IIC_isIn_MST} The IIC contains loops in lattices in dimension
  $d$ below 6. However, for networks ($d = \infty$), in the IIC
  loops can be neglegted and in this case for large $N$ and therefore the IIC must be a subset of the
  MST. In our simulations, we  found that more than $99\%$
  links of IIC belong to MST. For lattices, we only choose the part of the
  IIC that belongs to the MST.
\bibitem{FN_sd_mst} This $C$ measurement is equivalent to counting the number
  of times a node (link) is used by the set of optimal paths linking all
  pairs of nodes, in the limit of strong disorder.
\bibitem{FN_highest_bc_IIC} The overlap between the two groups is about
  $30\%$ for ER networks of size $N= 8192$
\bibitem{footnote1} This criterion can be regarded as the definition of strong
  disorder
\bibitem{footnote_uncorr}By
  uncorrelated we mean that the weights are not correlated with the topology,
  such as the degree of nodes,
\bibitem{footnote} This is actually the bombing algorithm that determines the
  optimal path in strong disorder~\cite{Cieplak}.
\bibitem{text4} MST's on scale-free networks with $\lambda=2.5$ were found to retain the original network's
degree distribution.
\bibitem{text5} Similar results can also be obtained for graphs
  embedded in two or three dimensions, with different power law
  exponents.
\end{thebibliography}
\end{document}